\pdfoutput=1
\documentclass[journal=acscii,layout=twocolumn, manuscript=article]{achemso}

\usepackage[english]{babel}
\usepackage[utf8]{inputenc}
\usepackage[colorinlistoftodos, color=green!40, prependcaption]{todonotes}
\usepackage{hyperref}
\usepackage{multirow}
\usepackage{textgreek}
\usepackage{placeins}
\usepackage{float}

\usepackage{graphicx}
\usepackage{booktabs}
\usepackage{makecell}

\usepackage{array}
\usepackage{threeparttable}
\usepackage{ragged2e}
\usepackage{placeins}

\addto{\captionsenglish}{} 

\usepackage{amsthm}
\usepackage{mathtools}
\usepackage{physics}
\usepackage{xcolor}
\usepackage{graphicx}
\usepackage{graphics}
\usepackage[fontsize=10]{fontsize}
\usepackage{amsmath}
\usepackage{hyperref}

\newcommand{\mathspace}{\hspace*{0.07cm}}

\renewcommand{\arraystretch}{1.15}
 \SectionNumbersOn
 \usepackage{caption}
 
\author{Simon Axelrod}
    \email{simonaxelrod83@gmail.com}
    \altaffiliation{These authors contributed equally to this work.}
    \affiliation{Department of Chemistry and Chemical Biology, Harvard University, Cambridge, MA, 02138}
    \alsoaffiliation{Department of Materials Science and Engineering, Massachusetts Institute of Technology, Cambridge, MA, 02139}

\author{Miroslav Kašpar}
    \email{dr.mirakaspar@gmail.com}
    \altaffiliation{These authors contributed equally to this work.}
    \affiliation{Institute of Organic Chemistry and Biochemistry of the Czech Academy of Sciences, Flemingovo nám. 2, 160 00 Prague 6, Czech Republic}

\author{Kristýna Jelínková}
    \affiliation{Institute of Organic Chemistry and Biochemistry of the Czech Academy of Sciences, Flemingovo nám. 2, 160 00 Prague 6, Czech Republic}

\author{Markéta Šmídková}
    \affiliation{Institute of Organic Chemistry and Biochemistry of the Czech Academy of Sciences, Flemingovo nám. 2, 160 00 Prague 6, Czech Republic}

\author{Erika Bartůňková}
    \affiliation{Institute of Organic Chemistry and Biochemistry of the Czech Academy of Sciences, Flemingovo nám. 2, 160 00 Prague 6, Czech Republic}

\author{Sille Štěpánová}
    \affiliation{Institute of Organic Chemistry and Biochemistry of the Czech Academy of Sciences, Flemingovo nám. 2, 160 00 Prague 6, Czech Republic}

\author{Eugene Shakhnovich}
    \affiliation{Department of Chemistry and Chemical Biology, Harvard University, Cambridge, MA, 02138}

\author{Václav Kašička}
    \affiliation{Institute of Organic Chemistry and Biochemistry of the Czech Academy of Sciences, Flemingovo nám. 2, 160 00 Prague 6, Czech Republic}

\author{Martin Dračínský}
    \affiliation{Institute of Organic Chemistry and Biochemistry of the Czech Academy of Sciences, Flemingovo nám. 2, 160 00 Prague 6, Czech Republic}

\author{Zlatko Janeba}
    \affiliation{Institute of Organic Chemistry and Biochemistry of the Czech Academy of Sciences, Flemingovo nám. 2, 160 00 Prague 6, Czech Republic}
    
\author{Rafael Gómez-Bombarelli}
    \affiliation{Department of Materials Science and Engineering, Massachusetts Institute of Technology, Cambridge, MA, 02139}

\date{\today} 

\newcommand{\PaperTitle}{Computational Design and Experimental Validation of Photoactive PARP1 Inhibitors}
\title[]{\PaperTitle}

\begin{document}

\begin{abstract}
Light-activated drugs are a promising way to treat localized diseases for which existing treatments have severe side effects. However, their development is complicated by the set of photophysical and biological properties that must be simultaneously optimized. Here we used computational techniques to find a set of promising candidates for the photoactive inhibition of the poly(ADP-ribose) polymerase 1 (PARP1) cancer target. Using our recently developed methods based on atomistic simulation and machine learning (ML), we screened a set of 5 million hypothetical photoactive ligands. Our workflow used protein-ligand docking to identify candidates with differential PARP1 binding under light and dark conditions; ML force fields and quantum chemistry calculations to predict p$K_\mathrm{a}$, absorption spectra, and thermal half-lives; graph-based surrogate models to screen additional compounds; excited-state nonadiabatic dynamics with ML force fields to estimate quantum yields; and free energy perturbation (FEP) to refine binding predictions. From these predictions, we prioritized a small set of synthetically feasible candidates expected to have red-shifted absorption spectra, thermal half-lives on the order of seconds to minutes, and isomer-dependent PARP1 binding under visible-light control. We synthesized 10 candidates and experimentally characterized their photobehavior and PARP1 inhibition constants. Among the validated compounds, \textbf{1} showed a 15-fold increase in inhibition of PARP1 upon green-light irradiation at 519 nm (208.8 $\pm$ 28.3 $\mu$M vs 14.4 $\pm$ 1.9 $\mu$M). These results validate the computation-guided screening strategy for identifying red-shifted PARP1 photoinhibitors, while also underscoring current limitations such as rapid thermal relaxation in aqueous media.

\end{abstract}
\maketitle

\section*{Introduction}
\begin{figure*}[t!]
    \centering
    \includegraphics[width=\textwidth]{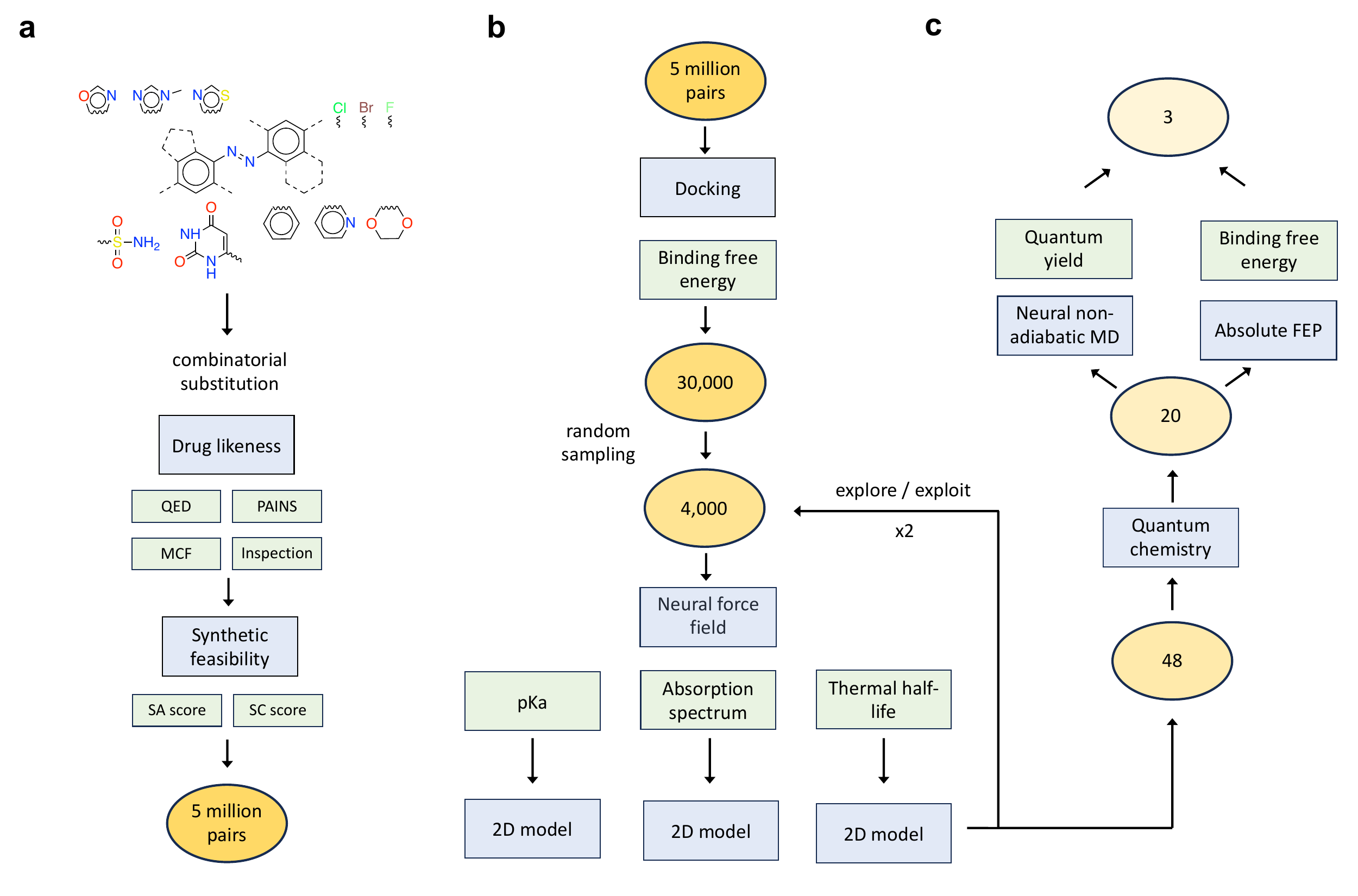}
    \caption{Virtual screening workflow. Molecular libraries are shown in yellow ovals, calculation types in blue rectangles, and computed properties in green rectangles. (a) Molecules were generated through combinatorial substitution and filtered based on drug likeness and synthetic feasibility. QED: quantitative estimate of drug-likeness \cite{bickerton2012quantifying}; PAINS: pan-assay interfering compounds \cite{capuzzi2017phantom, senger2016filtering}; MCFs: medicinal chemistry filters \cite{polykovskiy2020molecular}; SA score: synthetic accessibility score \cite{ertl2009estimation}; SC score: synthetic complexity score \cite{coley2018scscore}. (b) 5 million \textit{cis}-\textit{trans} pairs (10 million molecules) were docked into PARP1, and 30,000 were selected based on these docking scores. Next, 4,000 pairs were randomly chosen to receive thermal half-life and absorption spectrum calculations. These calculations involved atomistic simulations using our neural network force field (NFF). Some also received p$K_\mathrm{a}$ and azonium half-life calculations. These results were used to train a fast 2D model, which was in turn used to select 4,000 new molecules based on favorable properties (exploitation) or uncertainty (exploration). (c) Compounds with the best chemical and photophysical properties were selected for further validation. The predicted p$K_\mathrm{a}$ values, absorption spectra, and thermal half-lives were validated with quantum chemistry. The remaining compounds were tested for strong differential binding through absolute free energy perturbation (FEP) calculations, and non-zero quantum yields through NFF simulations of excited-state molecular dynamics (MD). }
    \label{fig:workflow}
\end{figure*}

Photopharmacology is an emerging field based on the optical control of drug activity \cite{lerch2016emerging, broichhagen2015roadmap}. By activating a drug at a specific location or time, one can minimize off-target effects and increase the maximum deliverable dose, thereby improving quality of life \cite{lerch2016emerging}. These therapeutics have been proposed as a treatment modality for cancer \cite{szymanski2015light} and blindness \cite{bonardi2010light}, among other conditions \cite{lerch2016emerging}. They are typically built around photoswitches such as azobenzene (Fig. \ref{fig:workflow}(a)), which undergoes \textit{trans} $\leftrightarrow$ \textit{cis} isomerism in response to light. Because bioactivity depends strongly on ligand shape, the large structural change induced by azobenzene photoisomerization can alter the drug’s biological effect. 

Despite its promise, photopharmacology is hindered by a difficult multi-objective optimization problem. Photoactive drugs should ideally absorb light in the near-infrared (IR) range between 700 and 900 nm \cite{dong2017near}, since only light in this range can penetrate human tissue. Yet unsubstituted \textit{trans} azobenzene absorbs in the UV and visible range \cite{knie2014ortho}. The thermal lifetime of the active drug, typically the \textit{cis} isomer generated by excitation of the \textit{trans} form, should be long enough to allow accumulation and the onset of a biological effect. Yet it should be short enough to prevent transport through the bloodstream and hence biological activity elsewhere \cite{dong2017near}. Lastly, the active isomer should exhibit strong biological activity, whereas the inactive isomer should exhibit weak or no activity.

Further, the numerous properties that must be optimized for standard pharmaceuticals also need to be optimized for photoactive ones. These include solubility, permeability, metabolic stability, and toxicity. Balancing these objectives leads to constraints on orally absorbed drugs, most famously quantified by Lipinski's rule of five \cite{lipinski2012experimental}.
This further complicates the design of light-activated drugs. For example, constraints on molecular mass and the number of aromatic rings \cite{bickerton2012quantifying} limit simple strategies for red-shifting azobenzene absorption. 

In this work, we tackle this multi-objective optimization problem in photopharmacology by focusing on photoswitchable inhibition of poly(ADP-ribose) polymerase 1 (PARP1), a cancer-relevant target that we previously predicted to be photodruggable (i.e., a protein for which \textit{cis} azobenzene derivatives bind more strongly than \textit{trans}) \cite{axelrod2023mapping}. Cancer therapy is one of the most promising applications for photopharmacology, because the side effects of many existing treatments are severe \cite{lerch2016emerging}.

To identify optimized photoactive inhibitors of PARP1, we applied computational methods of increasing accuracy and decreasing speed to a large virtual library of azobenzene derivatives. Our methods combined computational docking, machine learning (ML), quantum chemistry, and atomistic simulation to predict properties of millions of azobenzene derivatives. We identified promising compounds with red-shifted absorption spectra, thermal half-lives in the range of seconds to minutes, and strong differential binding affinities. 
We then synthesized a small set of compounds and measured these properties experimentally.

\section*{Results and discussion}

\subsection*{Virtual screening}

To generate a large virtual library for screening, we substituted azobenzene with common literature functional groups and patterns (Fig. \ref{fig:workflow}(a) and Computational Supplementary Information (CSI) Figs. \ref{fig:fused_rings}-\ref{fig:dioxane_iteration}). We then filtered these molecules by drug likeness and synthetic feasibility. Filters consisted of the quantitative estimate of drug likeness (QED) \cite{bickerton2012quantifying}, medicinal chemistry filters (MCFs) \cite{polykovskiy2020molecular}, removal of non-drug-like groups through inspection, and removal of pan-assay interfering compounds (PAINS) \cite{capuzzi2017phantom, senger2016filtering}. Synthetic feasibility was measured through the synthetic accessibility and complexity scores (SA and SC scores, respectively) \cite{ertl2009estimation, coley2018scscore}. This resulted in a library of 5 million \textit{cis}-\textit{trans} pairs. Further details of this process are given in CSI Sec. \ref{sec:molgen}.

We then filtered the library based on computed chemical, photophysical, and biological properties. We used the approach of a computational funnel, reducing the library size with calculations of increasing accuracy and decreasing speed (Fig. \ref{fig:workflow}(b)). We began with protein-ligand docking to predict the binding affinity of each isomer for each compound. This was the fastest but least accurate type of calculation. We used AutoDock Vina \cite{trott2010autodock} to dock all 10 million molecules into PARP1. We kept \textit{cis}-\textit{trans} pairs for which the \textit{cis} isomer was in the best 15\% of scores, and the \textit{trans} isomer was in the worst 50\%. This left 30,000 pairs.

Next we computed chemical and photophysical properties for 12,000 \textit{cis}-\textit{trans} pairs. We did so using a set of ML and simulation tools that we recently developed \cite{axelrod2022excited,axelrod2022thermal,axelrod2023mapping}. We used a neural network force field (NFF) trained on quantum chemical data to compute \textit{cis} thermal half-lives and \textit{trans} absorption spectra \cite{axelrod2022thermal}. NFFs are orders of magnitude faster than quantum chemistry, and hence can be used to screen far more candidates. 

We were also interested in the spontaneous protonation of azobenzene to form azonium in water. Protonation of one of the azo nitrogens leads to red-shifted absorption spectra and shorter half-lives \cite{dong2015red, dong2017near}. Appropriate substitution of azobenzene can lead to \textit{trans} azonium formation at physiological pH, and hence to the desired red-shift \cite{dong2015red, dong2017near}. We therefore computed the azonium p$K_\mathrm{a}$ for compounds with \textit{ortho} hydrogen bond acceptors, since these can stabilize the hydrogen atom and hence elevate the p$K_\mathrm{a}$ \cite{dong2015red, dong2017near}. This was combined with thermal half-life and absorption spectrum calculations for the azonium species.

These calculations involved molecular dynamics simulations and millions of geometry optimizations. This made them rather costly, even with an NFF. We therefore could not use them to screen all 30,000 isomer pairs. Instead we screened a portion of the library, and used the resulting data to train models that predicted these properties from the 2D molecular graph. We used these fast 2D models to screen the entire library and identify the most promising compounds for further calculation.

Compounds with thermal half-lives between seconds and minutes, and either red-shifted \textit{trans} absorption spectra or high \textit{trans} p$K_\mathrm{a}$ values, were further validated with quantum chemistry. The remaining species received the most intensive calculations. We used an NFF to perform excited-state, non-adiabatic molecular dynamics (NAMD) to check that each species had a non-zero photoisomerization quantum yield \cite{axelrod2022excited}. We also performed absolute free energy perturbation (FEP) calculations to compute the binding affinity of each compound to PARP1. These were the most computationally demanding calculations, taking five days per compound with four graphics processing units (GPUs). 


\begin{figure*}[t!]
    \centering
    \includegraphics[width=\textwidth]{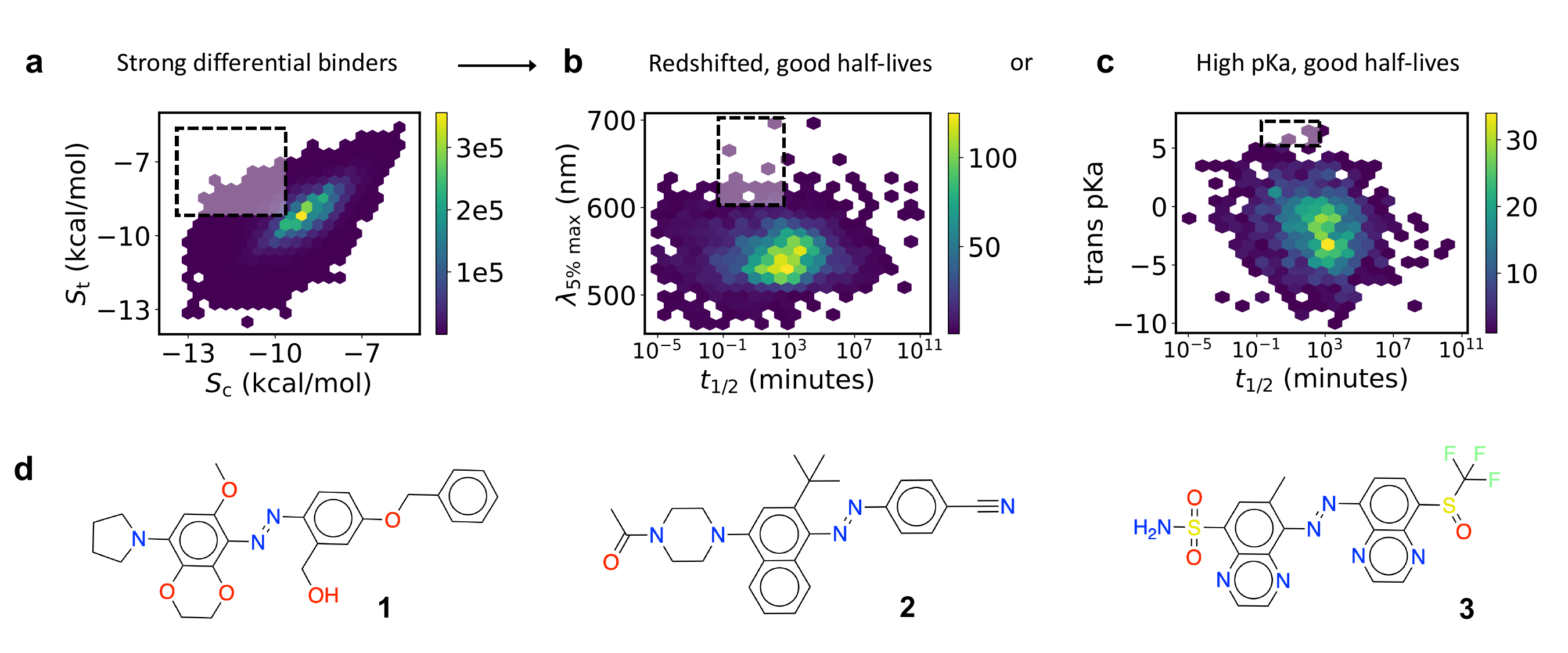}
    \caption{Results of virtual screening. (a) We first identified strong differential binders through docking. We then filtered remaining molecules based on (b) absorption spectra and thermal half-lives ($t_{1/2}$), or (c) high p$K_\mathrm{a}$ values and thermal half-lives. (d) Final hits after validation with quantum chemistry and FEP.}
    \label{fig:funnel_and_hits}
\end{figure*}

\begin{figure*}[t!]
    \centering
    \includegraphics[width=\textwidth]{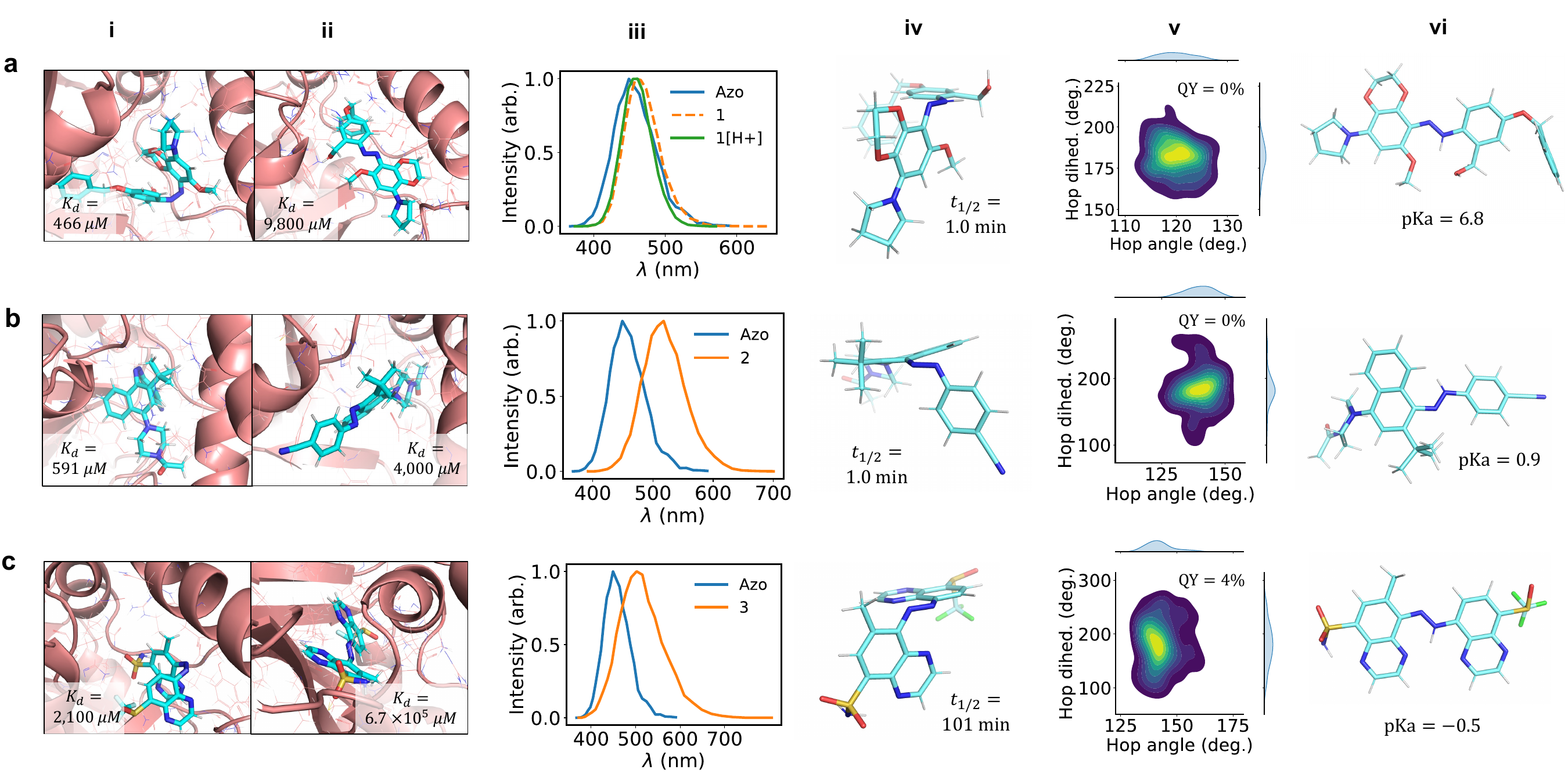}
    \caption{Properties, plots, and graphics for the top candidates. (a) Compound \textbf{1}. (i) \textit{Cis} binding pose from the last frame of FEP. (ii) As in (i), but for \textit{trans}. (iii) Absorption spectrum of \textit{trans} azobenzene, \textit{trans} \textbf{1}, and \textit{trans} \textbf{1}-azonium (denoted by 1[H+]). (iv) Transition state (TS) for thermal isomerization of \textbf{1}. Note that the lowest free energy TS involves azonium. (v) Kernel density estimation plot of the CNNC dihedrals and NNC angles at which hopping occurs in NAMD. NAMD was performed with \textbf{1}-azonium. (vi) Azonium isomer with the lowest free energy. (b) As in (a), but for compound \textbf{2}. Unlike in (a), the TS does not involve azonium, and NAMD was performed with the azobenzene form. (c) As in (b), but for compound \textbf{3}.}
    \label{fig:fep_snapshots}
\end{figure*}

\subsection*{Hits}
Virtual screening results are shown in Fig. \ref{fig:funnel_and_hits}. Panels (a) to (c) show the results of the computational funnel, starting with docking and followed by NFF calculations of absorption spectra, half-lives, and p$K_\mathrm{a}$ values. As we found in previous work \cite{axelrod2023mapping}, the docking scores of the \textit{cis} and \textit{trans} isomers are in fact highly correlated (Pearson $R=0.72$, Spearman $\rho=0.69$). Nevertheless, we identified 30,000 compounds with fairly strong \textit{cis} binding affinities and fairly weak \textit{trans} affinities (bottom 15\% of scores and top 50\% of scores, respectively). From these compounds we selected species with half-lives between 5 seconds and 5 hours, and absorption wavelengths above 600 nm (Fig. \ref{fig:funnel_and_hits}(b)). Absorption wavelengths were computed as $\lambda_{5\% \ \mathrm{max}}$, defined as the maximum absorption wavelength at 5\% of the maximum intensity. We also selected compounds with \textit{trans} p$K_\mathrm{a}$ values above 5 and half-lives in the desired range (Fig. \ref{fig:funnel_and_hits}(c)). As with docking, we see that compounds with the desired properties in the right two panels made up only a small portion of the overall library. This reinforces the difficulty of the optimization problem and the utility of large-scale screening.

The top three candidates are shown in Fig. \ref{fig:funnel_and_hits}(d). Visualizations and computed properties are shown in Fig. \ref{fig:fep_snapshots}, and further data can be found in CSI Sec. \ref{sec:top_candidate_data}. 
The first two compounds were predicted to be moderate differential binders of PARP1, with affinities that are typical for early-stage hit finding (Fig. \ref{fig:fep_snapshots}, columns (i) and (ii)). Compound \textbf{1} was predicted to form an azonium ion at physiological pH, with a predicted \textit{trans} p$K_\mathrm{a}$ of 6.8. Based on experimental results in the literature, this suggests a red-shifted absorption spectrum, though the azonium spectrum generated by TDDFT was not in fact red-shifted (Fig. \ref{fig:fep_snapshots}, row (a), column (iii)). This is discussed further below. Compound \textbf{2} was predicted to have a red-shifted absorption spectrum with no azonium formation (Fig. \ref{fig:fep_snapshots}, row (b), columns (iii) and (vi)). Both compounds \textbf{1} and \textbf{2} were predicted to have half-lives of 1.0 min in water at $37^{\circ}$C (Fig. \ref{fig:fep_snapshots}, column (iv)). Each compound had a predicted quantum yield of 0, and hence no photoswitching.

Compound \textbf{3} was predicted to be a weaker binder than the other two compounds, but with a more red-shifted spectrum, a non-zero quantum yield, and a longer (but perhaps too long) thermal half-life (Fig. \ref{fig:fep_snapshots}, row (c)). Although compound \textbf{3} exhibited a favorable predicted photophysical profile, including the only non-zero predicted quantum yield, we did not pursue its synthesis or the synthesis of its analogues due to practical resource constraints. Our synthetic efforts were instead directed toward derivatives of \textbf{1} and \textbf{2}, for which building blocks were commercially available, azo coupling was expected to be robust, and handling was straightforward. As part of this focus, we removed the bulky \textit{tert}-butyl group from \textbf{2} to improve synthetic accessibility, and to increase the probability of a nonzero quantum yield. The latter was motivated by the predicted zero yield, which we attributed to the bulky \textit{ortho} substituent that can often inhibit photoswitching \cite{axelrod2022excited}. We also hypothesized, and subsequently confirmed, that the azonium form of \textbf{1} would in fact be red-shifted, despite TDDFT predictions to the contrary.

\subsection*{Synthesis}

\begin{scheme}[t!]
\centering
\includegraphics[width=\columnwidth]{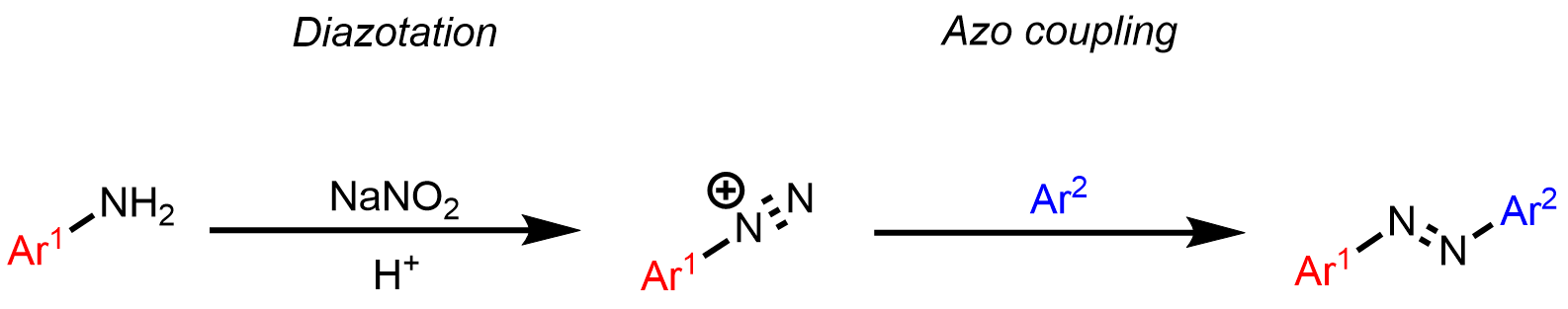}
\caption{General synthetic strategy for the azobenzene library. Ar\textsuperscript{1} denotes the electron-poor aryl partner, and Ar\textsuperscript{2} denotes the electron-rich aryl partner.}
\label{scheme:synthesis}
\end{scheme}

The photoswitches were prepared by late-stage assembly of the azobenzene linkage. The key disconnection forms the Ar--N=N--Ar motif by diazotization of an aniline precursor, followed by azo coupling with an electron-rich arene/heteroarene nucleophile (Scheme~\ref{scheme:synthesis}). Diazotization was typically conducted with sodium nitrite at 0--4$^\circ$C using either p-toluenesulfonic acid or tetrafluoroboric acid as a proton source. Overall, we attempted to synthesize 20 compounds and successfully created 10, with isolated yields ranging from 9--59\%. Experimental procedures, characterization, and unsuccessful attempts are reported in the Experimental Supplementary Information (ESI).

\subsection*{Photophysical Characterization}
Absorption spectra, thermal relaxation kinetics, quantum yields, and photostationary state (PSS) compositions were measured in acetonitrile and in phosphate-buffered saline (PBS) containing 0.01\% Tween~80 to approximate assay-relevant conditions and prevent sample precipitation. Compounds, experimental results, and computational predictions can be found in Table~\ref{tab:photoswitching}.

\textbf{Protonation state.} To establish compound protonation states during switching experiments, we determined thermodynamic p$K_\mathrm{a}$ values by capillary electrophoresis. Most compounds showed no evidence of protonation at pH~2, indicating p$K_\mathrm{a}$~$< 0$. Three exceptions were compounds \textbf{1} (p$K_\mathrm{a}$ = 3.78), \textbf{1a} (4.29), and \textbf{2b} (4.15). Of these, compound \textbf{1} was specifically chosen because of its predicted near-neutral p$K_\mathrm{a}$ of 6.8. The experimental value was significantly lower, indicating that our computational approach predicted an over-stabilization of the azonium form. The experimental results indicate negligible azonium formation at physiological pH for all tested compounds, meaning the observed photoswitching behavior arises from the neutral azobenzene chromophore rather than the protonated azonium species.

\textbf{Absorption spectra.} To compare UV-Vis absorption profiles, we report $\lambda_{5\%}$ values, defined as the wavelength at 5\% of maximum intensity relative to the maximum absorbance of the most red-shifted peak. Our computational pipeline correctly prioritized compounds with red-shifted absorption spectra: All compounds exhibited red-shifted absorption relative to unsubstituted azobenzene, with $\lambda_{5\%}$ values in acetonitrile ranging from 557~nm (\textbf{2h}) to 616~nm (\textbf{2b}). This red-shift arises from the push--pull electronic configuration imparted by electron-donating piperazine or pyrrolidine groups and electron-withdrawing nitrile or amide substituents. \cite{aleotti2020spectral} The effect of the electron-withdrawing group is clearly illustrated by the ortho-dihalogenated series: decreasing halogen electronegativity (F~$>$~Cl~$>$~Br) correlates with decreasing $\lambda_{5\%}$ values of 580~nm, 562~nm, and 557~nm, respectively, for compounds \textbf{2f}, \textbf{2g}, and \textbf{2h}. Compounds generally showed modest additional red-shifts of 2--8~nm when measured in aqueous buffer compared to acetonitrile. For example, compound \textbf{2a} shifted from $\lambda_{5\%}$~=~576~nm in acetonitrile to 578~nm in PBS/Tween~80.

\begin{figure}[t!]
    \centering
    \includegraphics[width=\columnwidth]{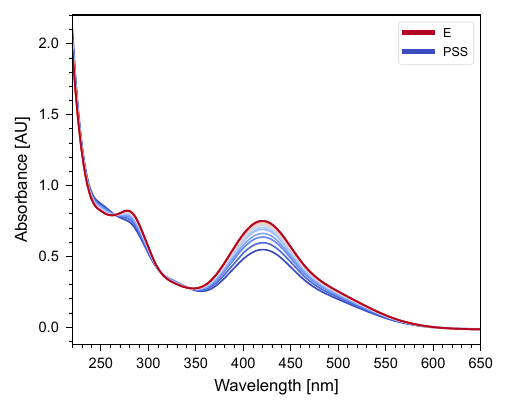}
    \caption{Absorbance spectrum change of compound \textbf{2a} upon irradiation by 519 nm LED in acetonitrile at $22^{\circ}$C, at concentration 4.452e-5 M.}
    \label{fig:spectral_evolution_2a}
\end{figure}

\begin{figure}[t!]
    \centering
    \includegraphics[width=\columnwidth]{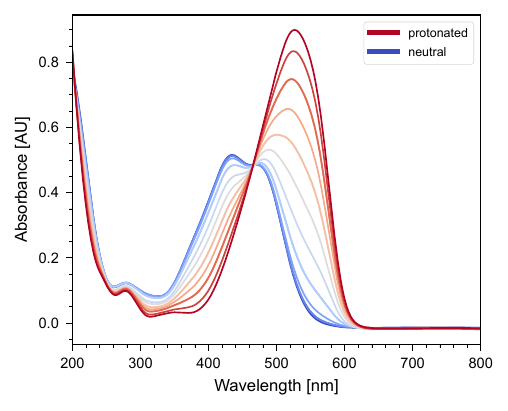}
    \caption{Titration of compound \textbf{1} with trifluoroacetic acid in acetonitrile. Each step is the addition of 10 $\mu$L of 0.01\% TFA in ACN into 2.0~mL of azobenzene dissolved in ACN at starting concentration 2.609e-5 M.}
    \label{fig:protonation_1}
\end{figure}

Compounds \textbf{1} and \textbf{1a}, which possess measurable p$K_\mathrm{a}$ values, exhibited pH-dependent absorption spectra. As shown for compound \textbf{1} in Fig. \ref{fig:protonation_1}, the addition of trifluoroacetic acid (5 ppm) red-shifted $\lambda_{5\%}$ to 606 nm, consistent with partial azonium formation, while triethylamine (5 ppm) blue-shifted the spectrum to $\lambda_{5\%} = 556$ nm. Compound \textbf{1a} displayed analogous behavior (acid: 627 nm; base: 567 nm; see ESI Figures~S31--S38).  The azonium red-shift agrees with our expectations, but not with TDDFT predictions (Fig. \ref{fig:fep_snapshots}, row (a), column (iii)). As shown in CSI Table \ref{tab:dg_data}, we observed a surprising trend for asymmetrically substituted azobenzenes: the azonium isomer with the higher p$K_\mathrm{a}$, and thus the greater likelihood of forming, often had a \textit{blue-shifted} TDDFT spectrum. The isomer with the lower p$K_\mathrm{a}$, and hence the lower likelihood of forming, usually had a red-shifted spectrum. 
We hypothesized that the predicted blue-shift was a TDDFT artifact, and therefore prioritized compounds with high p$K_\mathrm{a}$ values regardless of their predicted spectra. The experimental azonium spectra for compounds \textbf{1} and \textbf{1a} support this hypothesis.

We had targeted compounds \textbf{1} and \textbf{1a} because of their high predicted p$K_\mathrm{a}$ values, which meant that they would undergo a red-shift upon azonium formation under physiological conditions. Although the azonium forms were indeed experimentally red-shifted, the measured p$K_\mathrm{a}$ values were too low for this effect to be relevant under physiological conditions. This reflects a shortcoming of our computational predictions. Moreover, even if the p$K_\mathrm{a}$ values had been high enough, the spectral red-shifts would have been insufficient to reach the near-IR absorption of 720 nm achieved by the best azonium compounds in Ref. \cite{dong2017near}. So while our computational approach produced red-shifted azobenzene derivatives, it did not produce compounds with the desired near-IR absorption achieved in Ref. \cite{dong2017near}.

\begin{figure}[t!]
    \centering
    \includegraphics[width=\columnwidth]{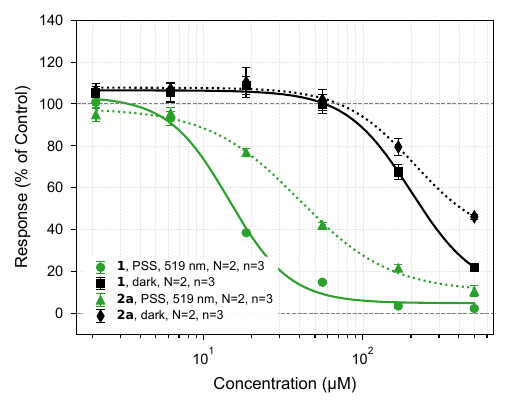}
    \caption{Dose--response curves for PARP1 inhibition by compounds \textbf{1} and \textbf{2a} in the dark (predominantly \textit{trans} isomer) and under continuous 519~nm irradiation (PSS). Compound \textbf{1} exhibits an approximately 15-fold potency enhancement upon illumination (IC$_{50}$ = 208.8~$\pm$~28.3~$\mu$M in the dark vs.\ 14.4~$\pm$~1.9~$\mu$M in the light), while compound \textbf{2a} shows an approximately 5-fold enhancement (209.4~$\pm$~51.9~$\mu$M in the dark vs.\ 39.7~$\pm$~4.7~$\mu$M in the light). Data points represent mean~$\pm$~SD of technical triplicates ($n = 3$) from two independent biological replicates ($N = 2$).}
    \label{fig:dose_response}
\end{figure}

\textbf{Thermal relaxation.} Thermal half-lives ($t_{1/2}$) in acetonitrile at 22$^{\circ}$C ranged from $<$ 0.01~min (\textbf{2b}) to 5.95~$\pm$~0.89~min (\textbf{2a}), placing most compounds within the computationally predicted range of seconds to minutes. Increasing the temperature from 22$^{\circ}$C to 37$^{\circ}$C reduced half-lives by approximately 40--50\%, as expected from Eyring theory \cite{eyring1935activated} (e.g., compound \textbf{2a}: $t_{1/2}$~=~3.30~$\pm$~0.19~min at 37$^{\circ}$C).

A critical finding was that thermal relaxation in aqueous buffer was dramatically accelerated: all compounds exhibited $t_{1/2}$~$<$~0.01~min ($<$0.6 seconds) in PBS/Tween~80 at 22$^{\circ}$C, faster than our instrumental time resolution (ESI Figures~S39--S48). This aqueous acceleration, likely driven by hydrogen bonding and solvent polarity effects on the rotational isomerization barrier \cite{dokic2009quantum, garcia2010kinetic, dunn2009ph}, was not captured by our implicit-solvent computational model and represents a significant limitation for \textit{in vivo} applications. However, Herges and co-workers recently demonstrated that an azo photoswitch with millisecond aqueous half-life could nonetheless exhibit prolonged biological activity ($>$3~h) when bound to its receptor, presumably because the protein environment stabilizes the active isomer \cite{Ewert2022photoswitchable}. This precedent suggests that fast aqueous relaxation may not preclude functional photopharmacology if target binding is sufficiently strong.

\textbf{Quantum yields.} Photoisomerization quantum yields ($\Phi_{trans \rightarrow cis}$) were determined at 519~nm in acetonitrile using indolyl-fulgide actinometry \cite{Reinfelds2019robust} (ESI Figures~S66--S73). An example of the spectral evolution upon irradiation is shown in Fig. \ref{fig:spectral_evolution_2a}. Measured quantum yields ranged from 1.2~$\pm$~0.2\% (\textbf{1}) to 5.2~$\pm$~1.7\% (\textbf{1a}). 

However, no photoswitching was observed for compounds \textbf{1} or \textbf{1a} under acidic conditions, where both exist in the \textit{trans} azonium form (ACN + 5 ppm TFA, 22$^\circ$C; denoted by superscript h in Table \ref{tab:photoswitching}). This could be due to rapid relaxation of the \textit{cis} isomer, leading to very little accumulation under steady-state illumination, or the complete absence of excited-state switching, and hence zero quantum yield. We cannot say for certain what the root cause is. However, the latter would be consistent with our prediction of zero quantum yield for compound \textbf{1}, which we had considered might be a simulation artifact.

Derivatives of compound \textbf{2} (Fig. \ref{fig:funnel_and_hits}) exhibited photoswitching under all conditions, and had low but nonzero quantum yields in acetonitrile. The yields varied from 1.7\% to 4.5\%, which is up to 18$\times$ lower than the yield of 31\% for unsubstituted \textit{trans}-azobenzene in acetonitrile \cite{bortolus1979cis}. It is interesting that the parent compound bearing a bulky \textit{tert}-butyl group had a predicted yield of zero, while its closest derivative, \textbf{2a}, had a measured yield of 1.7\%. The two values are not directly comparable, but it is still notable that replacing the \textit{tert}-butyl group with a methyl group did not produce a high yield. Complete removal of the methyl group in \textbf{2e} likewise did not alter the yield. Together, these results suggest a complex and somewhat unintuitive picture of the excited-state dynamics, one that nevertheless appears to be captured by our simulations.

\textbf{Photostationary state composition.} PSS compositions were determined by 519~nm irradiation of the sample at $-$78~$^{\circ}$C in CDCl$_3$ and subsequent measurement with low-temperature $^1$H~NMR. The \textit{trans}:\textit{cis} ratios varied substantially across the compound series. Compound \textbf{1} achieved a favorable ratio of 1:2.03, indicating that approximately 67\% of molecules occupy the \textit{cis} state at PSS. Compound \textbf{1a} similarly showed good switching (1:1.59, $\sim$61\% \textit{cis}). 

In contrast, ortho-dihalogenated derivatives \textbf{2g} (dichloro) and \textbf{2h} (dibromo) exhibited poor PSS enrichment (1:0.15 and 1:0.14, respectively). The origin of this poor switching is not entirely clear. While the larger van der Waals radii of Cl and Br relative to F may play a role, ortho-tetrafluoroazobenzenes are known to achieve excellent PSS ratios despite bearing four ortho substituents.\cite{knie2014ortho} Electronic effects, such as differences in how these halogens modulate the n$\rightarrow\pi^*$ band separation of the \textit{trans} and \textit{cis} isomers, may also contribute.

\textbf{Photostability.} Fatigue resistance was assessed through repeated irradiation--relaxation cycles in acetonitrile (ESI Figures~S74--S86). Most compounds showed excellent photostability over 10 cycles with no detectable degradation. However, compounds \textbf{1} and \textbf{1a} exhibited gradual baseline drift without irradiation in the presence of 5~ppm triethylamine, suggesting possible side reactions under prolonged basic conditions. After background subtraction, the irradiation--relaxation cycles revealed only minor drift, indicating that the neutral forms are likely to be photochemically robust. For biological applications under physiological pH, this base-mediated degradation pathway is unlikely to be relevant.

\subsection*{Biological Evaluation}

To assess whether the computationally predicted differential binding translates into measurable light-dependent PARP1 inhibition, we evaluated the synthesized compounds using a chemiluminescent PARP1 activity assay (BPS Bioscience). All measurements were made under green light illumination at 519 nm. This is an 80 nm red-shift relative to typical measurements under blue light, and was made possible by the compounds' red-shifted absorption spectra. Given the rapid thermal relaxation observed in aqueous buffer ($t_{1/2} < 0.01$~min for all compounds in PBS/Tween~80), we employed continuous \textit{in situ} illumination throughout the assay incubation to maintain a steady-state population of the \textit{cis} isomer. A custom 3D-printed bio-photoreactor was designed to uniformly irradiate half of a 384-well plate while leaving the remaining wells dark as internal controls (ESI Figures~S6--S11).

Solubility posed a significant challenge under assay conditions. At the screening concentration of 500~$\mu$M in PARP1 assay buffer containing 1\% DMSO, several compounds exhibited visible precipitation. Addition of 0.01\% Tween~80 substantially improved solubility for most compounds (ESI Figures~S99--S101), and this formulation was adopted for all subsequent IC$_{50}$ determinations.

Among the ten synthesized compounds, \textbf{1} and \textbf{2a} exhibited the most pronounced light-dependent inhibition (Fig. \ref{fig:dose_response}). Under dark conditions, compound \textbf{1} (predominantly the \textit{trans} isomer) displayed an IC$_{50}$ of 208.8~$\pm$~28.3~$\mu$M. Upon continuous 519~nm irradiation, the IC$_{50}$ decreased to 14.4~$\pm$~1.9~$\mu$M, representing an approximately 15-fold enhancement in potency. This result qualitatively aligns with the computational prediction of stronger \textit{cis} binding ($K_\mathrm{d}$~=~466~$\mu$M) relative to \textit{trans} ($K_\mathrm{d}$~=~9.8~mM), though the absolute values differ. 

Direct numerical comparison between the computed $K_\mathrm{d}$ values and the experimentally measured IC$_{50}$ values is not straightforward, as these are related but distinct quantities. The IC$_{50}$ depends not only on the intrinsic binding affinity but also on assay-specific conditions such as substrate concentration, enzyme concentration, and the mechanism of inhibition, as formalized by the Cheng--Prusoff relationship \cite{cheng1973relationship}. Consequently, IC$_{50}$ values are generally expected to exceed $K_\mathrm{d}$ values, and differences in absolute magnitude between the two metrics do not necessarily indicate poor predictive accuracy. The relevant comparison is therefore the \textit{directional} agreement---namely, that the \textit{cis} isomer is a substantially stronger inhibitor than the \textit{trans} isomer---which is consistent between the computational predictions and experimental measurements.

Compound \textbf{2a}, the second computationally prioritized hit, showed more modest photoswitching: IC$_{50}$ values of 209.4~$\pm$~51.9~$\mu$M (dark) and 39.7~$\pm$~4.7~$\mu$M (519~nm), corresponding to an approximately 5-fold potency enhancement. Compound \textbf{1a} exhibited a similar trend with IC$_{50}$ (PSS)~=~167.8~$\pm$~23.0~$\mu$M under illumination, although the dark IC$_{50}$ could not be reliably determined due to poor curve fitting at the accessible concentration range. 

\begin{table*}[t]
\centering
\caption{Predicted and experimental physicochemical, photophysical, and biological properties of synthesized compounds.}
\label{tab:photoswitching}

\scriptsize
\newcolumntype{C}[1]{>{\centering\arraybackslash}m{#1}}

\renewcommand{\arraystretch}{1.0}
\setlength{\tabcolsep}{1.0pt}

\newcommand{\calc}{\textcolor{red}{calc}}
\newcommand{\fillup}{\textcolor{red}{fill}}

\newcommand{\thickhline}{%
  \noalign{\global\arrayrulewidth=1.2pt}\hline
  \noalign{\global\arrayrulewidth=0.4pt}
}

\begin{threeparttable}
\scriptsize
\resizebox{\textwidth}{!}{
\setlength{\arrayrulewidth}{0.6pt}
\begin{tabular}{|C{3.5cm}|C{0.85cm}|C{0.85cm}|C{0.85cm}|C{0.85cm}|C{0.85cm}|C{1.85cm}|C{0.85cm}|C{0.85cm}|C{1.2cm}|C{0.85cm}|C{0.85cm}|C{0.85cm}|C{0.85cm}|}
\thickhline
\multirow{2}{*}{\textbf{Structure}} &
\multicolumn{2}{C{1.90cm}|}{\textbf{$\mathrm{p}K_{\mathrm{a}} \pm$ SD\tnote{g}}} &
\multicolumn{2}{C{1.70cm}|}{\textbf{\textit{trans} $\,\lambda_{5\%,\max}$ (nm)}} &
\multicolumn{2}{C{2.70cm}|}{\textbf{$t_{1/2}$ (min)}} &
\multicolumn{2}{C{1.70cm}|}{\textbf{QY (\%)\tnote{p}}} &
\multicolumn{1}{C{1.2cm}|}{\textbf{PSS (\textit{trans}: \textit{cis})\tnote{q} }} &
\multicolumn{2}{C{1.70cm}|}{\textbf{PARP1 Inhibition (\textit{trans}; $\mu$M)}} &
\multicolumn{2}{C{1.70cm}|}{\textbf{PARP1 Inhibition (PSS; $\mu$M)}} \\

\cline{2-14}
&
{pred.} & {exp.} &
{pred.} & {exp.} &
{pred.} & {exp.} &
{pred.} & {exp.} &
{exp.} &
{pred.}\tnote{r} & {exp.}\tnote{s} &
{pred.}\tnote{r} & {exp.}\tnote{s} \\
\thickhline

\raisebox{-10pt}{\includegraphics[width=3.2cm]{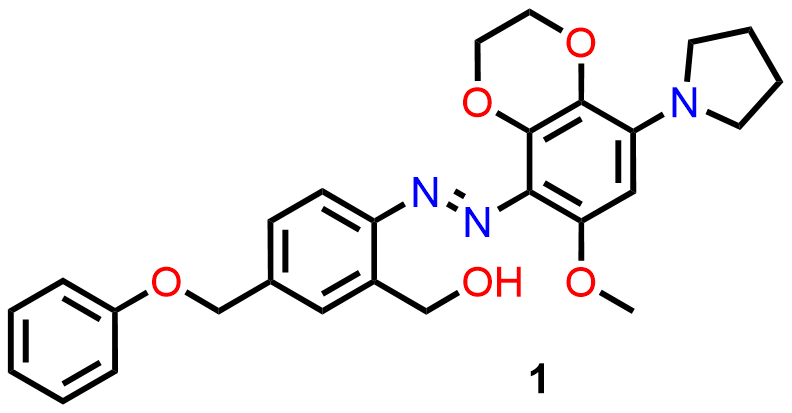}} &
6.8 &
3.78 $\pm$ 0.09 &
514 &
\begin{tabular}[c]{@{}c@{}}606$^{\mathrm{h}}$\\556$^{\mathrm{j}}$\\594$^{\mathrm{L}}$\end{tabular} &
1.0 &
\begin{tabular}[c]{@{}c@{}}N/A$^{\mathrm{b,h}}$\\0.03 $\pm$ 0.01$^{\mathrm{i}}$\\0.02 $\pm$ 0.01$^{\mathrm{j}}$\\$<0.01^{\mathrm{a,L}}$\end{tabular} &
0 $\pm$ 0 &
1.2 $\pm$ 0.2$^{\mathrm{j}}$ &
1:2.03 &
9.8 $\times 10^3$ &
208.8 $\pm$ 28.3 &
466 &
14.4 $\pm$ 1.9 \\
\hline

\raisebox{-10pt}{\includegraphics[width=3.2cm]{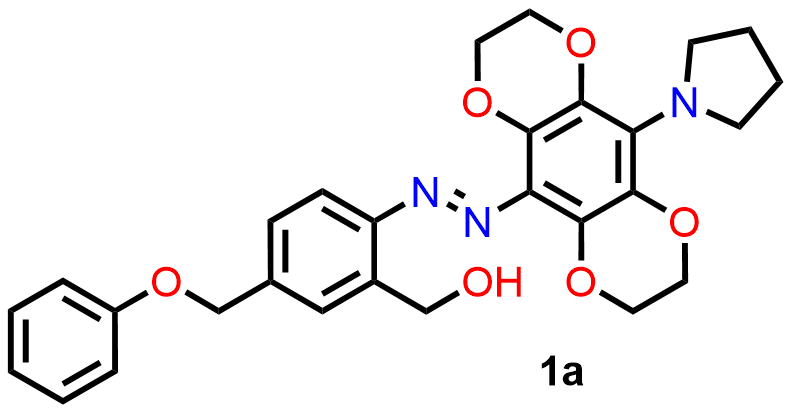}} &
5.0 &
4.29 $\pm$ 0.09 &
513 &
\begin{tabular}[c]{@{}c@{}}627$^{\mathrm{h}}$\\567$^{\mathrm{j}}$\\572$^{\mathrm{L}}$\end{tabular} &
1.8 &
\begin{tabular}[c]{@{}c@{}}N/A$^{\mathrm{b,h}}$\\0.17 $\pm$ 0.08$^{\mathrm{j}}$\\0.08 $\pm$ 0.01$^{\mathrm{k}}$\\$<0.01^{\mathrm{a,L}}$\end{tabular} &
-- &
5.2 $\pm$ 1.7$^{\mathrm{j}}$ &
1:1.59 &
5.0 $\times 10^3$ &
ND$^{\mathrm{e}}$ &
1.1 $\times 10^3$ &
167.8 $\pm$ 23.0 \\
\hline

\raisebox{-10pt}{\includegraphics[width=3.2cm]{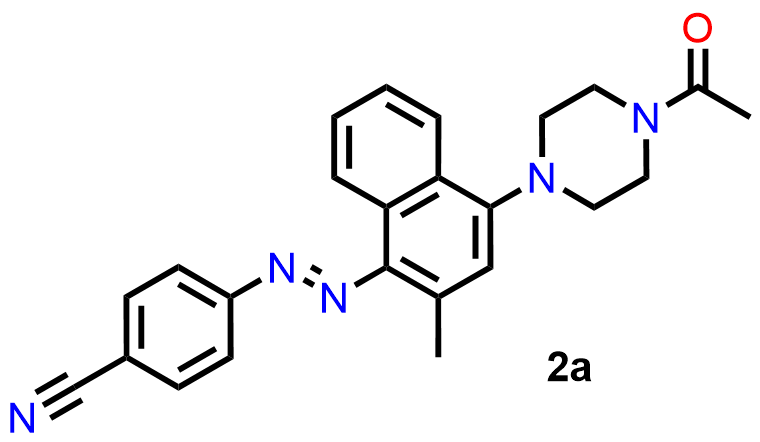}} &
1.3 &
$<0^{\mathrm{f}}$ &
601 &
\begin{tabular}[c]{@{}c@{}}576$^{\mathrm{n}}$\\578$^{\mathrm{L}}$\end{tabular} &
17.2 &
\begin{tabular}[c]{@{}c@{}}5.95 $\pm$ 0.89$^{\mathrm{n}}$\\3.30 $\pm$ 0.19$^{\mathrm{o}}$\\$<0.01^{\mathrm{a,L}}$\end{tabular} &
-- &
1.7 $\pm$ 0.1$^{\mathrm{n}}$ &
1:0.78 &
44.6 &
209.4 $\pm$ 51.9 &
2.9 &
39.7 $\pm$ 4.7 \\
\hline

\raisebox{-10pt}{\includegraphics[width=2.7cm]{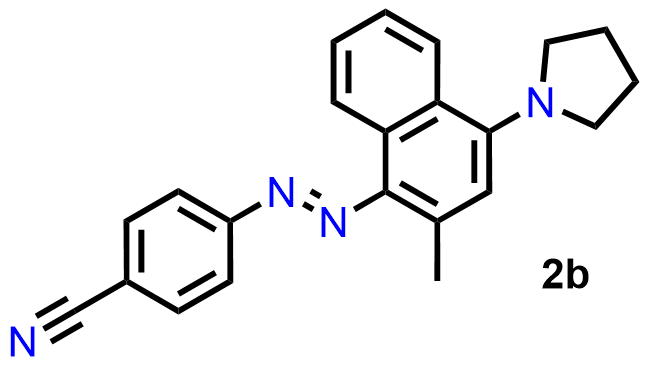}} &
4.6 &
4.15 $\pm$ 0.10 &
603 &
\begin{tabular}[c]{@{}c@{}}616$^{\mathrm{n}}$\\621$^{\mathrm{L}}$\end{tabular} &
0.2 &
\begin{tabular}[c]{@{}c@{}}$<0.01^{\mathrm{a,m}}$\\$<0.01^{\mathrm{a,L}}$\end{tabular} &
-- &
ND$^{\mathrm{c}}$ &
ND$^{\mathrm{c}}$ &
-- &
ND$^{\mathrm{e}}$ &
1.8 $\times 10^5$ &
ND$^{\mathrm{e}}$ \\
\hline

\raisebox{-10pt}{\includegraphics[width=3.2cm]{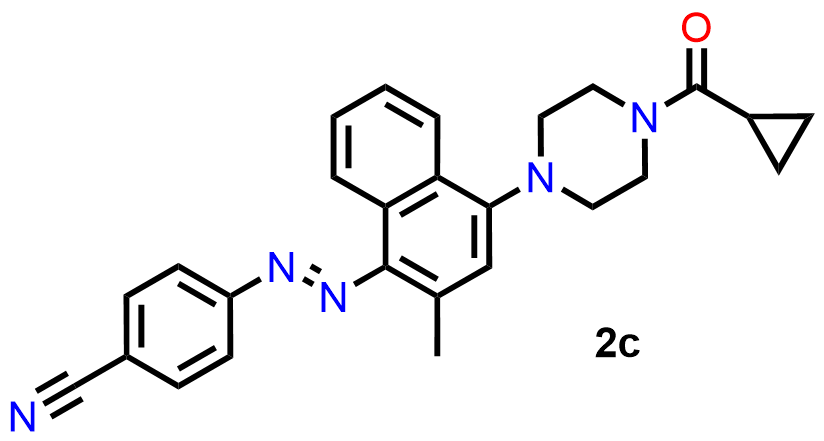}} &
$-0.1$ &
$<0^{\mathrm{f}}$ &
613 &
\begin{tabular}[c]{@{}c@{}}577$^{\mathrm{n}}$\\579$^{\mathrm{L}}$\end{tabular} &
20.7 &
\begin{tabular}[c]{@{}c@{}}1.80 $\pm$ 0.22$^{\mathrm{n}}$\\1.37 $\pm$ 0.18$^{\mathrm{o}}$\\$<0.01^{\mathrm{a,L}}$\end{tabular} &
-- &
4.5 $\pm$ 0.1$^{\mathrm{n}}$ &
ND$^{\mathrm{b}}$ &
356 &
ND$^{\mathrm{e}}$ &
1.5 $\times 10^4$ &
ND$^{\mathrm{e}}$ \\
\hline

\raisebox{-10pt}{\includegraphics[width=3.2cm]{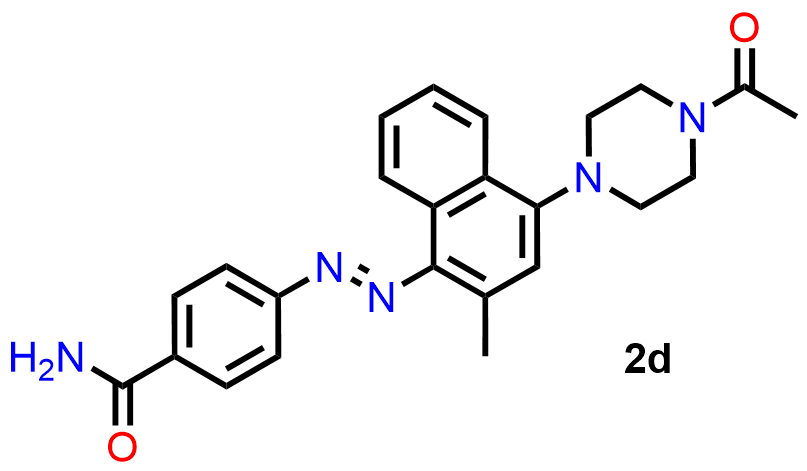}} &
0.4 &
$<0^{\mathrm{f}}$ &
582 &
\begin{tabular}[c]{@{}c@{}}562$^{\mathrm{n}}$\\562$^{\mathrm{L}}$\end{tabular} &
9.3 &
\begin{tabular}[c]{@{}c@{}}1.62 $\pm$ 0.23$^{\mathrm{n}}$\\0.93 $\pm$ 0.39$^{\mathrm{o}}$\\$<0.01^{\mathrm{a,L}}$\end{tabular} &
-- &
ND$^{\mathrm{d}}$ &
1:0.48 &
-- &
232.7 $\pm$ 20.3 &
-- &
209.2 $\pm$ 7.9 \\
\hline

\raisebox{-10pt}{\includegraphics[width=3.2cm]{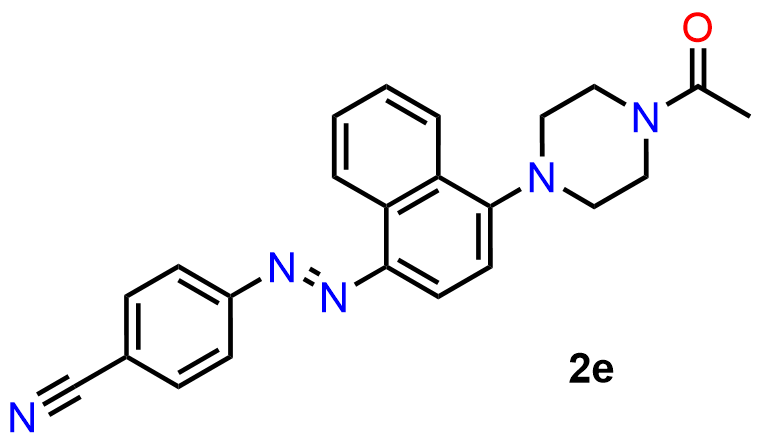}} &
$-0.4$ &
$<0^{\mathrm{f}}$ &
584 &
\begin{tabular}[c]{@{}c@{}}563$^{\mathrm{n}}$\\570$^{\mathrm{L}}$\end{tabular} &
36.3 &
\begin{tabular}[c]{@{}c@{}}3.32 $\pm$ 0.42$^{\mathrm{n}}$\\1.29 $\pm$ 0.24$^{\mathrm{o}}$\\$<0.01^{\mathrm{a,L}}$\end{tabular} &
-- &
1.7 $\pm$ 0.3$^{\mathrm{n}}$ &
1:0.74 &
297 &
ND$^{\mathrm{e}}$ &
6.4 $\times 10^{-7}$ &
78.6 $\pm$ 9.7 \\
\hline

\raisebox{-10pt}{\includegraphics[width=3.2cm]{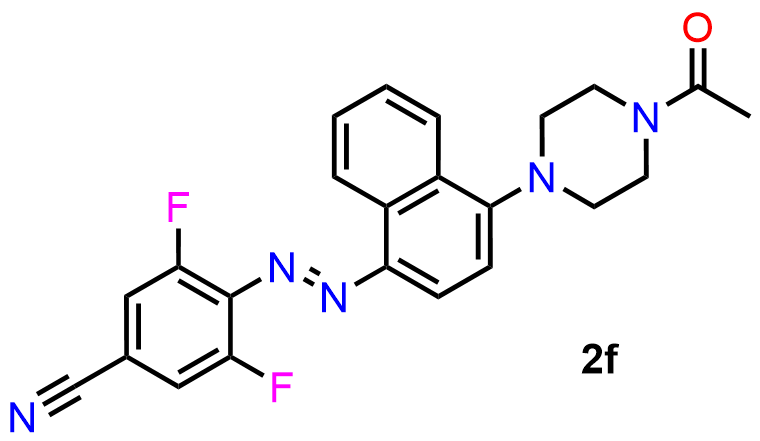}} &
$-1.8$ &
$<0^{\mathrm{f}}$ &
603 &
\begin{tabular}[c]{@{}c@{}}580$^{\mathrm{n}}$\\614$^{\mathrm{L}}$\end{tabular} &
1.0 &
\begin{tabular}[c]{@{}c@{}}0.45 $\pm$ 0.01$^{\mathrm{n}}$\\0.16 $\pm$ 0.01$^{\mathrm{o}}$\\$<0.01^{\mathrm{a,L}}$\end{tabular} &
-- &
2.6 $\pm$ 0.2$^{\mathrm{n}}$ &
1:0.47 &
2.4 $\times 10^5$ &
ND$^{\mathrm{e}}$ &
4.5 &
93.2 $\pm$ 11.2 \\
\hline

\raisebox{-10pt}{\includegraphics[width=3.2cm]{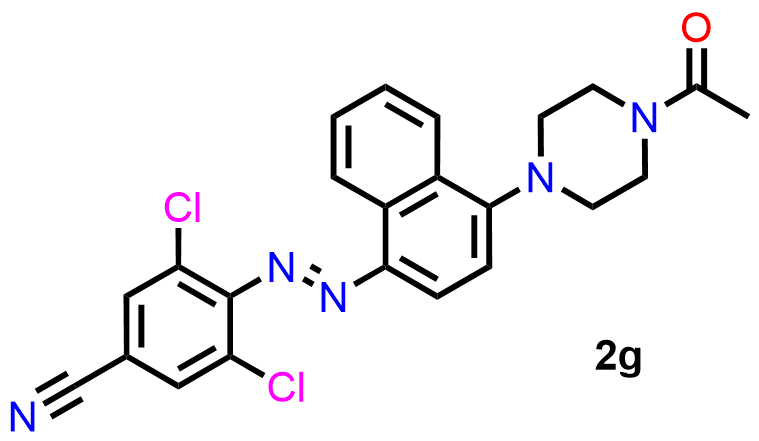}} &
$-1.5$ &
$<0^{\mathrm{f}}$ &
644 &
\begin{tabular}[c]{@{}c@{}}562$^{\mathrm{n}}$\\570$^{\mathrm{L}}$\end{tabular} &
-- &
\begin{tabular}[c]{@{}c@{}}0.06 $\pm$ 0.01$^{\mathrm{n}}$\\$<0.01^{\mathrm{a,L}}$\end{tabular} &
-- &
4.3 $\pm$ 0.3$^{\mathrm{n}}$ &
1:0.15 &
3.5 $\times 10^4$ &
ND$^{\mathrm{e}}$ &
2.0 $\times 10^4$  &
ND$^{\mathrm{e}}$ \\
\hline

\raisebox{-10pt}{\includegraphics[width=3.2cm]{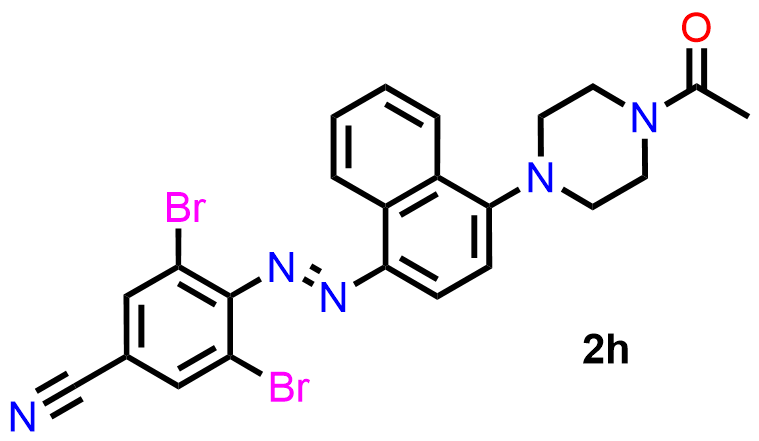}} &
$-1.3$ &
$<0^{\mathrm{f}}$ &
640 &
\begin{tabular}[c]{@{}c@{}}557$^{\mathrm{n}}$\\560$^{\mathrm{L}}$\end{tabular} &
-- &
\begin{tabular}[c]{@{}c@{}}0.07 $\pm$ 0.01$^{\mathrm{n}}$\\$<0.01^{\mathrm{a,L}}$\end{tabular} &
-- &
4.0 $\pm$ 0.1$^{\mathrm{n}}$ &
1:0.14 &
3.0 $\times 10^4$ &
ND$^{\mathrm{e}}$ &
13.8 &
144.3 $\pm$ 28.4 \\
\hline

\end{tabular}
}

    \begin{justify}
    \textsuperscript{a}Cannot be determined precisely with our experimental setup.
    \textsuperscript{b}No switching observed.
    \textsuperscript{c}Not determined; switching too fast to reliably measure with our setup.
    \textsuperscript{d}Not determined due to observed precipitation.
    \textsuperscript{e}Not determined due to unreliable fit.
    \textsuperscript{f}Estimated value.
    \textsuperscript{g}Thermodynamic acidity constant (related to zero ionic strength) determined by capillary electrophoresis; see ESI for full details.
    \textsuperscript{h}ACN + 5 ppm TFA, 22~$^\circ$C.
    \textsuperscript{i}ACN + 5 ppm Et$_3$N, 10~$^\circ$C.
    \textsuperscript{j}ACN + 5 ppm Et$_3$N, 22~$^\circ$C.
    \textsuperscript{k}ACN + 5 ppm Et$_3$N, 37~$^\circ$C.
    \textsuperscript{L}PBS/Tween~80, 22~$^\circ$C.
    \textsuperscript{m}ACN, 4~$^\circ$C.
    \textsuperscript{n}ACN, 22~$^\circ$C.
    \textsuperscript{o}ACN, 37~$^\circ$C.
    \textsuperscript{p}Quantum yields were determined at 519~nm spectrophotometrically against fulgide actinometer.
    \textsuperscript{q}PSS (\textit{trans}/\textit{cis}) ratios were determined by irradiation of 1.5~mM CDCl$_3$ solutions with 519~nm diode at -78~$^\circ$C, and subsequent measurement in low-temperature $^1$H NMR.
    \textsuperscript{r}Kd values.   
    \textsuperscript{s}IC$_{50}$ values were determined against recombinant human PARP1 using a chemiluminescent activity assay
    (BPS Bioscience PARP1 Chemiluminescent Assay Kit, cat.\ no.\ 80551) according to the manufacturer's protocol
    (see ESI). IC$_{50}$ (\textit{trans}) corresponds to dark-adapted samples (predominantly \textit{trans}), and IC$_{50}$ (PSS) corresponds to samples irradiated \textit{in situ} at 519~nm. IC$_{50}$ values represent the mean \(\pm\) SD of technical triplicates (\(n = 3\)) from two independent biological replicates (\(N = 2\)). Irradiation was performed in a custom-built 3D-printed photoreactor; see ESI for reactor details and irradiation parameters. Photophysical values are mean \(\pm\) SD of technical triplicates (\(n = 3\)) from a single synthesized batch (\(N = 1\)). Abbreviations: exp, experimental; pred, predicted; ACN, acetonitrile; TFA, trifluoroacetic acid; Et$_3$N, triethylamine; PBS, phosphate-buffered saline;
    PSS, photostationary state; QY, quantum yield. Predictions: The quantum yield was only predicted for compound \textbf{1}, since predicting it for the other compounds would require training a new set of models with new quantum chemical data generated through active learning. Dashes for other predictions indicate failed calculations. PSS prediction was not part of the computational pipeline, and predicted PSS PARP1 inhibition is simply the predicted \textit{cis} $K_{\mathrm{d}}$. 
    \end{justify}
    
\end{threeparttable}
\end{table*}

\FloatBarrier

Several compounds displayed minimal or no measurable differential activity.

For compound \textbf{2g}, reliable IC$_{50}$  values could not be determined under either condition. For \textbf{2h}, the dark IC$_{50}$  could not be reliably determined, although an illuminated IC$_{50}$ of $144.3 \pm 28.4$ $\mu$M was obtained, precluding quantitative assessment of light-dependent potency shifts. Moreover, their poor PSS enrichment (\textit{trans}:\textit{cis} $\approx$ 1:0.15) means that irradiation generated only a small increase in the proportion of the \textit{cis} isomer. It is therefore difficult to determine whether both isomers are weak inhibitors, or whether the \textit{cis} isomer is a moderate inhibitor whose activity is masked by its low proportion in the PSS.

It is important to note that the measured IC$_{50}$ values for all compounds remain in the micromolar range, substantially weaker than clinically used PARP1 inhibitors such as olaparib (IC$_{50}$~$\approx$~5~nM under our assay conditions; ESI Figure~S102). This modest potency is typical of early-stage hits identified through computational screening and would require further optimization in a hit-to-lead campaign. Nevertheless, the demonstration of photoswitchable PARP1 inhibition under green light---particularly the 15-fold potency enhancement observed for compound \textbf{1}---validates the core photopharmacological concept and the utility of the computational screening funnel for identifying differentially binding photoswitches under red-shifted illumination.

\section*{Conclusions}
In this work, we combined large-scale computational screening with experimental synthesis and validation to identify photoresponsive PARP1 inhibitors. From an initial library of 5 million azobenzene derivatives, our computational funnel---comprising docking, neural network force field simulations, and free energy perturbation calculations---prioritized candidates predicted to exhibit red-shifted absorption, appropriate thermal half-lives, and isomer-dependent binding. We synthesized 10 compounds and experimentally characterized their photophysical and biological properties. The photophysical measurements validated several computational predictions: all compounds exhibited red-shifted absorption relative to unsubstituted azobenzene ($\lambda_{5\%} = 557$--$616$ nm), and thermal half-lives in acetonitrile fell within the computationally predicted range of seconds to minutes. 

However, we also observed critical discrepancies. The experimental p$K_\mathrm{a}$ value for compound \textbf{1} was substantially lower than predicted (3.78 vs. 6.8). This indicates that our computational model can overestimate azonium stability for critical compounds, even when results for other derivatives are quite accurate (Table \ref{tab:photoswitching}). This highlights an inherent risk of computational screening: by screening for exceptional properties, one can unintentionally prioritize artifacts generated by model error over truly exceptional candidates.

More significantly, thermal relaxation in aqueous buffer was dramatically accelerated ($t_{1/2} < 0.6$ s for all compounds in PBS/Tween 80), a limitation not captured by implicit-solvent calculations. Despite this rapid aqueous relaxation, compound \textbf{1} exhibited a $\sim$15-fold enhancement in PARP1 inhibition upon green light irradiation at 519 nm (IC$_{50} = 208.8 \pm 28.3$ $\mu$M in the dark vs.\ $14.4 \pm 1.9$ $\mu$M under illumination), demonstrating that functional photopharmacology remains achievable under continuous irradiation conditions.

Together, these results validate computation-guided multi-property optimization as a viable strategy for photopharmacology, while highlighting the need to explicitly model aqueous-phase dynamics and to develop scaffolds with improved thermal stability in physiological media. Future efforts should focus on extending \textit{cis}-isomer lifetimes in aqueous environments, improving target potency into the nanomolar range, and exploring additional red-shifted chromophores compatible with deeper tissue penetration.

\section*{Computational Methods}
\subsection*{Library Generation}
The initial molecular library was generated using in-house scripts. As described in CSI Sec. \ref{sec:molgen}, we used a set of common literature substitution patterns with substituents from both the literature and the Enamine REAL catalogue \cite{enamine}. We also expanded the library using a set of random fragment additions, and a set of modifications of promising motifs. QED and SA scores were computed with RDKit \cite{rdkit}, PAINS and MCF calculations were performed with MOSES (\url{https://github.com/molecularsets/moses}), and SC scores were computed with code from \url{https://github.com/connorcoley/scscore}.

\subsection*{Binding Free Energy Calculation}
Docking was performed with Dockstring version 0.2.0 \cite{garcia2022dockstring}. We removed all compounds that did not maintain their \textit{cis}-\textit{trans} isomerism during docking. Absolute FEP was performed with Yank version 0.25.2 \cite{yank}. For FEP we used the ff14SB force field for the protein \cite{maier2015ff14sb}, the TIP4P-Ew model for water \cite{horn2004development}, and the GAFF2 force field \cite{wang2004development} with re-parameterized azo terms for the ligands (CSI Sec. \ref{subsec:fep_force_fields}). To generate starting structures for FEP, we began with the docked pose and then performed a minimization in Gromacs \cite{abraham2015gromacs}, followed by 250 ps of MD in the NVT ensemble and 500 ps in the NPT ensemble with restrained protein heavy atoms. We then performed FEP for 10 ns. An equilibration time period was determined by Yank and removed from the analysis \cite{chodera2016simple}. $\lambda$ windows for FEP were also determined automatically; on average about 70 windows were used for uncharged species and 100 windows for charged ones. Replica exchange among the windows was performed automatically \cite{chodera2011replica}. The total binding free energy was computed with MBAR \cite{shirts2008statistically}. FEP calculations took an average of 4.8 days per uncharged molecule using four GeForce RTX 2080 Ti GPUs. 

CSI Table \ref{tab:fep_statistics} provides a benchmark of FEP and docking against experimental data. Both perform rather well, with FEP outperforming docking (e.g. $R^2$ and Spearman $\rho$ are 0.32 and 0.52 for docking, and 0.69 and 0.81 for FEP). FEP predictions were shifted by $+$4.81 kcal/mol to match the mean of the experimental data. This shift was also applied to virtual screening results, such that $K_{\mathrm{d}} = \mathrm{exp}((\Delta G_{\mathrm{FEP}}  \ + \ \mathrm{shift}) / k_{\mathrm{B}} T)$, where $k_{\mathrm{B}}$ is Boltzmann's constant and $T$ is the temperature. Further details of docking can be found in CSI Sec. \ref{sec:docking}, and details of FEP can be found in CSI Sec. \ref{sec:fep}.

\subsection*{Chemical Property Calculations}
The ground state NFF was trained through active learning (CSI Fig. \ref{fig:barriers_al}).  To make use of previously generated
data, we first pre-trained the models with 680,736 gas-phase spin-flip (SF) TDDFT \cite{shao2003spin} calculations from Ref. \cite{axelrod2022excited}. We then refined the model with implicit solvent SF-TDDFT data from Ref. \cite{axelrod2022thermal}, plus newly generated data from active learning. Calculations were performed with Q-Chem 5.3 \cite{epifanovsky2021software}. This procedure yielded 90,460 calculations in total. However, we found that spin contamination was severe for azonium TSs, and so we fine-tuned the model with data from spin-complete MRSF-TDDFT. We performed MRSF-TDDFT calculations for 17,027 geometries using GAMESS \cite{GAMESS}. 

Activation free energies and p$K_\mathrm{a}$ values were computed as described in CSI Secs. \ref{sec:thermal} and \ref{sec:azonium}. The p$K_\mathrm{a}$ predictions are compared to experimental data in CSI Fig. \ref{fig:azonium_fit}(a). The predictions are quite good, with an $R^2$ value of 0.81, and a mean absolute error (MAE) of only 0.62. 

Half-life performance statistics for species outside the training set are given in CSI Table \ref{tab:model_accuracy}. The model is quite accurate for azobenzene derivatives, with an MAE of 0.77 kcal/mol relative to MRSF-TDDFT for the activation energy ($R^2=0.88$). It is less accurate for azonium, with an MAE of 1.92 kcal/mol. This is mainly due to a subset of geometries with very large errors. Removing errors above 5 kcal/mol, which make up 8\% of the test set, yields an MAE of 1.20 kcal/mol ($R^2=0.90$). Note also that we performed a benchmark against experiment in Ref. \cite{axelrod2022thermal}. 

\subsection*{Photophysical Property Calculations}
The NFF for absorption spectra was trained on excitation energies, gradients, and oscillator strengths from TDDFT with the CAM-B3LYP functional. Geometries were selected from metadynamics simulations of non-TS geometries, which were performed during active learning for half-life predictions. The initial training set was selected randomly. A model was then trained and used to select geometries with high predicted absorption wavelengths. These geometries were added to the training set and the model was re-trained. 83,481 geometries were used in total.

Performance statistics for species outside the training set are given in CSI Table \ref{tab:excited_model_accuracy}. The model is quite accurate: the MAE for absorption wavelengths is 3.51 nm for azobenzene compounds and 7.03 nm for azonium derivatives ($R^2=0.98$ and $0.97$, respectively). The MAE of the transition dipole moment is 0.05 Debye for azobenzene and 0.16 Debye for azonium ($R^2=0.98$ for both). 

The predicted spectrum is in good agreement with experiment for unsubstituted azobenzene. The predicted peak and extent are approximately 450 and 550 nm, respectively, which agree well with the spectrum measured in Ref. \cite{knie2014ortho} in acetonitrile. However, the performance is significantly worse for azonium derivatives. For example, the predicted peak and extent are approximately 509 nm and 604 nm for the azonium compound \textbf{22} in Ref. \cite{dong2017near}; the experimental values in water are 597 nm and 750 nm \cite{dong2017near}. By contrast, the predicted p$K_\mathrm{a}$ for this compound is in good agreement with experiment (predicted and experimental values of 6.6 and 6.7, respectively). This suggests that the problem comes from TDDFT, not from incorrect sampling of ground state geometries.

A final NFF was trained on ground and excited state energies and forces to perform NAMD simulations. This model was trained on MRSF-TDDFT data generated through active learning. We used a similar active learning setup to the one shown in CSI Fig. \ref{fig:barriers_al}, with slight differences in the sampling strategies (see CSI Sec. \ref{sec:qy}). Surface hopping was performed with the Zhu-Nakamura method \cite{yu2014trajectory}, which does not require nonadiabatic coupling vectors. We performed 15 ps of ground state MD, followed by 200 independent excited state simulations. Excited state simulations were initialized from randomly sampled MD geometries and propagated for 4 ps. The quantum yield was computed as the proportion of trajectories that ended in a different isomer. The uncertainty in the yield was computed through statistical bootstrapping with 1,000 samples. The predicted yield of \textit{trans} azobenzene was 43\%, in good agreement with the experimental value of 35\% in an 80-20 ethanol-water mixture \cite{bortolus1979cis}.

\subsection*{Graph-to-Property Model Training}
Graph-based models were trained to predict isomerization barriers, absorption wavelengths, and p$K_\mathrm{a}$ values from NFF simulations. We used the Attentive FP architecture \cite{xiong2019pushing} implemented in the DeepChem library \cite{Ramsundar-et-al-2019}. Model performance is given in CSI Table \ref{tab:graph_to_property}. The models are reasonably accurate, with $R^2$ values between 0.54 and 0.74 depending on the property and training generation. 

\section*{Experimental Methods}
\subsection*{Photostationary State Ratio Measurement}
The sample ($\sim$1~mg, weighed on a microbalance) was weighed into a 1.5~mL HPLC screw cap vial and diluted with CDCl$_3$ through a dilution series, resulting in a final analyte concentration of 1.5~mM in approximately 500~\textmu L of CDCl$_3$. The tube was placed in an NMR-photoreactor, cooled to $-78$~$^\circ$C, and irradiated with an appropriate LED. After a defined time, the tube was quickly transferred to a pre-cooled NMR at $-20$~$^\circ$C. The collected spectra were immediately evaluated, and the irradiation/measurement procedure was repeated until equilibrium was achieved. The custom-made, 3D-printed NMR photoreactor is described in more detail in the ESI, section NMR-photoreactor, pages S38--S39.

\subsection*{UV-Vis Spectrophotometry and Measurement of Thermal Isomerization}
A stock solution ($\sim$3~$\times$~10$^{-4}$~M) was prepared as follows: the sample ($\sim$1~mg, weighed on a microbalance) was placed into a 10~mL volumetric flask and diluted with dry spectroscopy-grade acetonitrile. The stock solution was transferred into a 10~mL amber screw-cap vial, sealed with vinyl duct tape, and stored under argon in a freezer at $-20$~$^\circ$C. Samples were prepared by diluting the stock solution directly into a fluorescence quartz cuvette just before measurement. The cuvette, equipped with a Teflon-coated stirring bar, was capped and placed into a spectrometer-photoreactor. Spectra were recorded while irradiating the sample with an LED until a photostationary equilibrium was reached. Subsequently, the LED was turned off, and the thermal relaxation process was monitored spectroscopically until thermal equilibrium was attained.

Absorption coefficients for the pure (\textit{trans}) isomer were calculated from the initial spectral scan. Thermal relaxation half-lives ($t_{1/2}$) and rate constants ($k$) were determined by fitting the spectral data obtained after LED irradiation was ceased to first-order kinetic models. 

Values are reported as mean \(\pm\) SD of technical triplicates (\(n = 3\)) from a single synthesized batch (\(N = 1\)). Detailed mathematical procedures, schematics of spectrometer-photoreactor, and representative graphs illustrating the thermal isomerization behavior for each compound are available in the ESI, section spectrophotometer-photoreactor, pages S39--S40.

\subsection*{Spectrophotometric Determination of Quantum Yield}
Material ($\sim$1--2~mg) was placed into a 4~mL amber screw-cap vial and dissolved in spectroscopy-grade acetonitrile (4~mL). The sample was transferred into a fluorescence quartz cuvette. The cuvette was equipped with a Teflon-coated stirring bar, capped, and placed into a spectrometer-photoreactor. Spectra were accumulated, and the cuvette was irradiated with an LED until at least 10 full scans were obtained ($\sim$0.5~minute). LEDs were then turned off and the sample was left to thermally re-equilibrate (at least 10~$\times$~$t_{1/2}$).

Values are reported as mean \(\pm\) SD of technical triplicates (\(n = 3\)) from a single synthesized batch (\(N = 1\)). For mathematical details and representative graphs, see the ESI, section Calculation of Experimental Photophysical Properties, pages S183--S185.

\subsection*{Chemiluminescent Assay for the Determination of PARP1 Inhibition Constants}
The PARP1 chemiluminescent assay was optimized for screening in 384-well plates. White 384-well plates with transparent bottoms were used for this purpose. Prior to performing the assay, plates were prepared by charging the wells with Ni$^{2+}$ ions to capture 6His-tagged histones. Briefly, 20~\textmu L of 10~mM BCML solution in 0.1~M sodium phosphate buffer, pH~8.0, was added to each well and incubated overnight at room temperature. After incubation, each well was washed three times with 200~\textmu L of 0.05\% Tween~20.

Subsequently, the plates were blocked by incubating with 3\% BSA in 50~mM Tris-HCl buffer (pH~7.5, containing 150~mM NaCl and 0.05\% Tween~20) for 2~hours at room temperature. After blocking, the plates were sequentially washed with a series of buffers in the following order: first with 50~mM Tris-HCl (pH~7.5, containing 500~mM imidazole and 0.05\% Tween~20), then with 0.05\% Tween~20, followed by 100~mM EDTA (pH~8.0), and finally again with 0.05\% Tween~20. Next, the plates were incubated with 10~mM NiSO$_4$ solution for 20~minutes at room temperature, followed by washing first with 0.05\% Tween~20 and then with 50~mM Tris-HCl buffer containing 500~mM NaCl, pH~7.5. Finally, plates were washed three times with PBS and then immediately used for the PARP1 assay.

PARP1 activity was evaluated using the PARP1 Chemiluminescent Assay Kit according to the manufacturer's instructions, with certain modifications. First, 12.5~\textmu L of 1X histone mixture was added to each well, and the plate was incubated overnight at 4~$^\circ$C. After incubation, the plate was washed three times with PBS-T. Next, wells were blocked by adding 50~\textmu L of blocking buffer per well and incubating at room temperature for 90~minutes.

After blocking, 6.25~\textmu L of master mix (containing PARP1 assay buffer, PARP substrate mixture~1, activated DNA, 0.5~mM DTT solution, and 0.01\% Tween~80) was added to each well. At this step, the plate was placed in a custom-made, 3D-printed bio-photoreactor (described in more detail in the ESI, section Bio-photoreactor, pages S35--S38), which illuminated half of the plate at either 519~nm or 590~nm, while the other half of the plate remained in the dark. After a 10-minute illumination period, 1.25~\textmu L of compound solution at the desired concentrations (ranging from 2~\textmu M to 500~\textmu M) was added to wells containing the master mix. The concentration of DMSO was adjusted to 1\% in all samples, including the blank and positive controls.

The reaction was initiated by adding 5~\textmu L of PARP1 enzyme solution (final reaction concentration: 0.132~ng/mL). Positive controls corresponded to reactions containing enzyme but no tested compound, while blanks corresponded to reactions without enzyme or compound. After incubation for 1~hour at room temperature under either illuminated or dark conditions, the plate was washed three times with PBS-T.

Detection was performed using streptavidin-HRP and an ELISA ECL substrate. Chemiluminescence was measured using a plate reader (Tecan Spark, Switzerland). Blank values were subtracted from all measured signals, and resulting values were normalized to the positive control, defined as 100\%. IC$_{50}$ values were calculated using a custom Python script by fitting data to a four-parameter sigmoid curve using the \texttt{curve\_fit} function from the SciPy module, and plots were generated using the Matplotlib library. Data are reported as mean \(\pm\) SD of technical triplicates (\(n = 3\)) from two or one independent biological replicates ($N = 2$ for measurements at 519~nm and $N = 1$ for measurements at 590~nm). All dose response graphs can be found in the ESI, Figures S102--S112.

\subsection*{Determination of Solubility in PARP1 Buffer}
Compounds (500~\textmu M final concentration; 12.5~\textmu L per well in a 384-well plate) were prepared in 1$\times$ PARP1 assay buffer supplied with the BPS Bioscience chemiluminescent PARP1 activity kit (cat.\ \#80551). According to the manufacturer, the exact buffer formulation is proprietary; however, from our experience such buffers typically contain 50~mM Tris-HCl, 150~mM NaCl, 5~mM MgCl$_2$, and 1~mM DTT at pH~$\approx$~7.5. All samples contained 1\% (v/v) DMSO; where indicated, Tween~80 was added to a final concentration of 0.01\% (v/v). Samples were incubated for 1~hour at room temperature, irradiated at 519~nm, and imaged with a Zeiss Axio microscope (50$\times$ magnification) using AxioVision SE64 Rel.\ 4.9.1. The resulting photomicrographs are shown in the ESI, Figures S99--S101.

\subsection*{Capillary Electrophoretic Determination of Acidity Constants}
Compounds were dissolved in methanol to achieve final concentrations of 0.16--0.25~mg/100~\textmu L. Samples were introduced into the capillary via hydrodynamic injection using pressures of 13.8--20.7~mbar for 10--20~seconds, resulting in injection volumes of 5.1--15.4~nL. Detection was performed using a PDA (200~nm).

Initial screening of all compounds was conducted in acidic background buffer (pH~2.01) using dimethyl sulfoxide (DMSO) as an electroneutral osmotic flow marker. Compounds exhibiting positive charge under these conditions were selected for further analysis.

Selected compounds were subsequently analyzed in ten individually prepared background electrolytes: covering a pH range from 1.71 to 12. All buffers, except the neat NaOH solution (pH = 12), were adjusted to an ionic strength of 25~mM and equilibrated at 25~$^\circ$C before use. From the pH-dependent effective electrophoretic mobilities ($\mu_\mathrm{eff}$), mixed acidity constants (p$K_\mathrm{a,mix}$) and actual ionic mobilities of the cationic forms ($\mu_{1+}$) were determined. 

These mixed constants, which relate to the activity of hydronium ions at the experimental ionic strength of 25~mM, were subsequently converted to thermodynamic acidity constants (p$K_\mathrm{a,th}$) using the Debye--H\"uckel theory as described.\cite{Solinova2013determination}

Values are reported as mean \(\pm\) SD of technical triplicates (\(n = 3\)) from a single synthesized batch (\(N = 1\)). Electropherograms, pH dependence of effective mobilities, p$K_\mathrm{a1,th}$, p$K_\mathrm{a2,th}$, p$K_\mathrm{a1,mix}$, p$K_\mathrm{a2,mix}$, and $\mu_{1+}$ are presented in the ESI, Table S2, Figures S87--S98.

\section*{Acknowledgments}
This work was financially supported by DARPA (Award HR00111920025) and the Institute of Organic Chemistry and Biochemistry of The Czech Academy of Sciences (RVO 61388963). Harvard Cannon cluster, MIT Engaging cluster, and MIT Lincoln Lab Supercloud cluster at MGHPCC are gratefully acknowledged for computational resources and support. We also thank Dr.\ Ji\v{r}\'i Kaleta, Dr.\ Tom\'a\v{s} Slanina, and Dr.\ Milan Ma\v{s}\'at for their valuable photochemical consultations. We gratefully acknowledge Dr. Lucie Bednárová for her consultations regarding vibrational spectra; Dr.\ Tom\'a\v{s} Ma\v{s}ek for his assistance with retrosynthetic design; Dr.\ Robert Reiberger for his deep Buchwald--Hartwig amination expertise; Dr. Karel Kudlá\v{c}ek for assisting with solubility studies.  Dr.\ Veronika Vet\'y\v{s}kov\'a and Dr.\ Jan Vold\v{r}ich for their consultations regarding the PARP1 assay; Dr.\ Ema Chaloupeck\'a for assistance during NMR measurements; Dr.\ Ond\v{r}ej Tich\'a\v{c}ek for his expert consultation on mathematical implementation in Python scripts; Dr.\ Jakub Copko and Dr.\ Eva Bedn\'a\v{r}ov\'a for their assistance with UV-Vis spectroscopy; Dr. Jaroslava Hniličková for measuring Karl Fischer titrations; and Jakub Bro\v{z} for his help with CAD modeling and 3D printing.

\clearpage
\newpage

\setcounter{page}{1}
\setcounter{section}{0} 
\setcounter{subsection}{0} 
\setcounter{subsubsection}{0} 
\setcounter{equation}{0} 
\setcounter{table}{0}
\setcounter{figure}{0}
\renewcommand\thefigure{S\arabic{figure}}
\renewcommand\thetable{S\arabic{table}}

\clearpage
\FloatBarrier

\renewcommand\tablename{Supplementary Table}
\renewcommand\figurename{Supplementary Figure}

\onecolumn

\begin{center}
\section*{Computational Supplementary Information} 
\end{center}

\section{Properties of top candidates}

\label{sec:top_candidate_data}
\begin{figure}[H]
    \centering
    \includegraphics[width=\textwidth]{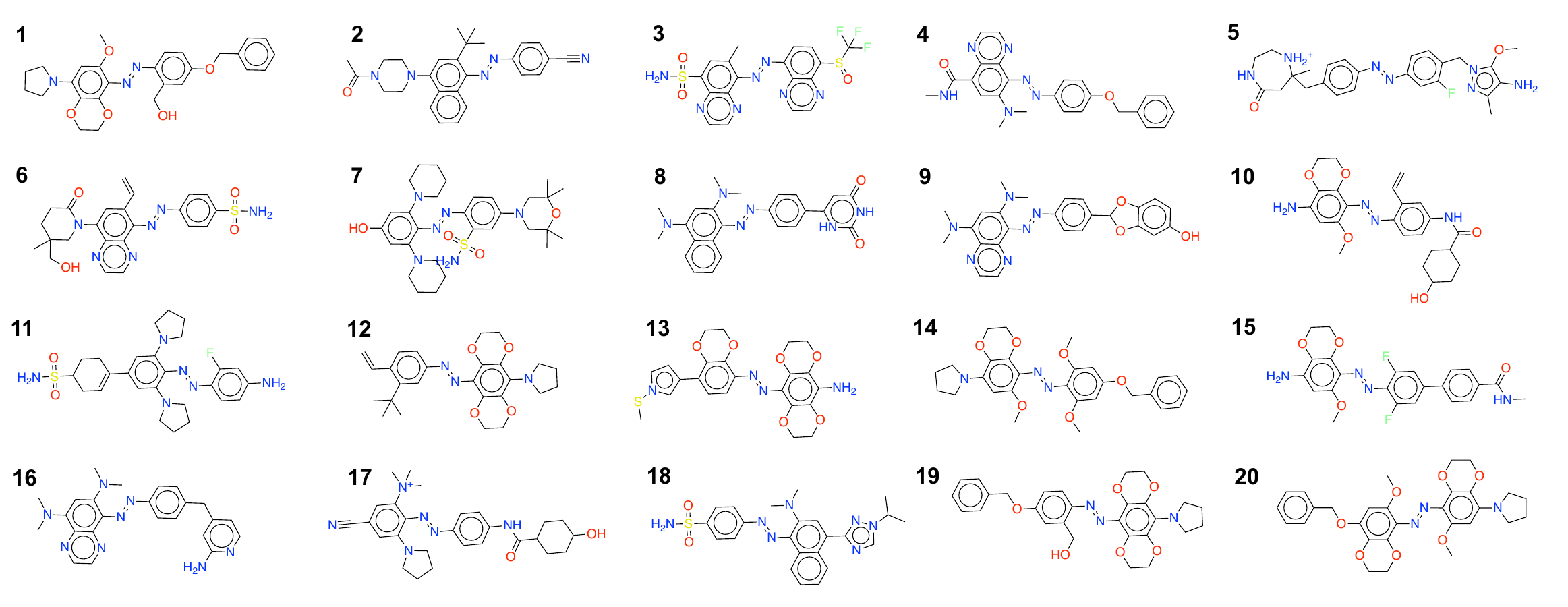}
\caption{Candidates that reached the final stage of screening (NAMD and FEP).}
\label{fig:final_candidates}
\end{figure}

Here we provide FEP results and quantum chemistry data for the top candidates. CSI Fig. \ref{fig:final_candidates} shows the candidates that reached the final stage of screening, before being filtered based on NAMD and FEP results. Compounds \textbf{14} and \textbf{20} were generated manually based on \textbf{1}. CSI Table \ref{tab:fep_data} shows the FEP results for molecules \textbf{1} to \textbf{20}. CSI Table \ref{tab:dg_data} contains chemical and photophysical data for the candidates with predicted $K_{\mathrm{d}} < $ 5 mM.

\section{Molecule generation}
\label{sec:molgen}
\subsection{Base library}
\begin{figure}[t]
    \centering
    \includegraphics[width=0.5\columnwidth]{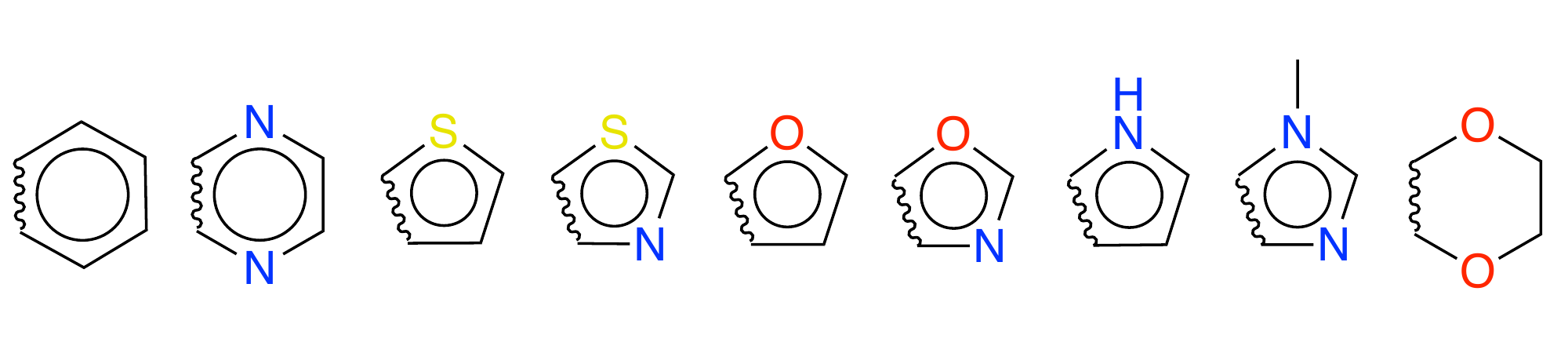}
\caption{Groups fused to the azobenzene core. The fused bond is shown with a wavy line.}
\label{fig:fused_rings}
\end{figure}

Following our previous work \cite{axelrod2022thermal}, we combined common literature functional groups and substitution patterns to create a virtual library of azobenzene derivatives. We also expanded the library and filtered the molecules based on drug-likeness and synthetic accessibility (Fig. \ref{fig:workflow}(a)). To expand the library, we fused five- and six-member rings to the \textit{ortho} and \textit{meta} positions of benzene in azobenzene (CSI Fig. \ref{fig:fused_rings}). We used the conjugated rings in Ref. \cite{mukadum2021efficient}, plus the six-member di-methoxy ring in Ref. \cite{dong2017near}. Fusing conjugated rings to benzene can red-shift the absorption spectrum by expanding the $\pi$ conjugated system. The di-methoxy ring can red-shift absorption by stabilizing the azonium species \cite{dong2017near}. The oxygens act as hydrogen bond acceptors, which stabilize azonium and hence increases its presence in solution \cite{dong2017near}. We further expanded our library by adding drug-like fragments from the Enamine REAL catalogue \cite{enamine}. We used all compounds with 11 heavy atoms or fewer in the Essential Fragment Library and the High Fidelity Fragment Library. We also used the basic fragments from Fig. 2-4 in Ref. \cite{cheron2016opengrowth}.

\begin{table}[H]
\begin{tabular}{c|cccccc|ccc} 
  & \multicolumn{6}{c|}{\textit{cis}} & \multicolumn{2}{c}{\textit{trans}} \\ 
 \hline \hline
  Compound & $\Delta G^{\dagger}$ & $\Delta G^{\dagger}_{\mathrm{calib}}$ & p$K_\mathrm{a}$ & $\Delta G_{\mathrm{prot}}$ & $\Delta G^{\dagger, \mathspace \mathrm{eff}}_{\mathrm{calib}}$ & $t_{1/2}$ & p$K_\mathrm{a}$& $\lambda_{5\% \ \mathrm{max}}$  \\
 \hline
  \textbf{1}, azobenzene & 26.2 & 23.6 & -- & -- & 23.6 & 84 & -- & 534 \\
  \textbf{1}, azonium a & 15.3 & 12.7 & 1.4 & 8.2 & \underline{20.9} & \underline{1.0} & 2.1 & 667 \\
  \textbf{1}, azonium b & 26.9 & 24.2 & 3.9 & 4.8 & 29.1 & $5.9 \times 10^5$ & 6.8 & 514 \\
 \hline
  \textbf{2}, azobenzene & 23.5 & 20.9 & -- & -- & \underline{20.9} & \underline{1.0} & -- & 617 \\
  \textbf{2}, azonium a & $23.7^*$ & 21.1 & 3.4 & 5.5 & 24.3 & $2.8 \times 10^2$ & $-5.6$ & 657 \\
  \textbf{2}, azonium b & 23.7 & 21.1 & $-3.7$ & 15.2 & 34.2 & $2.5 \times 10^9$ & 0.9 & 530 \\
 \hline
 \textbf{3}, azobenzene & 26.3 & 23.7 & -- & -- & \underline{23.7} & \underline{101} & -- & 649  \\
  \textbf{3}, azonium a & 20.0 & 17.4 & $-5.6$ & 17.7 & 35.1 & $1.0 \times 10^{10}$ &  $-4.8$ & 534 \\
  \textbf{3}, azonium b & 19.7 & 17.1 & $-7.4$ & 20.2 & 37.3 & $3.9 \times 10^{11}$ & $-0.5$ & 510 \\
 \hline
  \textbf{7}, azobenzene & 23.8 & 21.2 & -- & -- & \underline{21.2} & \underline{1.7} & -- & 619  \\
  \textbf{7}, azonium a & $21.7^*$ & 19.1 & 3.4 & 5.4 & 24.5 & $3.3 \times 10^2$ & 0.8  & 531 \\
  \textbf{7}, azonium b & 21.7 & 19.1 & 3.5 & 5.3 & 24.3 & $2.8 \times 10^2$ & 1.5  & 712 \\
 \hline
  \textbf{8}, azobenzene & 26.2 & 23.6  & -- & -- & 23.6 & 87 & -- & 600  \\
  \textbf{8}, azonium a & $20.1^*$ & 17.5 & 4.4 & 4.1 & \underline{21.6} & \underline{3.4} & 0.3 & 643 \\
  \textbf{8}, azonium b & 20.1 & 17.5 & $-0.8$ & $11.2$ & 28.7 & $3.2 \times 10^5 $ & 4.8 & 460 \\
 \hline
\end{tabular}
\caption{Chemical and photophysical predictions for compounds with predicted \textit{cis} binding affinities below 5 mM. $\Delta G^{\dagger}$ is the activation free energy for thermal isomerization, and $\Delta G^{\dagger}_{\mathrm{calib}}$ is the calibrated activation free energy, shifted so that the value for azobenzene matches experiment. $\Delta G_{\mathrm{prot}}$ is the free energy cost of protonating the molecule, and $\Delta G^{\dagger, \mathspace \mathrm{eff} }_{\mathrm{calib}}$ is the effective value of $\Delta G^{\dagger }_{\mathrm{calib}}$ after adding the protonation cost. Compounds are separated into azobenzene and their two azonium forms. We have underlined the lowest value of $\Delta G^{\dagger, \mathrm{eff}}_{\mathrm{calib}}$ and its associated half-life. An asterisk indicates that TS optimization did not converge, and so the azonium barrier was simply taken as the barrier of the other azonium form. Units are kcal/mol for free energies, minutes for half-lives, and nm for absorption wavelengths. Calculations were performed using MRSF-TDDFT with BHHLYP applied to NFF critical points for all quantities other than $\lambda$. For $\lambda$ we used TDDFT with the CAM-B3LYP functional. Further details can be found in CSI Secs. \ref{sec:thermal}, \ref{sec:azonium}, and \ref{sec:absorption}. }
\label{tab:dg_data}
\end{table}

\begin{table}[H]
\begin{tabular}{c|cccccccc} 
  Compound & $\Delta G^{\mathrm{FEP}}$ (kcal/mol) & $\Delta G^{\mathrm{FEP}}_{\mathrm{calib}}$ (kcal/mol) & $K_{\mathrm{d}}$ ($\mu M$) \\
  \hline
  \textbf{1} \textit{cis} & $-9.4$ & $-4.5$ & \underline{466} \\
  \textbf{1} \textit{trans} & $-7.6$ & $-2.7$ & $9.8\times 10^3$ \\
  \hline
  \textbf{2} \textit{cis} & $-9.2$ & $-4.4$ & \underline{591} \\
  \textbf{2} \textit{trans} & $-8.1$ & $-3.3$ & \underline{$4.0 \times 10^3$}\\
  \hline
  \textbf{3} \textit{cis} & $-8.5$ & $-3.7$ & \underline{$2.1 \times 10^3$}  \\
  \textbf{3} \textit{trans} & $-5.0$ & $-0.2$ & $6.7 \times 10^5$ \\
  \hline
  \textbf{4} \textit{cis} & $-5.1$ & $-0.3$ & $6.2 \times 10^5$   \\
  \hline
  \textbf{5} \textit{cis} & $-6.9$ & $-2.1$ & $3.0 \times 10^4$ \\
  \hline
  \textbf{6} \textit{cis} & $-3.8$ & $1.0$ & $5.3 \times 10^6$\\
  \hline
  \textbf{7} \textit{cis} & $-9.3$ & $-4.5$ & \underline{534} \\
  \textbf{7} \textit{trans} & $-10.0$ & $-5.2$ & \underline{163} \\
  \hline
  \textbf{8} \textit{cis} & $-10.0$ & $-5.2$ & \underline{154} \\
  \textbf{8} \textit{trans} & $-10.5$ & $-5.7$ & \underline{70} \\
  \hline
  \textbf{9} \textit{cis} & $-6.2$ & $-1.4$ & $1.0 \times 10^5$ \\
  \hline
  \textbf{10} \textit{cis} & $-3.6$ & $1.2$ & $7.6 \times 10^6$ \\
  \hline
  \textbf{11} \textit{cis} & $-7.5$ & $-2.7$ & $9.9 \times 10^3$ \\
  \hline
  \textbf{12} \textit{cis} & $-4.8$ & $0.0$ & $1.1 \times 10^6$ \\
  \hline
  \textbf{13} \textit{cis} & $-4.5$ & $0.3$ & $1.6 \times 10^6$ \\
  \hline
  \textbf{14} \textit{cis} & $-2.6$ & $2.2$ & $4.2 \times 10^7$ \\
  \hline
  \textbf{15} \textit{cis} & $-0.6$ & $4.2$ & $1.2 \times 10^9$ \\
  \hline
  \textbf{16} \textit{cis} & $-7.8$ & $-3.0$ & $6.6 \times 10^3$ \\
  \hline
  \textbf{17} \textit{cis} & $-3.9$ & $0.9$ & $4.9 \times 10^6$ \\
  \hline
  \textbf{18} \textit{cis} & $-7.1$ & $-2.3$ & $2.0 \times 10^4$ \\
  \hline
  \textbf{19} \textit{cis} & $-8.8$ & $-4.0$ & \underline{$1.2\times 10^3$} \\
  \textbf{19} \textit{trans} & $-7.9$ & $-3.1$ & \underline{$5.0\times 10^3$} \\
  \hline
  \textbf{20} \textit{cis} & $0.4$ & $5.2$ & $6.7 \times 10^9$ \\
  \hline
\end{tabular}
\caption{FEP binding predictions for candidates that reached the final stage of screening (CSI Fig. \ref{fig:final_candidates}). $\Delta G^{\mathrm{FEP}}$ is the binding free energy computed with FEP. $\Delta G^{\mathrm{FEP}}_{\mathrm{calib}}$ is the calibrated binding free energy, computed by adding a shift of 4.81 kcal/mol to match experiment (see CSI Table \ref{tab:fep_statistics}). $K_{\mathrm{d}}$ is the dissociation constant computed from $\Delta G^{\mathrm{FEP}}_{\mathrm{calib}}$. Binding affinities below 5 mM are underlined. \textit{Trans} binding scores are only shown for \textit{cis} compounds with binding affinities below 5 mM.}
\label{tab:fep_data}
\end{table}

\begin{figure}[t]
    \centering
    \includegraphics[width=0.8\textwidth]{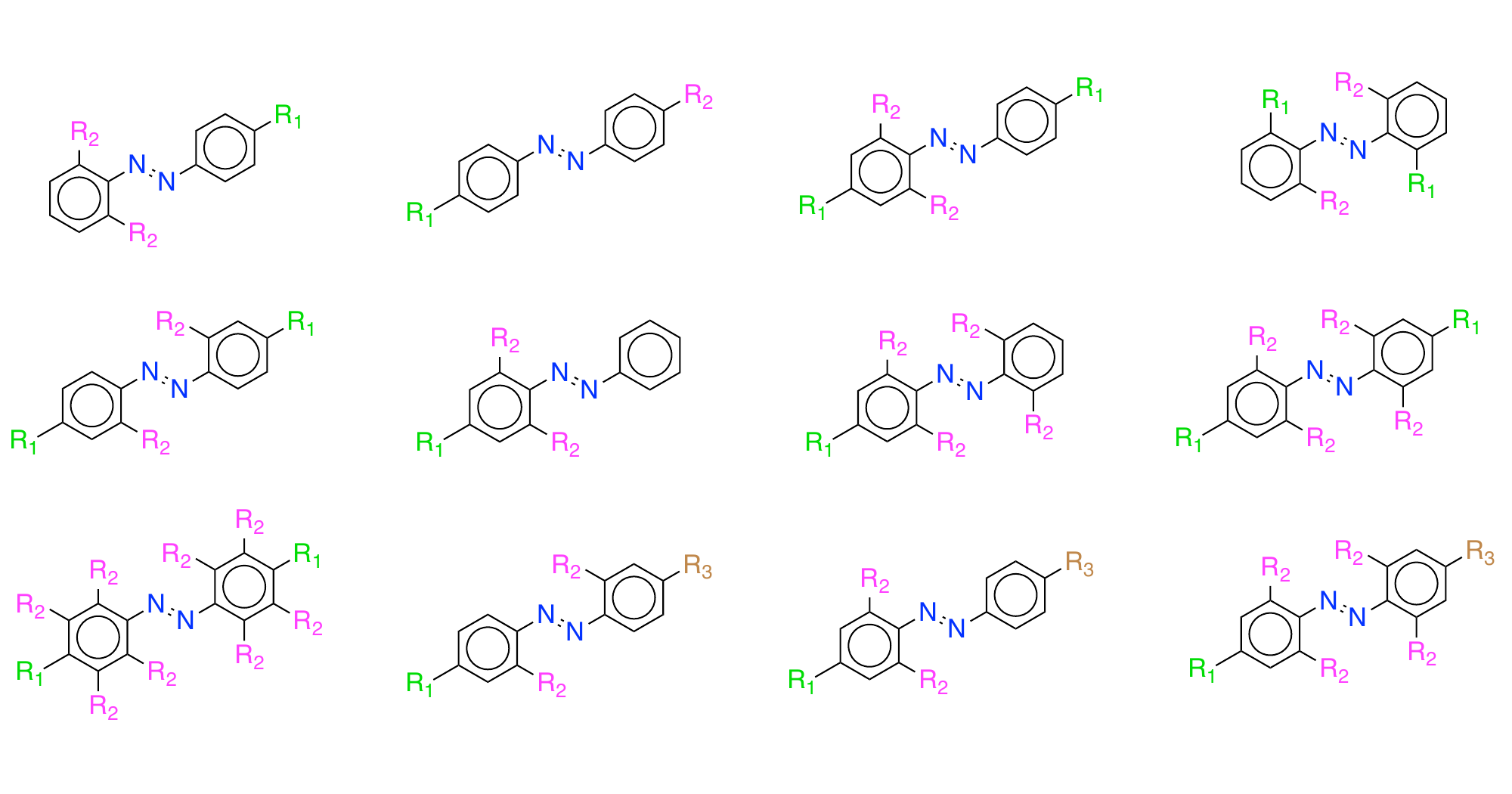}
\caption{Substitution patterns used for combinatorial molecule generation.}
\label{fig:substitution_patterns}
\end{figure}

The substitution patterns are shown in CSI Fig. \ref{fig:substitution_patterns}. We enumerated all possible combinations of patterns and fragments. For each generated molecule, we optionally added a small substituent with $\leq$ 4 atoms to a random carbon on the azobenzene scaffold. When adjacent \textit{ortho} and \textit{meta} positions were unsubstituted, we fused one of the rings in CSI Fig. \ref{fig:fused_rings} with benzene at up to two different sites. More fusions would bring the number of aromatic rings above four, which is already quite high for drug-like molecules (see Fig. 1g in Ref. \cite{bickerton2012quantifying}). 

We then narrowed our search space to synthetically accessible drug-like molecules with a reasonable chance of having good pharmacokinetic properties. To do so we used several different filters. First we used the quantitative estimate of drug-likeness (QED) \cite{bickerton2012quantifying}, a continuous generalization of Lipinski's rule of 5 \cite{lipinski2012experimental} that estimates drug-likeness on a scale of 0 to 1 (higher is better). We filtered out all molecules with QED $< 0.4$. We then removed compounds that failed any of the medicinal chemistry filters (MCFs) in Ref. \cite{polykovskiy2020molecular}, apart from the N=N filter and the limitation of $\leq$ 3 halogens. The first filter was removed because N=N is present in azobenzene. This bond can be cleaved by azoreductase \cite{misal2018azoreductase}, but may be metabolically stable if the molecule is appropriately substituted \cite{brown1993predicting}. The halogen filter was removed because tetra-halogenation is a good red-shifting strategy \cite{gelabert2023predicting}. We also removed groups containing pan-assay interfering compounds (PAINS) \cite{capuzzi2017phantom, senger2016filtering}. These groups often lead to false positives through promiscuous binding. Code for filtering by PAINS and MCFs was accessed from Ref. \cite{polykovskiy2020molecular}. We also inspected a few hundred compounds by hand. We found a small number of non drug-like groups, such as aldehydes and nitroso groups, and removed all molecules that contained them.

Lastly, we filtered molecules by synthetic accessibility. We used the synthetic accessibility (SA) score \cite{ertl2009estimation} and synthetic complexity (SC) score \cite{coley2018scscore}. The SA score uses a number of chemistry-informed rules to estimate accessibility, and its scores are well-correlated with those of synthetic organic chemists \cite{ertl2009estimation}. The SC score uses a neural network trained on published reactions. It is trained to implicitly recognize the number of steps needed to synthesize a compound \cite{coley2018scscore}. While less accurate than the SA score \cite{coley2018scscore}, it is nevertheless a useful second filter for reducing the number of compounds. We removed molecules with SA scores above 5 (scale of 1 to 10; lower is better) and SC scores above 4 (scale of 1 to 5; lower is better).

After generating the candidate pool and filtering by drug-likeness and synthetic accessibility, we were left with 9.5 million molecules in total. Half of these were \textit{cis} and half were \textit{trans}. We then generated two azonium compounds for each species, one for each nitrogen protonation site. In some cases the two azonium compounds were equivalent by symmetry. Adding the azonium species yielded 26.8 million molecules in total. These molecules were sampled randomly in each round of active learning, while azobenzenes were sampled together with their two azonium species during screening.

\subsection{Library expansion}

\begin{figure}[t]
    \centering
    \includegraphics[width=0.8\textwidth]{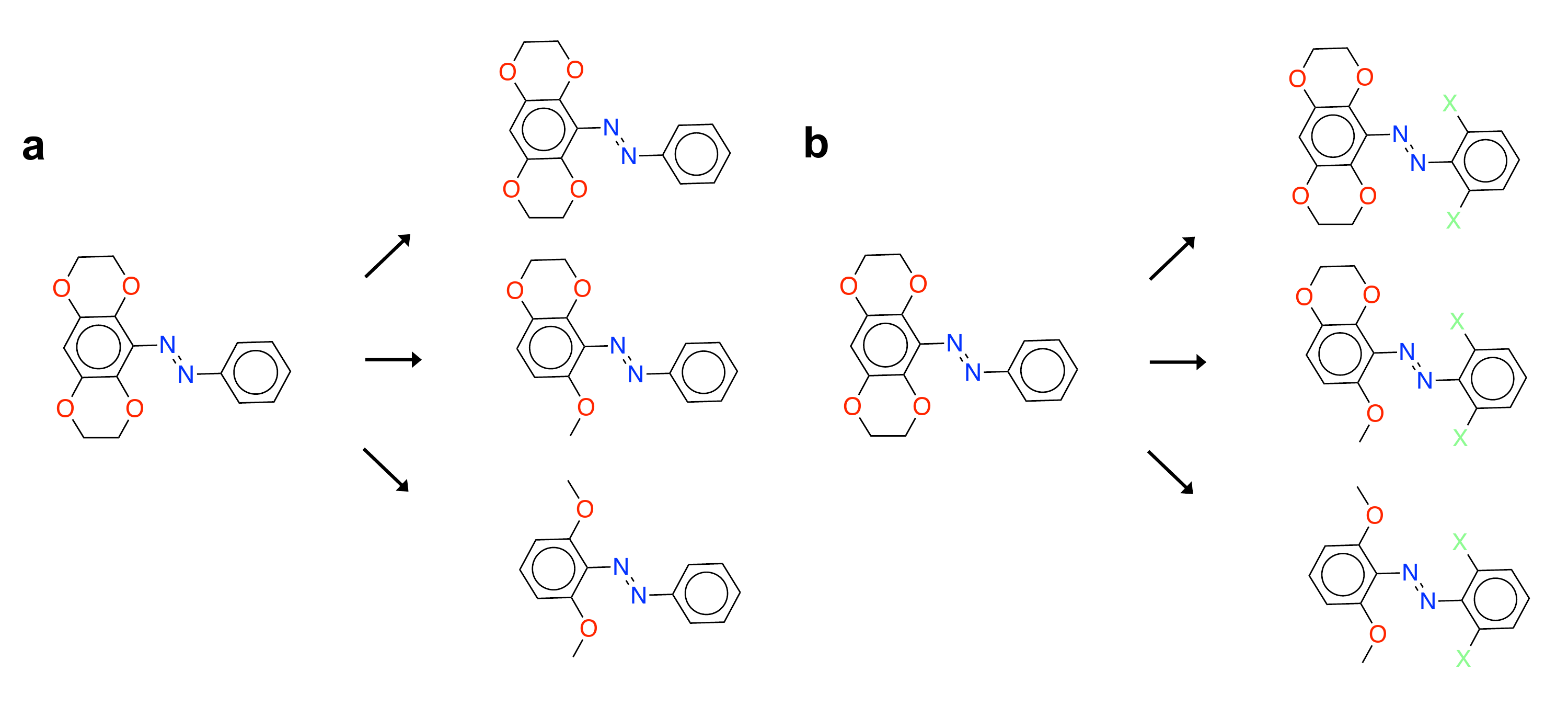}
    \caption{Strategies used to expand the library after docking. (a) For all docking hits with at least one fused dioxane ring, all permutations of (dioxane $\to$ methoxy) and (dioxane $\to$ dioxane) were generated. (b) For a hit with one or two dioxanes fused to the same benzene, two halogens were added in the \textit{ortho} positions on the other benzene. All permutations of (dioxane $\to$ methoxy) and (dioxane $\to$ dioxane) were also generated. }
    \label{fig:dioxane_iteration}
\end{figure}

After docking all 10 million compounds, we expanded our virtual library in three ways. First, we identified docking hits with at least one fused dioxane ring. These scaffolds are of interest because they were used in Ref. \cite{dong2017near} to increase the \textit{trans} p$K_\mathrm{a}$, and hence red-shift the absorption spectrum. We generated new species by converting some dioxane rings to methoxy groups in the \textit{ortho} position (CSI Fig. \ref{fig:dioxane_iteration}(a)). This strategy was also used in Ref. \cite{dong2017near}. We performed this step to increase the pool of hits with the potential for achieving near-neutral p$K_\mathrm{a}$.

Second, for hits with one or two dioxanes fused to the same ring, we added two halogens to the \textit{ortho} position of the other benzene ring (CSI Fig. \ref{fig:dioxane_iteration}(b)). We also generated all combinations of methoxy and dioxane for these compounds. We used fluorine, chlorine, and bromine as the halogen substituents. We added halogens because azonium compounds tend to have low barriers \cite{dong2017near}, while \textit{ortho} halogenation tends to increase barriers \cite{knie2014ortho}. \textit{Ortho} halogenation red-shifts the absorption spectrum, with the size of the effect increasing when moving down the periodic table \cite{gelabert2023predicting}.

Third, for hits with two \textit{ortho} acetylpiperazine substituents, we added two \textit{ortho} halogens on the opposite benzene ring. In our first round of screening we found some hits with this piperazine motif, similarly to our previous work \cite{axelrod2023mapping}, and so we generated more candidates with this scaffold. We used \textit{ortho} halogenation for the same reasons described above.

We filtered the new molecules by drug-likeness and synthetic accessibility, as described above, and docked the remaining compounds. After filtering by docking score, we were left with 898 \textit{cis}-\textit{trans} pairs in total. All 898 molecules were then screened for chemical and photophysical properties, independent of the predictions of the graph-to-property model.

\section{Virtual screening}

In the first NFF stage of screening, we began by randomly sampling 4,000 molecules. 1,000 were chosen to have hydrogen bond acceptors in the \textit{ortho} position, or one atom away from it. These compounds received p$K_\mathrm{a}$ and azonium barrier calculations. After performing NFF calculations on these molecules, we trained 2D models to predict the p$K_\mathrm{a}$, the maximum absorption wavelength at 5\% of the spectrum peak, and the thermal half-life. We trained an ensemble of three models for each property. The models were then used to select 4,000 new species. A maximum of 50\% were chosen based on optimal chemical and photophysical properties or high p$K_\mathrm{a}$ values. This was called exploitation. 25\% were chosen to maximize the model uncertainty, as measured by the variance of the ensemble predictions. 25\% were chosen randomly. The latter two were called exploration; their purpose was to explore areas of chemical space in which the model was not well-trained. The selected molecules received new NFF calculations. Molecules chosen because of high \textit{trans} p$K_\mathrm{a}$ values, and molecules with \textit{ortho} hydrogen bond acceptors, received p$K_\mathrm{a}$ and azonium barrier calculations.

We performed three rounds of screening and 2D model training. Only exploitation was used in the final round. We then validated the chemical and photophysical predictions with quantum chemistry. We kept all molecules with predicted half-lives between 5 seconds and 5 hours (raw $\Delta G^{\dagger}$ between 22 and 27 kcal/mol), and $\lambda_{5\% \ \mathrm{max}}$ above 600 nm. 

\section{ML-accelerated simulation}

Properties were computed through ML-accelerated simulation, using methods that we developed in previous work \cite{axelrod2022excited,axelrod2022thermal,axelrod2023mapping}. To compute p$K_\mathrm{a}$ values, we used an NFF to compute the free energies of the azonium and azo forms of the molecules. We then used a linear regression to fit free energy differences to experimental p$K_\mathrm{a}$ values, as described in CSI Sec. \ref{sec:azonium}. To compute absorption spectra, we performed MD with the same NFF, and computed oscillator strengths and absorption energies for each frame with a second NFF. The histogram of the absorption energies weighed by the oscillator strengths gave the spectrum, as described in CSI Sec. \ref{sec:absorption}. For the quantum yield, we used a third NFF to perform NAMD on the top candidates. This model was trained on both $S_0$ and $S_1$ energies and forces, and was used with the Zhu-Nakamura surface-hopping method to predict isomerization quantum yields \cite{yu2014trajectory}. 

\begin{figure}[t]
    \centering
    \includegraphics[width=\textwidth]{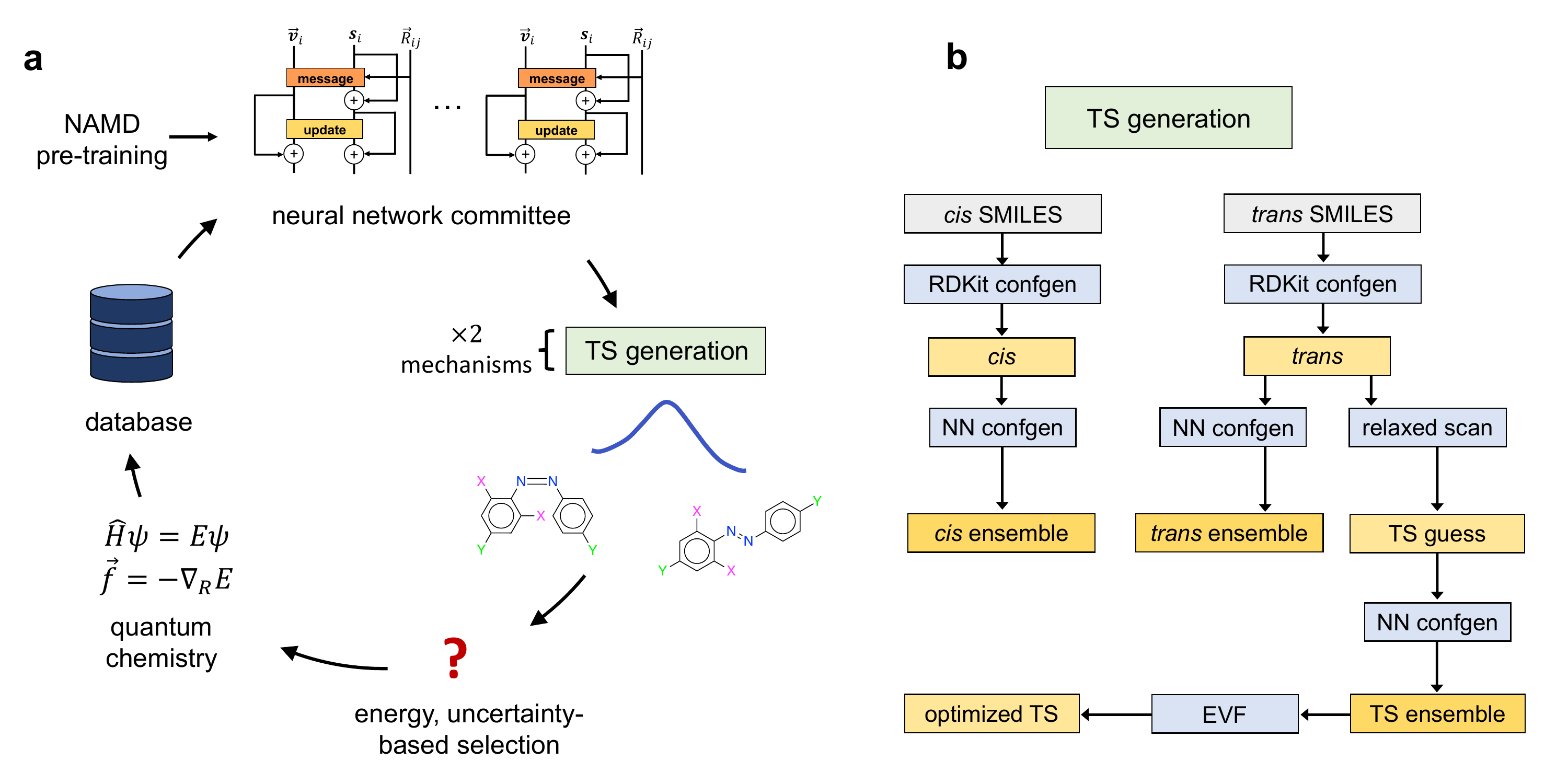}
    \caption{Approach to active learning for TS generation. (a) Active learning loop for training the NN. (b) Workflow for generating equilibrium and TS geometries. ``NN confgen'' means conformer generation with the NFF using metadynamics. Initial conformations were generated with RDKit \cite{rdkit} and refined with NN conformer generation. TSs were generated through relaxed scans followed by conformer generation and eigenvector following (EVF). }
    \label{fig:barriers_al}
\end{figure}

Thermal half-lives were computed with the ground state NFF using the workflow developed in Ref. \cite{axelrod2022thermal} and shown in CSI Fig. \ref{fig:barriers_al}(b). In Ref. \cite{axelrod2022thermal} we argued that thermal isomerization in azobenzene occurs through rotation mediated by intersystem crossing (ISC).  However, we also found a strong correlation between rates from ISC and rates from Eyring transition state (TS) theory using a rotational TS. Therefore, to avoid computing triplet forces to train a separate NFF, we simply used the ground-state singlet NFF and computed rates from Eyring TS theory.
We report half-lives below by subtracting a constant shift in $\Delta G^{\dagger}$, equal to the difference between the computed and experimental values for azobenzene (28.0 kcal/mol and 25.4 kcal/mol in benzene \cite{talaty1967thermal}, respectively).

The NFFs used for ground-state properties and NAMD were trained on mixed-reference spin-flip time-dependent density functional theory (MRSF-TDDFT) data \cite{lee2018eliminating,lee2019efficient}. We used the BHHLYP functional \cite{becke1993new}, the 6-31G* basis \cite{francl1982self}, D3-BJ dispersion \cite{grimme2010consistent, grimme2011effect}, and a C-PCM description of water \cite{truong1995new, barone1998quantum, cossi2003energies}. MRSF-TDDFT can account for some multi-reference effects, which are critical for NAMD \cite{axelrod2022excited} and for TS searches in azobenzene \cite{axelrod2022thermal}. The excited-state NFF used for absorption spectra was trained with TDDFT using the CAM-B3LYP functional \cite{yanai2004new}, the def2-SVP basis \cite{weigend2005balanced}, and a C-PCM model of water \cite{truong1995new, barone1998quantum, cossi2003energies}. Further quantum chemistry details can be found in CSI Secs. \ref{sec:thermal}, \ref{sec:absorption}, and \ref{sec:qy}.


All NFF models used the PaiNN architecture \cite{schutt2021equivariant}, which predicts molecular properties through equivariant message-passing. This approach generates a feature vector for each atom that incorporates information from its surrounding environment. The initial feature vector is generated from the atomic number alone, and is then updated through a set of ``messages'' that incorporate the distance, orientation, and features of atoms within a cutoff distance. This process is repeated several times to gather information from increasing distances. 

Geometries for training the models were generated through active learning (CSI Fig. \ref{fig:barriers_al}(a)). Three ground state models were used to generate TSs for a set of sampled reactions, and geometries from the TS workflow were sampled based on energy and uncertainty to receive new quantum chemistry calculations. The new data was added to the training set and the models were retrained. A similar process was used for NAMD. The excited-state model used for absorption spectra was trained on metadynamics samples of \textit{cis} and \textit{trans} equilibrium geometries. Further details of the training process and model accuracy can be found in Methods and CSI Secs. \ref{sec:thermal}, \ref{sec:absorption}, and \ref{sec:qy}.

All models performed well relative to the quantum chemistry methods used for training (see Methods and CSI Tables \ref{tab:model_accuracy}, \ref{tab:excited_model_accuracy}, and \ref{sm_tab:namd_accuracy}). With the exception of azonium absorption wavelengths, the models also performed well relative to experiment (see Methods, CSI Table \ref{tab:fep_statistics}, and CSI Fig. \ref{fig:azonium_fit}(a)).

\section{Protein-ligand binding}
\subsection{PARP1}
Previously we docked 9,000 azobenzene derivatives into 58 targets using Dockstring \cite{axelrod2023mapping}. We showed that the docking scores of \textit{cis} and \textit{trans} isomers were highly correlated for most targets. We also found that \textit{trans} ligands bind more strongly than \textit{cis} to most proteins. Based on this work, we identified PARP1 as a promising photo-druggable protein for several reasons. The first is that it is a cancer target. The second is that \textit{cis} isomers bind more strongly to PARP1 than \textit{trans} isomers, which is uncommon in the proteins that we studied \cite{axelrod2023mapping}. This is desirable because one usually wants the inactive drug to be thermodynamically stable (\textit{trans}). If the inactive drug were thermodynamically unstable (\textit{cis}), then it would eventually become active throughout the whole body due to thermal isomerization. The last reason is that docking scores for PARP1 are fairly reliable \cite{garcia2022dockstring}. This makes it a good target for computational screening. It is especially important given that FEP is quite slow, and so it is important for a reasonable proportion of the docking hits to be actual hits.

\subsection{Docking}
\label{sec:docking}
Docking was performed with Dockstring \cite{garcia2022dockstring}. This open-source package performs docking into one of 58 different proteins, and requires only the ligand SMILES string and protein name as input. It performs (de)protonation at physiological pH using OpenBabel \cite{o2011open}, generates initial ligand conformers using RDKit \cite{rdkit}, and performs docking with AutoDock Vina \cite{trott2010autodock}. The protein structures come from the DUD-E database \cite{mysinger2012directory}. The search boxes were specified by hand by the authors of Ref. \cite{garcia2022dockstring} for each protein. We docked all 9.5 million azobenzene derivatives into PARP1.

\subsection{FEP}
\label{sec:fep}
\subsubsection{Theory}
Docking scores were improved through absolute FEP calculations \cite{chodera2011alchemical}. These calculations use a thermodynamic cycle with two stages. In the first stage, the ligand is decoupled from the protein and the solvent. A restraint is usually added to keep the uncoupled ligand close to the protein. In the second stage, the ligand is re-coupled to the solvent in the absence of the protein. The free energy change of the cycle, plus a correction for the restraint, is then the free energy difference between the solvated ligand and the bound ligand. 
\begin{figure}[t]
    \centering
    \includegraphics[width=0.25\columnwidth]{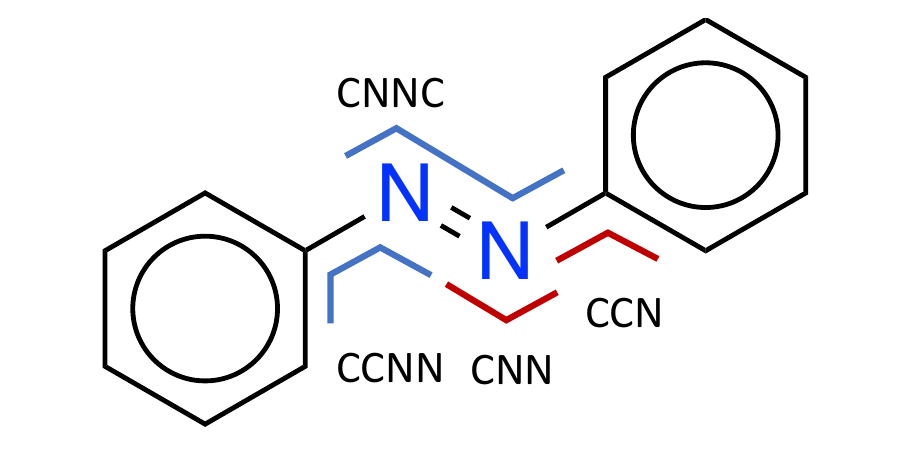}
    \caption{Internal coordinates of interest in azobenzene. Dihedrals are shown in blue and angles in red. The NN and CN bonds are those within the CNNC dihedral. }
    \label{fig:azo_indices}
\end{figure}
\begin{table*}[t!]
\begin{tabular}{c|ccc} 
  & Atoms & $r_0$ & $k$ \\
 \hline
 \multirow{2}{*}{\textit{cis}} & NN & 1.225 & 1792.543 \\
 &  CN & 1.420 & 669.216 \\
 \hline
 \hline
 \multirow{2}{*}{\textit{trans}} & NN & 1.234 & 1732.791 \\
 & CN & 1.405 & 717.017 \\
 \hline
\end{tabular}
\caption{New bond parameters for \textit{cis} and \textit{trans} azobenzene. Units are \AA \ for distances and kcal/mol/\AA\textsuperscript{2} for force constants. }
\label{tab:bond_parameters}
\end{table*}

\begin{table*}[t!]
\begin{tabular}{c|ccc} 
  & Atoms & $\theta_0$ & $k$ \\
 \hline
 \multirow{2}{*}{\textit{cis}} & CCN & 111.0 & 59.751 \\
 &  CNN & 116.0 & 143.403 \\
 \hline
 \hline
 \multirow{2}{*}{\textit{trans}} & CCN & 116.0 & 95.602 \\
 & CNN & 110.5 & 167.304 \\
 \hline
\end{tabular}
\caption{New angle parameters for \textit{cis} and \textit{trans} azobenzene. Units are degrees for angles and kcal/mol/$\mathrm{rad}^2$ for force constants. }
\label{tab:angle_parameters}
\end{table*}

\begin{table*}[t!]
\begin{tabular}{c|ccccc} 
  & Atoms & $\phi_1$ & $\phi_2$ & $k_1$ & $k_2$ \\
 \hline
 \multirow{2}{*}{\textit{cis}} & CNNC & 0.0 & 0.0 & 0.0 & $-9.560$ \\
 &  CCNN & 180.0 & 180.0 & 0.0 & 1.076   \\
 \hline
 \hline
 \multirow{2}{*}{\textit{trans}} & CNNC & 0.0 & 0.0 & 0.0 & $-8.485$  \\
 &  CCNN & 0.0 & 0.0 & 0.0 & $-1.649$ \\
 \hline
\end{tabular}
\caption{New dihedral parameters for \textit{cis} and \textit{trans} azobenzene. Units are degrees for dihedral angles and kcal/mol for force constants. }
\label{tab:dihedral_parameters}
\end{table*}

The free energy differences are computed with MD runs from a series of windows, called $\lambda$ windows. The Hamiltonians of these windows interpolate between the fully coupled system and the fully uncoupled system. An estimator such as the multi-state Bennett acceptance ratio (MBAR) \cite{shirts2008statistically} is used to determine the free energy change between the first and last window, using MD samples from all windows.

\subsubsection{Implementation}
Absolute FEP calculations were performed with Yank (version 0.25.2) \cite{yank}. Yank is an open-source package that uses OpenMM to run MD \cite{eastman2013openmm}, determines $\lambda$ windows and MD equilibration times automatically \cite{chodera2016simple}, performs replica exchange among windows to accelerate sampling \cite{chodera2011replica}, and computes the binding free energy with MBAR. It also provides convergence analysis and uncertainty estimates. We ran Yank for 10 ns per window, using an average of 70 windows for uncharged species (further details below). This took an average of 4.8 days per molecule using four GeForce RTX 2080 Ti GPUs.

Each run was divided into segments of 1 ps of MD, followed by displacement and rotation of the ligand with Monte Carlo moves. Replica exchange was then performed among all windows using Gibbs sampling \cite{chodera2011replica}. This process was repeated 10,000 times to generate equilibrium samples for all windows. For MD we used the NPT ensemble with the \texttt{LangevinSplittingDynamicsMove} function, which performs Langevin dynamics using the BAOAB split integrator \cite{leimkuhler2013rational}. We used a time step of 2 fs and a collision rate of 1 $\mathrm{ps}^{-1}$. Velocities were sampled from the Maxwell-Boltzmann distribution at the start of each segment. The pressure was enforced with the default Monte Carlo barostat in Yank. The temperature was set to 298.15 K and the pressure to 1 atmosphere. 

\begin{figure*}[t!]
    \centering
    \includegraphics[width=\textwidth]{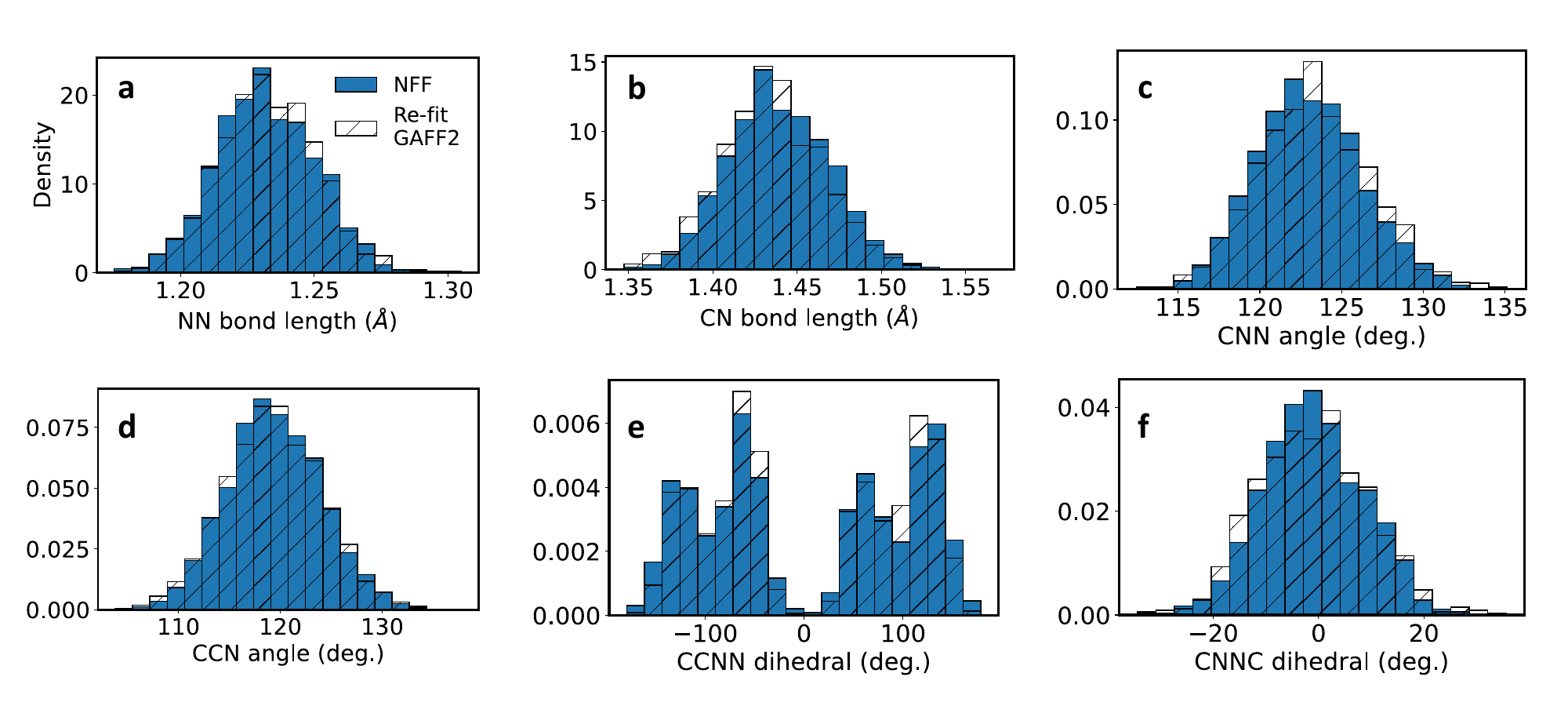}
    \caption{Distribution of internal coordinates for \textit{cis} azobenzene, computed with our NFF (solid blue) and our re-fit GAFF2 classical force field (hatched white). 
    }
    \label{fig:nff_vs_new_classical_cis}
\end{figure*}

\begin{figure*}[t!]
    \centering
    \includegraphics[width=\textwidth]{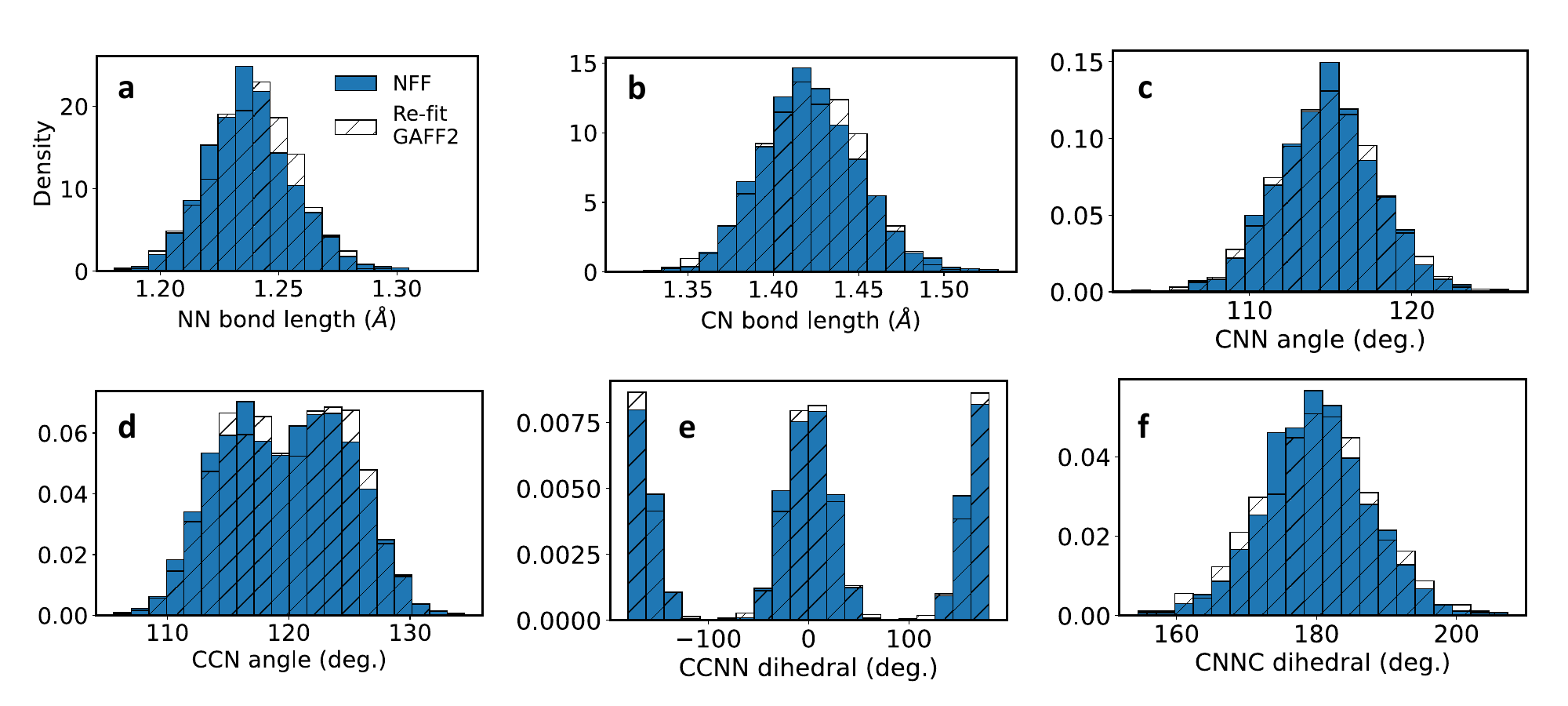}
    \caption{Distribution of internal coordinates for \textit{trans} azobenzene, computed with our NFF (solid blue) and our re-fit GAFF2 classical force field (hatched white).  }
    \label{fig:nff_vs_new_classical_trans}
\end{figure*}

\begin{figure*}[t!]
    \centering
    \includegraphics[width=\textwidth]{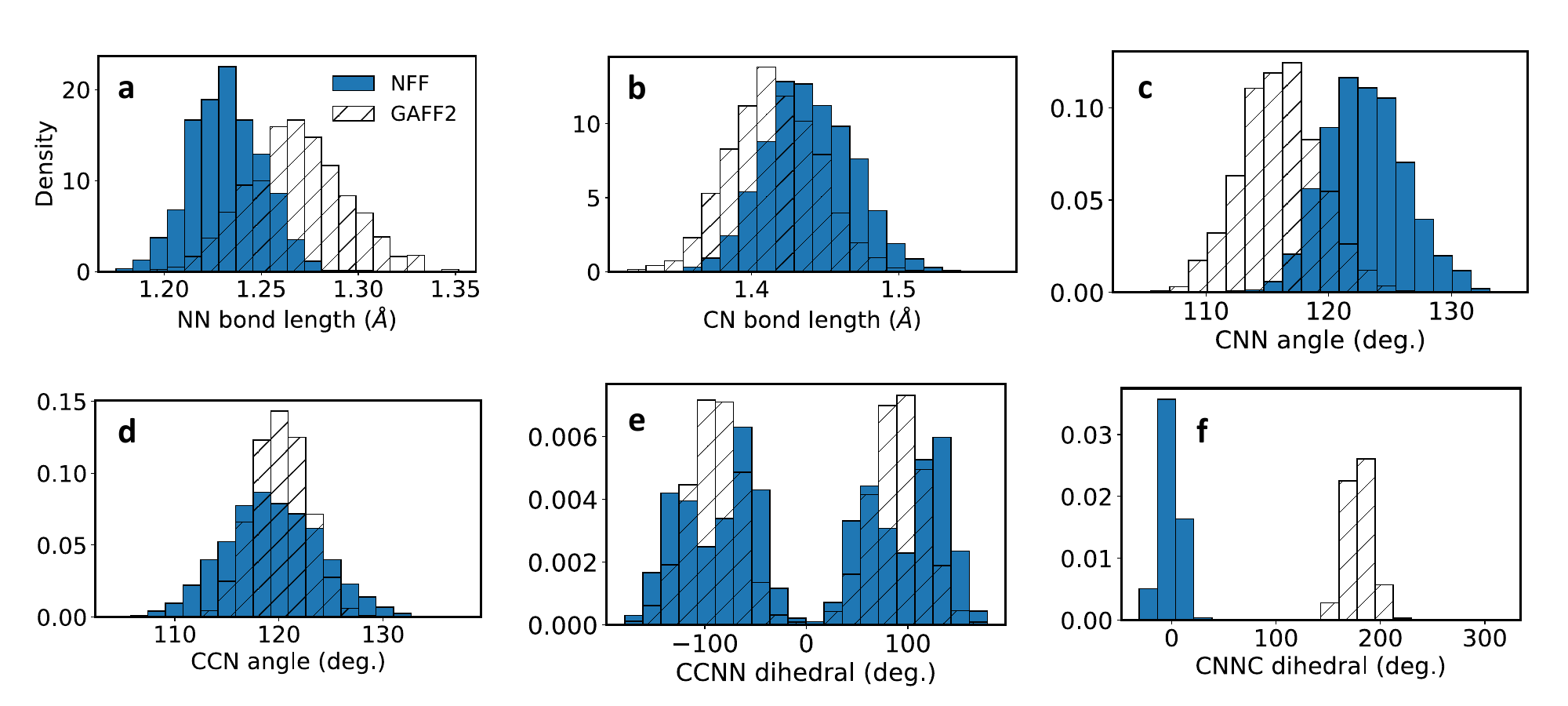}
    \caption{Distribution of internal coordinates for \textit{cis} azobenzene, computed with our NFF (solid blue) and the unmodified GAFF2 classical force field (hatched white).  }
    \label{fig:nff_vs_old_classical_cis}
\end{figure*}

\begin{figure*}[t!]
    \centering
    \includegraphics[width=\textwidth]{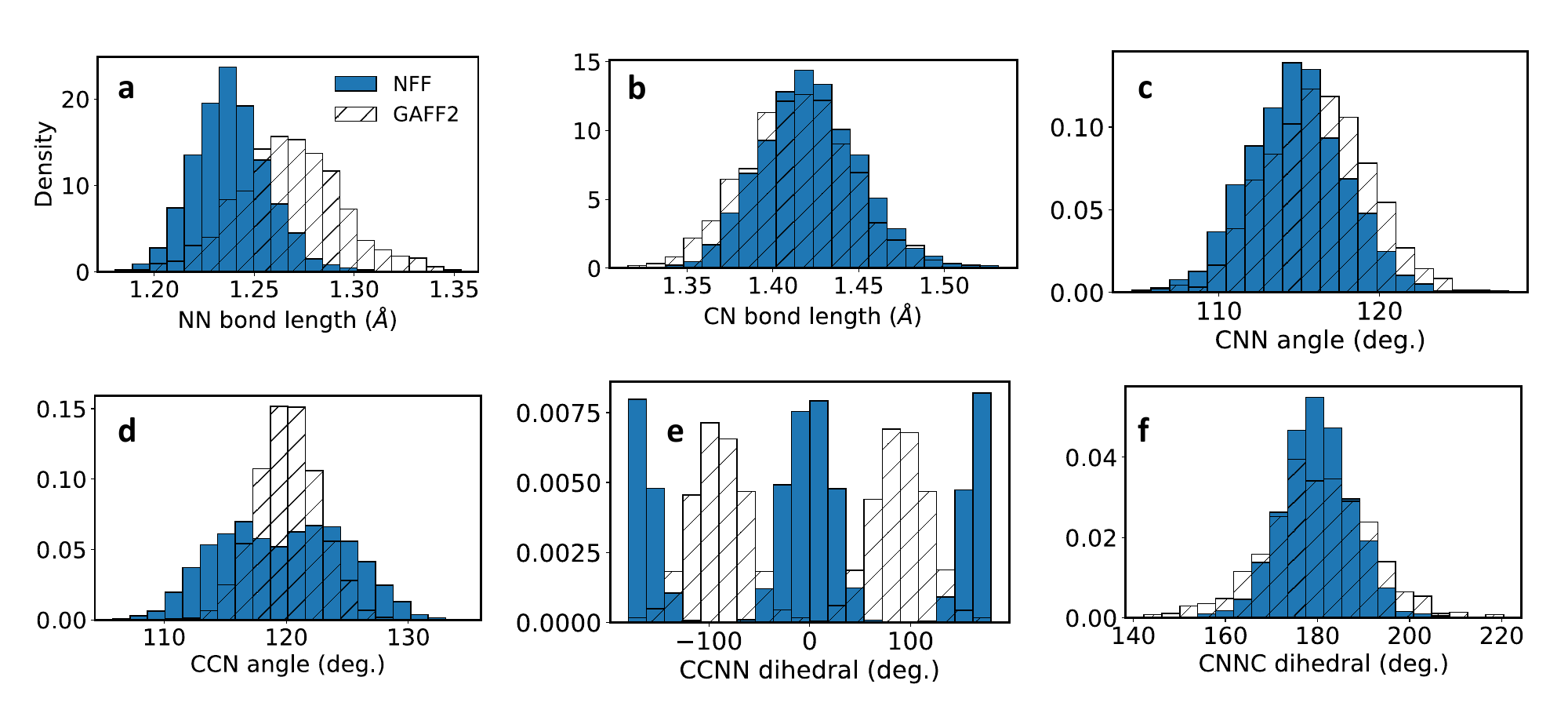}
    \caption{Distribution of internal coordinates for \textit{trans} azobenzene, computed with our NFF (solid blue) and the unmodified GAFF2 classical force field (hatched white).  }
    \label{fig:nff_vs_old_classical_trans}
\end{figure*}
Windows were chosen automatically by Yank. Yank determines a schedule for first turning off electrostatic interactions and then turning off van der Waals interactions. In the protein-ligand stage, it also determines a schedule for turning on the restraint, chosen here to be harmonic. The electrostatic interactions were turned off by linearly scaling all ligand charges to zero. The van der Waals interactions were turned off by scaling Lennard–Jones terms to zero using a soft-core schedule \cite{pham2011identifying}, with $(a, b, c, \alpha)$ in Ref. \cite{pham2011identifying} set to $(1, 1, 6, 0.5)$. All windows were chosen such that the standard deviation of energy differences between states were close to equal. These states were generated through an iterative algorithm that uses samples from short MD runs in each window. On average, about 70 windows were used in the protein-ligand stage for neutral species, and 100 were used for charged species. For the solvent stage, 60 and 90 windows were used for neutral and charged molecules, respectively. 

Long-range electrostatics were computed with the particle-mesh Ewald method \cite{essmann1995smooth}. We used a non-bonded cutoff of 9 \AA \ and a box clearance of 10 \AA. We used $\mathrm{Na}^+$ and $\mathrm{Cl}^-$ ions to form a salt buffer of 150 mM, and to neutralize any net charges in the system.

\subsubsection{Force fields}
\label{subsec:fep_force_fields}
We used the ff14SB force field \cite{maier2015ff14sb} for the protein and the TIP4P-Ew model \cite{horn2004development} for water. In our benchmark against experimental data, we used the GAFF2 force field \cite{wang2004development} with AM1-BCC partial charges \cite{jakalian2000fast}. We used GAFF atom types instead of GAFF2 atom types; surprisingly, we found that this significantly improved the agreement with retrospective experimental data. Charges were computed on the lowest-energy structure generated by a conformer search. We performed conformer searches with CREST \cite{pracht2020automated} for the benchmark and our NN version of CREST \cite{axelrod2022thermal} for azobenzene screening. 

For prospective calculations of azobenzene derivatives, we used GAFF2 with a re-parameterization of the azo terms (see below), again with GAFF atom types. We also re-ran FEP with GAFF2 atom types and compared to the GAFF/GAFF2 predictions (not shown). We found that the top GAFF/GAFF2 binders were also good GAFF2/GAFF2 binders. However, there were some strong GAFF2/GAFF2 binders that were not good GAFF/GAFF2 binders (e.g. compounds \textbf{11} and \textbf{13}). We chose not to focus on these compounds because of the lack of FEP consensus, and because of synthetic accessibility concerns.

We re-fit the azo terms because GAFF2 gave poor results for azobenzene. The internal coordinates of interest for azobenzene are shown in CSI Fig. \ref{fig:azo_indices}. We modified parameters for the NN and CN bonds, the CNN and CCN angles, and the CCNN and CNNC dihedrals. We used harmonic functions for bonds and angles, and periodic functions for dihedrals:
\begin{align}
& V_{\mathrm{bond}}(r) = \frac{1}{2} k (r - r_0)^2 \\
& V_{\mathrm{angle}}(\theta) = \frac{1}{2} k (\theta - \theta_0)^2 \\
& V_{\mathrm{dihedral}}(\phi) = \sum_{n=1}^{2} k_n (1 + \mathrm{cos}(\phi - \phi_n)).
\end{align}
Here $V$ is the potential energy, $k$ is the force constant, $r$ is the bond length, $\theta$ is the angle, and $\phi$ is the dihedral angle. The new parameters are given in CSI Tables \ref{tab:bond_parameters}-\ref{tab:dihedral_parameters}; they were determined by hand to reproduce distributions from NFF dynamics. 

NFF distributions are compared with re-fitted GAFF2 distributions in CSI Figs. \ref{fig:nff_vs_new_classical_cis} and \ref{fig:nff_vs_new_classical_trans}. Statistics were collected from 1 ns simulations in vacuum using the Nos\'e-Hoover thermostat \cite{nose1984unified, hoover1985canonical} at 298.15 K. Classical FF simulations were performed with Gromacs \cite{abraham2015gromacs}, and NFF simulations were performed with ASE \cite{ase-paper}. No bond constraints were used. Distributions from equivalent internal coordinates were aggregated in each plot (there are two equivalent CNN angles, four equivalent CCN angles, and four equivalent CCNN dihedrals). 
\begin{figure*}[t!]
    \centering
    \includegraphics[width=\textwidth]{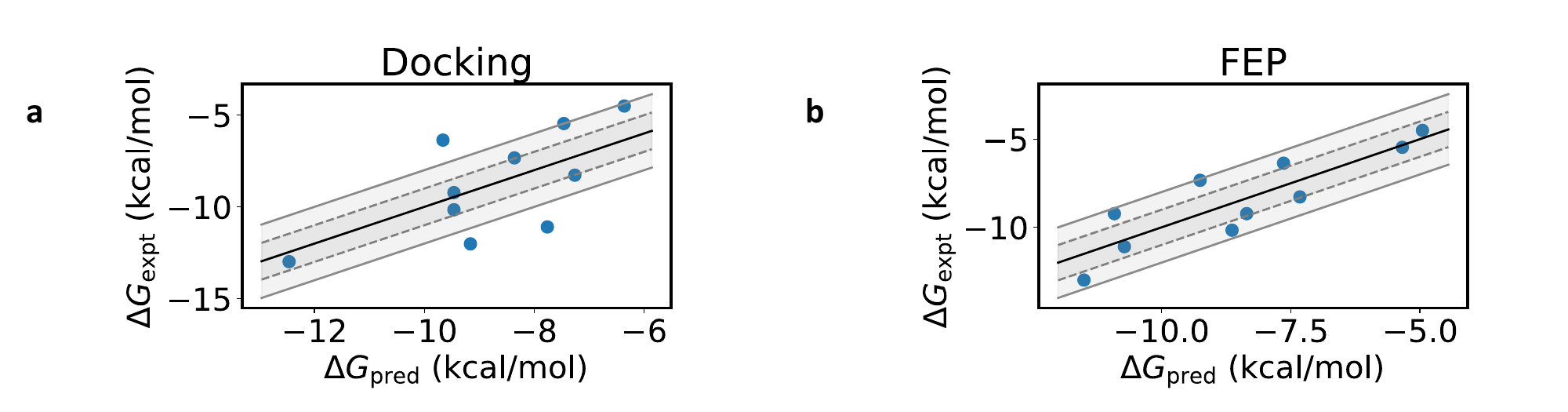}
    \caption{Docking and FEP results for the PARP1 benchmark. 1 kcal/mol and 2 kcal/mol ranges are shown with dashed and solid gray lines, respectively. All results are shifted by a method-dependent constant, $\Delta G_{\mathrm{shift}}$, equal to the mean of the experimental results minus the mean of the predictions. The constants and the prediction statistics are given in CSI Table \ref{tab:fep_statistics}. (a) Docking and (b) FEP run for 10 ns.}
    \label{fig:fep_results}
\end{figure*}

All distributions agree quite well. The optimized geometries also agree well: the \textit{cis} and \textit{trans} geometries have RMSDs of only 0.048 and 0.041 \AA \ with respect to NFF geometries. By contrast, the agreement is poor for standard GAFF2 (CSI Figs. \ref{fig:nff_vs_old_classical_cis} and \ref{fig:nff_vs_old_classical_trans}). \textit{Cis}-azobenzene isomerizes to \textit{trans}, while the CCNN dihedrals in \textit{trans} azobenzene are centered at the wrong angles. The latter leads to non-planarity of the optimized \textit{trans} molecule. The CCN distributions are also too narrow, and the CNNC distributions are too wide. 
\begin{table*}[t!]
\begin{tabular}{c|cccccc} 
  & $R$ & $R^2$ & $\rho$ & MAE & RMSE & $\Delta G_{\mathrm{shift}}$ \\
 \hline
 \multirow{2}{*}{Docking} &  0.63  & 0.32  &  0.52 & 1.68 & 1.98 & 1.64 \\
 & [0.18, 0.90] & [$-$0.21, 0.73] & [$-$0.12, 0.92] & [1.09, 2.27]  & [1.39, 2.52] \\
 \hline
 \multirow{2}{*}{FEP (10 ns)} & 0.86  & 0.69  & 0.81 & 1.05 & 1.19 & 4.81 \\
 & [0.68, 0.97] & [0.33, 0.89] & [0.49, 1.0] & [0.74, 1.34] & [0.92, 1.44] \\
 \hline
\end{tabular}
\caption{Statistics of docking and FEP predictions for the PARP1 benchmark. $R$ is the Pearson $R$ value, $R^2$ is the coefficient of determination, and $\rho$ is the Spearman rank correlation coefficient. 95\% confidence intervals are given in square brackets below; they were computed with statistical bootstrapping using 1,000 samples. $\Delta G_{\mathrm{shift}}$, MAE, and the root-mean-squared error (RMSE) are given in kcal/mol. }
\label{tab:fep_statistics}
\end{table*}

Modified ligand force field parameters were supplied to Yank through \texttt{frcmod} files. To check that we rendered these files correctly, we used Yank functions to load their parameters, and used the results to create OpenMM simulations of azobenzene. We then ran 1 ns simulations in OpenMM in the NVT ensemble, and confirmed that the distributions matched the NFF results.

\subsubsection{Initial structures}
To generate initial structures for FEP, we started with docked protein-ligand complexes, and then relaxed them with restrained MD. This was necessary to ensure stable Yank simulations, as starting with docked complexes often led to divergences. 

Relaxations were performed in Gromacs. We started with the docked structure of the ligand and a PARP1 structure from the PDB database (PDB ID 3L3M). This is the same structure used in DUD-E and hence in docking, apart from a few extra atoms that had been removed by the DUD-E curators. The structure was prepared using PDBFixer \cite{pdb_fixer}. We performed 50,000 minimization steps with steepest descent, followed by 50,000 steps with conjugate gradient descent, each with a force convergence tolerance of 5 kJ/mol/nm. We then performed 250 ps of MD in the NVT ensemble at 298.15 K, followed by 500 ps of MD in the NPT ensemble at 1 atm. Protein heavy atoms were restrained with a force constant of 1,000 kJ/mol/$\mathrm{nm}^2$.
\begin{figure*}[t]
	\centering
	\includegraphics[width=\textwidth]{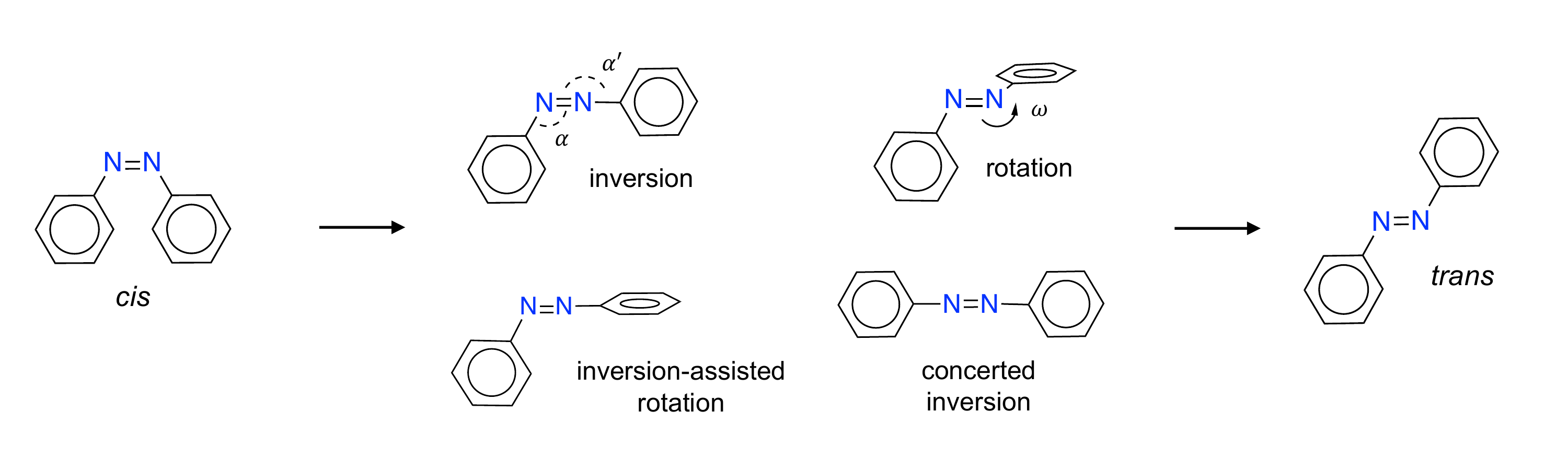}
	\caption{Possible mechanisms of thermal isomerization. The inversion TS has $\alpha$ or $\alpha' \approx 180^{\circ}$, while the rotational TS has $\omega \approx 90^{\circ}$. Inversion-assisted rotation combines inversion and rotation. Concerted inversion combines $\alpha$ inversion and $\alpha'$ inversion.  }
	\label{fig:mechanisms}
\end{figure*}

Gromacs simulations were performed with the ff14SB protein force field and the TIP3P water model \cite{jorgensen1983comparison}. The re-fitted GAFF2 force field was used with AM1-BCC charges for the ligand. Ligand files were converted from PDB to \texttt{mol2} using OpenBabel \cite{o2011open} followed by Antechamber \cite{wang2001antechamber}. The ligand \texttt{mol2} files were supplied to \texttt{acpype} \cite{sousa2012acpype}. The \texttt{acpype} script was locally modified to allow the GAFF2 force field to be used. We used \texttt{acpype} to produce Gromacs input files, which were then modified to account for changes in the GAFF2 force field. Long-range interactions were treated with PME, using a 10 \AA \ cutoff for long-range electrostatics and a 10 \AA \ cutoff for Lennard-Jones terms. All bonds were constrained with the LINCS algorithm \cite{hess1997lincs,hess2008p}. The Nos\'e-Hoover thermostat \cite{nose1984unified, hoover1985canonical} was used for NVT and NPT simulations, and the Berendsen barostat \cite{berendsen1984molecular} was used for NPT simulations. Separate heat baths were used for the solute and the solvent.

Frames from the NPT MD run were clustered using the single linkage method, with a clustering cutoff of 1 \AA. The centroid of the largest cluster was used as the starting point for FEP. The structure was split into protein and ligand PDB files; the protein PDB file was supplied to Yank, while the ligand PDB file was converted to \texttt{mol2} using OpenBabel and Antechamber. A \texttt{frcmod} file was supplied with the custom azobenzene parameters.

\subsubsection{Benchmark}
\label{subsec:fep_benchmark}
To evaluate the performance of FEP and docking, we performed a benchmark of 10 ligands with experimental PARP1 binding data. IC$_{50}$ values for human PARP1 were retrieved for 1,542 ligands from the ExCape database \cite{sun2017excape, excape_site}. We converted IC$_{50}$ values to binding free energies through $\Delta G_{\mathrm{bind}} = k_{\mathrm{B}} T \cdot \mathrm{log} \ K_{\mathrm{d}}$, where we approximated the binding dissociation constant as $K_{\mathrm{d}} \approx \mathrm{IC50}$. We selected the ten ligands with binding free energies that evenly interpolated between the highest and lowest values of $\Delta G$. All selected ligands were neutral.

Docking was performed with Dockstring, using the same setup as in azobenzene screening. FEP was performed as described above. Dockstring (de)protonates each ligand with OpenBabel, using a default pH of 7.4. However, this brings two complications when comparing to experiment. The first is that different experiments use different pHs, so care must be taken to choose the right pH for each data point. The second is that the ligand protonation state can change upon binding. In particular, for a ligand that is protonated at a given pH, the binding free energy is
approximately the lower of the following two options: \begin{enumerate}
    \item The binding free energy of the \textit{protonated} ligand.
    \item The binding free energy of the \textit{unprotonated} ligand, plus the free energy cost of deprotonation. The latter is equal to $(k_{\mathrm{B}} T \  \mathrm{log} \ 10 ) \cdot (\mathrm{p}K_{\mathrm{a}} - \mathrm{pH}) $, which follows from the Henderson-Hasselbalch equation.
\end{enumerate}
Therefore, for the most accurate results, one should compute the binding free energy for both protonation states. The lower of (protonated binding free energy, unprotonated binding free energy $+$ deprotonation cost) should then be compared to experiment. We used this approach in FEP for the two ligands in the benchmark that were protonated by OpenBabel. We also tried using it for docking, but found that performance was unchanged. For docking we therefore only used the protonated forms generated by OpenBabel. For azobenzene screening, we only ran FEP with the protonated forms from OpenBabel to save computational time. 

FEP and docking scores were adjusted by a constant shift to match the mean of the experimental data (using a linear regression did not improve results for either). 
Results are shown in CSI Fig. \ref{fig:fep_results} and CSI Table \ref{tab:fep_statistics}. The docking results are fairly good, which is consistent with the performance reported in Ref. \cite{garcia2022dockstring}. The Spearman rank correlation coefficient, which is the most important metric for ordering hits, is above 0.5. The Pearson correlation coefficient is above 0.6, while the $R^2$ value is only 0.32. The FEP results are better, with a Spearman rank correlation of 0.81, $R^2$ of $0.69$, and $R$ of $0.86$. The RMSE is under 1.2 kcal/mol.

\section{Chemical properties}
\subsection{Thermal half-lives}
\label{sec:thermal}
\subsubsection{Mechanisms}
The possible mechanisms for thermal azobenzene isomerization are given in CSI Fig. \ref{fig:mechanisms}. Previously we argued that isomerization occurs through partial rotation to $\omega \approx 70^{\circ}$, followed by intersystem crossing to a lower-energy triplet state, and then re-crossing to the singlet state at $\omega \approx 105^{\circ}$ \cite{axelrod2022thermal}. We also showed that the isomerization rate from singlet rotation is highly correlated with that of singlet-triplet-singlet rotation. Therefore, to avoid the extra cost of calculating triplet gradients to train a triplet model, we simply computed the barrier of singlet rotation during screening. Since rotation can occur in the clockwise or counter-clockwise direction, this led to two mechanisms per species.

While inversion might be favored over singlet- or triplet-mediated rotation for certain derivatives, preliminary work suggests that this is rare \cite{axelrod2022thermal}. Moreover, affordable quantum chemical methods are not accurate enough to determine when inversion is favored \cite{axelrod2022thermal}. Nevertheless, we also generated inversion TSs during training (but not screening), so that the model could be fine-tuned with more accurate methods in future work.
\begin{table*}[t!]
\begin{tabular}{c|cccccccc} 
  & Optimizer & LR & $\mathrm{LR}_{\mathrm{min}}$ & Patience & Factor & Train & Validation & Test \\
  \hline
  All models & Adam \cite{kingma2014adam} & 1e$-4$ & 5e$-6$ & 50 & 2 & 94\% & 3\% & 3\% \\
 \hline
\end{tabular}
\caption{Training parameters used for all models. LR is the initial learning rate, $\mathrm{LR}_{\mathrm{min}}$ is the minimum learning rate, and patience is the number of epochs without improvement in the validation loss before dropping the learning rate. Factor is the factor by which the learning rate is dropped. Train, validation, and test refer to the proportion of the total dataset used for the corresponding split. }
\label{tab:training_details}
\end{table*}

\begin{table*}[t!]
\begin{tabular}{c|cc|cc|cc|c} 
   & $E_0$ & $ \vec{F}_0$ & $E_1$ & $\vec{F}_1$ & $E_1 - E_0$  & $\vec{F}_{1} - \vec{F}_0$ & $\vec{\mu}_{01}$ \\
  \hline
  Thermal barrier & 0.4 & 1 & -- & -- & -- & -- & -- \\
  Absorption spectrum & -- & -- & -- & -- & 1.0 & 0.2 & 4.0  \\ 
  Quantum yield & 0.4 & 1.0 & 0.4 & 1.0 & 1.0 & 1.0 & -- \\ 
 \hline
\end{tabular}

\caption{Loss coefficients for different models. All models use an MAE loss, with different coefficients for different quantities. $E_i$ is the energy of state $i$, $\vec{F}_i$ are the forces in state $i$, and $\vec{\mu}_{ij}$ is the transition dipole moment between states $i$ and $j$. Training was performed in units of kcal/mol for energies, kcal/mol/\AA \ for forces, and Debye for transition dipoles.}
\label{tab:loss_coefs}
\end{table*}
\begin{table*}[t!]
\centering
\begin{tabular}{c|c|c||c|c|c}
    \hline
    Base compound & Dataset & Data points & $\Delta{E}^{\dagger}$ MAE ($\downarrow$) & $\Delta{E}^{\dagger}$ $R^2$ ($\uparrow$) & \ $\vec{F}$ MAE ($\downarrow$) \\
    \hline
    \multirow{2}{*}{Azobenzene} & All data & 57 & 0.77 & 0.88 & 0.61  \\ 
    & \ Remove $\ge 5$ kcal/mol errors \ & 57 & 0.77 & 0.88 & 0.61  \\
    \hline
    \multirow{2}{*}{Azonium} & All data & 101 & 1.92 & 0.59 & 1.13  \\ 
    & \ Remove $\ge 5$ kcal/mol errors \ & 93 & 1.20 & 0.90 & 1.12  \\
    \hline
    \multirow{2}{*}{All} & All data & 158 & 1.51 & 0.73 & 0.94  \\ 
    & \ Remove $\ge 5$ kcal/mol errors \ & 150 & 1.04 & 0.93 & 0.93  \\
    \hline
\end{tabular}
\caption{TS model performance for 158 species outside the training set. We report the accuracy of $\Delta E^{\dagger} = E_{\mathrm{TS}} - E_{\mathrm{cis}}$, and of the forces $\vec{F}$ for both TS and \textit{cis} geometries. Only rotational TSs are considered. Units are kcal/mol for energies and kcal/mol/\AA \ for forces. ``Remove $\ge 5$ kcal/mol errors'' means that we remove all species for which the prediction error in $\Delta E$ exceeds 5 kcal/mol. $R^2$ is not provided for forces, because, by construction, all predicted forces are zero at critical points. }
\label{tab:model_accuracy}
\end{table*}

\subsubsection{Data generation and training}
Data was generated with the active learning loop shown in CSI Fig. \ref{fig:barriers_al}. In each round we trained a committee of three NFFs, and then generated geometries for reactants, products, and transition states (TSs) for 1,000 \textit{cis}-\textit{trans} reactions. The isomer pairs were randomly sampled from our virtual library, which included both the azo and azonium forms of each compound. The TSs were generated with one of the three NFFs using the workflow in CSI Fig. \ref{fig:barriers_al}(b). The workflow involves NN conformer generation for the reactants and products, together with a relaxed scan, conformer generation, and eigenvector following (EVF) to optimize TSs. Activation free energies $\Delta G^{\dagger}$ were computed as $G_{\mathrm{TS}} - G_{\mathrm{cis}}$. Each free energy was computed with a modified rigid rotor-harmonic oscillator (mRRHO) approximation \cite{grimme2012supramolecular} applied to Hessian vibrational frequencies, plus a conformational entropy correction \cite{chang2007ligand}. Further details can be found in Ref. \cite{axelrod2022thermal}; our only change here was to flip all negative frequencies above $-30 \ \mathrm{cm}^{-1}$, and to discard equilibrium geometries with frequencies below $-30 \ \mathrm{cm}^{-1}$, rather than zero.

Geometries from the TS workflow were then sampled for quantum chemistry calculations. We selected 10,000 geometries in each round (15,000 in the final one), from a random sample of 100,000 configurations. As in Ref. \cite{axelrod2022thermal}, 50\% of the final geometries were selected based on uncertainty, 30\% by high energy, and 20\% randomly. Uncertainty was estimated as the standard deviation of the forces computed by the three models. These calculations were added to the training set, and the models were re-trained. This was repeated five times in total. Removing outliers and spin-contaminated geometries as in Ref. \cite{axelrod2022thermal} yielded 47,313 data points in total. Combining these with 43,147 calculations of Ref. \cite{axelrod2022thermal} led to a final training set of 90,460 geometries.

The models were based on the PaiNN architecture \cite{schutt2021equivariant}, with an analytical D3-BJ dispersion correction \cite{grimme2010consistent, grimme2011effect} added to the output energies. They were trained with the settings in CSI Tables \ref{tab:training_details} and \ref{tab:loss_coefs}. We first pre-trained the models with 680,736 gas-phase SF-TDDFT \cite{shao2003spin} calculations from Ref. \cite{axelrod2022excited}. These calculations were performed for non-adiabatic molecular dynamics (NAMD) simulations. 95\% of the data was used for training, 4\% was used for validation, and 1\% for testing. The models were then trained with SF-TDDFT calculations using an implicit solvent model of water, without freezing any parameters or changing the learning rate. We used SF-TDDFT with the BHHLYP functional \cite{becke1993new}, the 6-31G* basis \cite{francl1982self}, D3-BJ dispersion \cite{grimme2010consistent, grimme2011effect}, and a C-PCM description of water \cite{truong1995new, barone1998quantum, cossi2003energies} using Q-Chem 5.3 \cite{epifanovsky2021software}.

\subsubsection{MRSF-TDDFT refinement}
We chose SF-TDDFT because it can account for some multi-reference effects \cite{lee2019conical}, which are important for the rotation mechanism in azobenzene \cite{axelrod2022thermal}. However, we found that spin contamination was often severe for the rotational TSs of azonium compounds. About 50\% of these azonium TSs had square spin $\langle S^2 \rangle > 1$. Moreover, azonium NN-EVF calculations only converged 49\% of the time, while azo calculations converged 98\% of the time. 

To fix this problem, we refined the final models with MRSF-TDDFT calculations \cite{lee2018eliminating,lee2019efficient}. MRSF-TDDFT is a spin-complete variant of SF-TDDFT, so it does not suffer from spin contamination. While the method was developed several years ago, it only recently became publicly available, and so we had not used it previously. We ran MRSF-TDDFT on 17,027 geometries using GAMESS \cite{GAMESS} (version release Sept. 30, 2022), with the same basis, functional, solvation model, and dispersion correction as in SF-TDDFT. The models were refined using the same learning parameters in CSI Table \ref{tab:training_details}, and without freezing any parameters. EVF with this final model converged for 99.3\% of azo species and 86.5\% of azonium species.

\subsubsection{Model accuracy}
To evaluate the model's performance, we used it to generate equilibrium and TS geometries for 158 species that were not in the training set. If a \textit{cis} azobenzene derivative was in the training set, then neither the \textit{trans} form nor any azonium forms could be in the evaluation set. Since a maximum of 5 TSs per species were generated from EVF (see Ref. \cite{axelrod2022thermal}), we randomly selected one TS per species to receive quantum chemical calculations. We performed MRSF-TDDFT calculations to evaluate the predicted activation energy, $\Delta E^{\dagger}$, and the forces on each geometry.

The results are given in CSI Table \ref{tab:model_accuracy}. The performance is excellent for azobenzene derivatives: the MAEs of $\Delta E^{\dagger}$ and $\vec{F}$ are 0.77 kcal/mol and 0.61 kcal/mol/\AA, respectively. The $R^2$ value for the activation energies is 0.88, which is also quite good. The performance is much worse for azonium derivatives, with MAEs of 1.92 kcal/mol and 1.13 kcal/mol/\AA \ for energies and forces, respectively.

Interestingly, however, the high azonium error for $\Delta E^{\dagger}$ mainly comes from a small number of compounds. Indeed, 8\% of all azonium species have errors in $\Delta E^{\dagger}$ that exceed 5 kcal/mol. Removing these compounds leads to a 38\% reduction in error, with the energy MAE dropping from 1.92 to 1.20 kcal/mol. This new MAE is rather good when one considers the spread in azonium activation energies. As shown in CSI Fig. \ref{fig:azonium_figures}(a), the distribution of $\Delta G^{\dagger}$ is much wider for azonium than for azobenzene (standard deviations of 4.30 and 2.95 kcal/mol, respectively). This leads to an increase in $R^2$, since the $R^2$ metric measures the error relative to the width of the underlying distribution. Hence this new azonium subset has a higher $R^2$ than azobenzene (0.90 vs. 0.88), despite also having a higher MAE. We therefore conclude that the predictions are quite good for 92\% of azonium species and quite bad for 8\%. Further, combining this 92\% subset with all azobenzene species leads to a very good $R^2$ value of 0.93.

We inspected the eight species with high prediction errors, but could not find any obvious trends in the molecules or geometries. There were no common functional groups or abnormal configurations. However, we did find that the model systematically \textit{overestimated} the barriers, with a mean signed error of 7.45 kcal/mol. We also found that re-training the model with the outlier data did not remove all outlier predictions. Therefore, it appears that there is an effect that is quite difficult to predict.

\subsubsection{Quantum chemical validation}
Hits were validated through single-point energy calculations with MRSF-TDDFT. For each species we used the TS geometry and \textit{cis} geometry with the lowest NFF free energy. The resulting DFT energies were combined with the NFF-computed mRRHO and entropic corrections to obtain the free energy.

\subsection{Protonation}
\label{sec:azonium}
\begin{figure*}[t!]
    \centering
    \includegraphics[width=\textwidth]{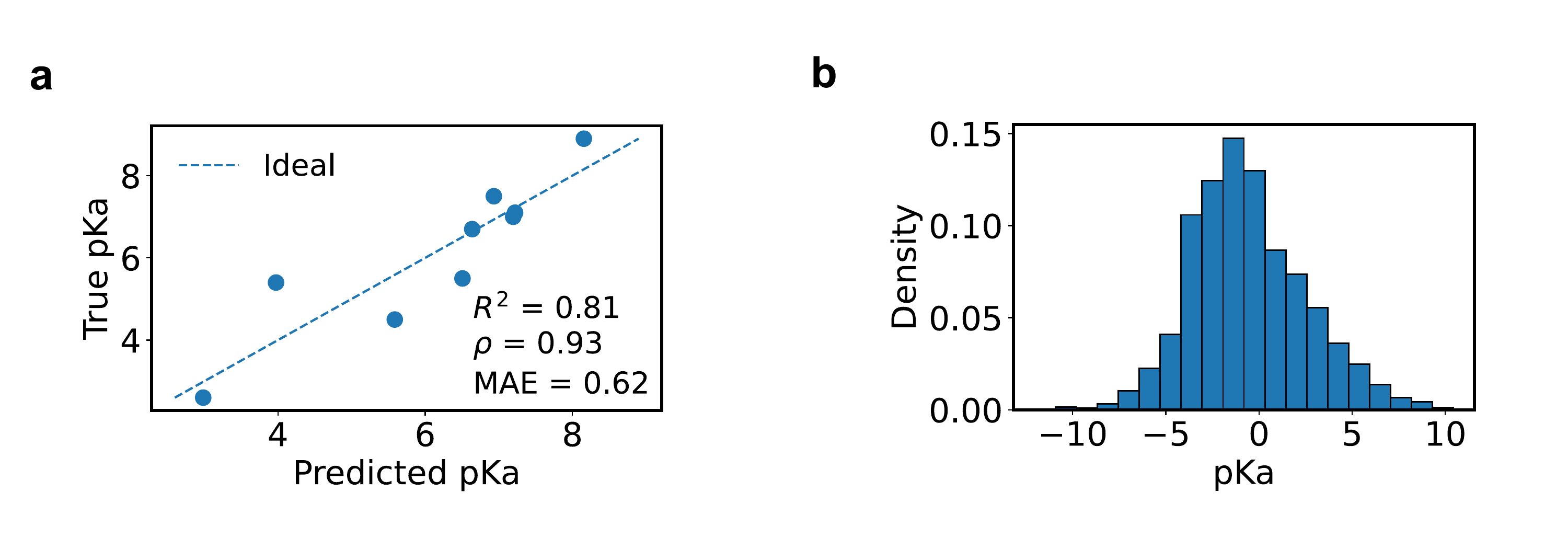}
    \caption{Accuracy of p$K_\mathrm{a}$ predictions and distribution of p$K_\mathrm{a}$ values during virtual screening. (a) Predicted vs. experimental p$K_\mathrm{a}$ values. (b) Distribution of computed p$K_\mathrm{a}$ values during virtual screening. The p$K_\mathrm{a}$ was only computed for compounds with hydrogen bond acceptors at \textit{ortho} positions.  }
    \label{fig:azonium_fit}
\end{figure*}

\subsubsection{Theory}
We computed the p$K_\mathrm{a}$ of azonium formation for both \textit{cis} and \textit{trans} isomers, as well as the free energy barrier to azonium isomerization. \textit{Trans} azonium is of interest because it is red-shifted relative to azobenzene. \textit{Cis} azonium is of interest because it is required to compute the free energy of azonium-mediated isomerization. To compute this activation barrier, we require both the azonium barrier and the \textit{cis} azonium p$K_\mathrm{a}$: 
\begin{align}
& \Delta G^{\dagger} = \left[ G_{\text{azonium TS}} + G_{(\mathrm{H}_2 \mathrm{O})_{n - 1} (\mathrm{OH}^{-}) } \right] - \left[ G_{\text{azo \textit{cis}}} + G_{(\mathrm{H}_2 \mathrm{O})_{n}  } \right] \\
& = \Delta G^{\dagger}_{\mathrm{azonium}} + \Delta G_{\text{\textit{cis} protonation}} \\
& = \Delta G^{\dagger}_{\mathrm{azonium}} + k_{\mathrm{B}} T \  \mathrm{log} [ 1 + 10^{\mathrm{pH} - \text{\textit{cis} }\mathrm{p}K_{\mathrm{a}}  }]. \label{eq:dg_eff_azonium}
\end{align}
The first bracketed term in the first line is the free energy of the azonium TS with $n - 1$ water molecules and one hydroxide. Subtracted from this term is the free energy of the \textit{cis} isomer with $n$ water molecules. The second term in the last line follows from the Henderson-Hasselbalch equation.

\subsubsection{p$K_\mathrm{a}$ implementation}
The p$K_\mathrm{a}$ for the $m$-fold protonation of a species $\mathrm{A}$ is given by \cite{pracht2018high}
\begin{align}
& \mathrm{p}K_{\mathrm{a}} = -\frac{\Delta G}{k_{\mathrm{B}} T \mathspace \mathrm{log} 10} \label{eq:ideal_pka} \\
& \Delta G = G_{\mathrm{products}} - G_{\mathrm{reactants}} \\ 
& \mathrm{reactants} = (\mathrm{H}_2 \mathrm{O})_{n} + \mathrm{A} \\
& \mathrm{products} = (\mathrm{H}_2 \mathrm{O})_{n - m} (\mathrm{OH}^{-})_{m} + \mathrm{A}(\mathrm{H}^{+})_m,
\end{align}
where $n$ is the number of water molecules. One can compute $\Delta G$ in two steps. First, one can compute the free energy of $n$ water molecules in a box, followed by the free energy of $n - m$ water molecules and $m$ hydroxide ions. One can then compute the free energy of the protonated and unprotonated forms of A. Supplying the free energy difference to Eq. (\ref{eq:ideal_pka}) then yields the p$K_\mathrm{a}$. However, quantum chemical free energy differences are typically not accurate enough to reproduce experimental data. Therefore, Eq. (\ref{eq:ideal_pka}) is usually passed through a linear regression that is fit to experiment \cite{pracht2018high}. Computing free energies with double-hybrid DFT applied to geometries from semi-empirical DFT, and fitting to experiment with a linear regression, has yielded $R^2$ values above 0.93 \cite{pracht2018high}.

In this work the calculation is further simplified because $m = 1$ for all compounds. This means that the change in free energy of water is constant for all species. Hence it can be absorbed into the intercept during linear regression, and does not need to be calculated. We simply calculated $\Delta G = G_{\mathrm{azonium}} - G_{\mathrm{azo}}$ and applied a linear regression to experiment to get the p$K_\mathrm{a}$.
\begin{figure*}[t!]
    \centering
    \includegraphics[width=\textwidth]{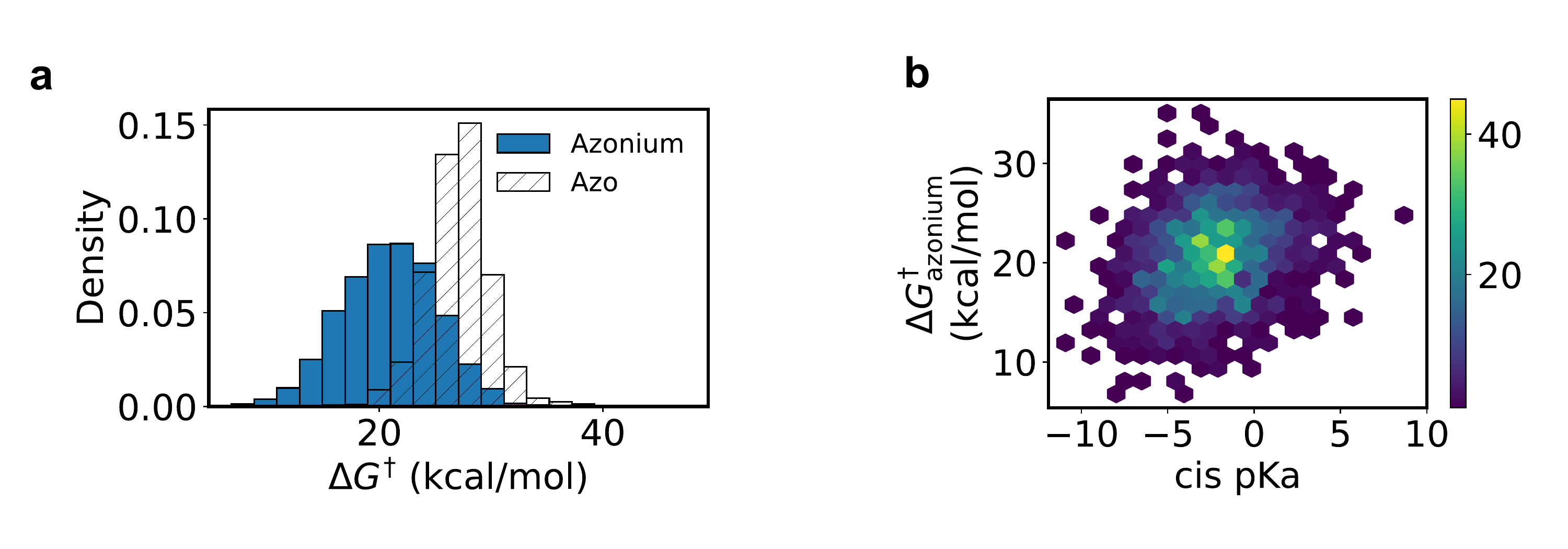}
    \caption{Distribution of computed azonium properties. These calculations were only performed for azonium compounds with \textit{ortho} hydrogen bond acceptors. (a) Comparison of the activation free energies of azo and azonium compounds. (b) Activation free energy vs. \textit{cis} p$K_\mathrm{a}$ of azonium species. }
    \label{fig:azonium_figures}
\end{figure*}

The free energies of the azo and azonium species were computed with our NN. We used the mRRHO approximation \cite{grimme2012supramolecular} applied to Hessian vibrational frequencies, plus a conformational entropy correction \cite{chang2007ligand}, as described in Ref. \cite{axelrod2022thermal}. Free energy differences were fit to nine experimental p$K_\mathrm{a}$ measurements of azobenzene derivatives. Five came from Ref. \cite{dong2017near} and four from Ref. \cite{dong2015red}. We excluded compound \textbf{20} from Ref. \cite{dong2017near}, since it was predicted by OpenBabel to have protonated substituents. We did not want this to cause problems in the p$K_\mathrm{a}$ fit if the OpenBabel prediction was incorrect. We also removed the glutathione groups attached to the ends of the functional groups in Ref. \cite{dong2015red}. This is because glutathione is very flexible, which can lead to unconverged free energy estimates. It is also not representative of molecules in the training set. Both of these issues could have led to inaccuracies in the ML p$K_\mathrm{a}$ predictions.

The p$K_\mathrm{a}$ predictions are compared to experiment in CSI Fig. \ref{fig:azonium_fit}(a). The results are quite good. $R^2$ is greater than 0.8, the Spearman rank correlation coefficient is above 0.9, and the MAE is only 0.62. Note that, while there is a fairly even distribution of p$K_\mathrm{a}$ values from 3 to 9 in the experimental data, this is unusual for the compounds screened here. Indeed, as shown in CSI Fig. \ref{fig:azonium_fit}(b), only 13\% of compounds screened had a p$K_\mathrm{a}$ value above 3.

\subsubsection{Barrier implementation}
Azonium activation free energies were computed as in Ref. \cite{axelrod2022thermal}, but with one notable change. In Ref. \cite{axelrod2022thermal}, we generated rotational TSs by performing a relaxed scan that forced the CNNC dihedral to $\pm 90^{\circ}$. We also forced the CNN angles to $122^{\circ}$, so that we would not generate an inversion TS. However, when we did this for azonium derivatives, we found that only 10\% of EVF jobs converged. In successful optimizations, we found that the azonium nitrogen formed a planar improper dihedral with its neighbors. In \textit{cis} and \textit{trans} equilibrium geometries, this improper dihedral was non-planar. We therefore added a third constraint to the scans, forcing the $\mathrm{N[NH^+]C}$ improper dihedral to $180^{\circ}$. Further, we held the azonium hydrogen atom fixed during TS conformer generation, in addition to the CNNC atoms held fixed in Ref. \cite{axelrod2022thermal}. The total activation free energy of proton-mediated isomerization was computed with Eq. (\ref{eq:dg_eff_azonium}).

\subsubsection{Virtual screening}
We computed azonium barriers and p$K_\mathrm{a}$ values for all species with hydrogen bond acceptors at the \textit{ortho} positions, or one atom away. We also did so for any species predicted to have p$K_\mathrm{a}$ $> 3$ by the graph-to-property model. To compute the p$K_\mathrm{a}$ for a given species, we first checked that free energies had been calculated for all azonium forms (i.e. the $\mathrm{N[NH^+]}$ and $\mathrm{[NH^+]N}$ forms for asymmetrically substituted compounds, which are equivalent for symmetric substitutions). We also checked that all vibrational frequencies were above $-30 \ \mathrm{cm}^{-1}$, and discarded the species if they were not. We flipped the signs of negative frequencies above this value. We also removed any geometries with broken bonds, as measured through the D3 coordination number described in Ref. \cite{axelrod2022thermal}. The azonium free energy was approximated as the lower of the free energies of the two azonium isomers. Compounds with simulated \textit{trans} p$K_\mathrm{a}$ $> 5$ and thermal half-lives in the target range were flagged as potential hits.

\subsubsection{Quantum chemical validation}
\label{subsec:pka_validation}
Hits were validated through single-point energy calculations with MRSF-TDDFT. For each species we used the geometry with the lowest NFF free energy. The resulting DFT energies were combined with the NFF-computed mRRHO and entropic corrections to obtain the free energy. This was in turn used to compute the p$K_\mathrm{a}$.

For some species, the EVF calculation failed to converge for one of the two azonium forms. 
Typically EVF failed for the azonium form with the higher p$K_\mathrm{a}$. To estimate the free energy cost of this mechanism [Eq. (\ref{eq:dg_eff_azonium})], we used the p$K_\mathrm{a}$ of this azonium isomer with the barrier of the other isomer (i.e., the one for which EVF succeeded) in CSI Table \ref{tab:dg_data}.

To handle the failed EVF calculations, we also tried replacing NFF conformer generation with CREST \cite{pracht2020automated} using GFN2-xTB \cite{bannwarth2019gfn2}. As in NFF conformer generation, we fixed all $\mathrm{CN[NH^+]C}$ atoms. We inspected the TS geometries generated by NFF and CREST, and found that both looked qualitatively wrong. EVF with the NFF failed for all TS guesses from CREST. Therefore, it seems that our general approach to TS generation caused the failure, not the NFF itself.

\section{Photophysical properties}
\subsection{Absorption spectrum}
\label{sec:absorption}
\subsubsection{Simulation}
We \cite{axelrod2023mapping} and others \cite{gelabert2023predicting} have shown that it is important to consider the full absorption spectrum of azobenzenes, not simply the absorption wavelength of an optimized geometry. Azobenzene derivatives that absorb in the near-IR have absorption peaks in the visible \cite{dong2017near}; it is the high-wavelength tail of their spectra that allows for near-IR absorption. Moreover, dynamical effects can lead to asymmetric absorption spectra with long tails that could not be anticipated by single-point calculations \cite{axelrod2023mapping}. 

We therefore simulated the full absorption spectrum of the azo and azonium forms of each \textit{trans} compound. To do so we performed ground-state MD, and computed the absorption spectrum $I$ as the weighted histogram of absorption frequencies at each frame \cite{marenich2015electronic}:
\begin{align}
& I(\omega) \propto \mathrm{histogram} ( \{ f_i \mathspace \omega_i \}_{i \sim \mathrm{MD}  }  ), \\
& f_i \propto \omega_i \vert \vec{\mu}_i \vert^2.
\end{align}
Here $\omega$ is the frequency, $\omega_i$ is the absorption frequency of frame $i$ sampled from MD, and $f_i$ is its oscillator strength. The latter is proportional to the absorption frequency and the square of the transition dipole moment $\vec{\mu}_i$. We discarded absorption energies outside of (1.3, 6.2) eV, and created histograms with 30 bins. Frequency spectra were then converted to wavelength spectra. This was done by first converting frequencies to wavelengths, and then multiplying by the Jacobian of the transformation, which is proportional to $\omega^2$.

MD was initialized with conformers from NN conformer generation \cite{axelrod2022thermal}. We performed 5 ps of MD per conformer and used 20 conformers per molecule, yielding 100 ps in total. Conformer $i$ was sampled from the conformer ensemble with probability $p_i \propto \mathrm{exp}(-E_i / k_{\mathrm{B}} T)$, where $E$ is the energy. We performed MD with the Nos\'e-Hoover thermostat \cite{nose1984unified, hoover1985canonical} and our trained NFF using the same MD parameters as in Ref. \cite{axelrod2022thermal}. 

\subsubsection{Model}
Absorption energies and transition dipole moment vectors were predicted with a single model. The model used the same PaiNN architecture as the ground state model. However, instead of outputting a single energy, it outputted the transition energy and a vector representing the transition dipole moment. The D3 correction was not applied to the output energy, since the D3 corrections of the ground- and excited-state energies cancel. The vector was produced by applying equivariant operations to the vector features $\vec{\mathbf{v}}_i$ generated in the message-passing stage, as described in Ref. \cite{schutt2021equivariant}.

\subsubsection{Training data}
Training data was produced with TDDFT using the CAM-B3LYP functional \cite{yanai2004new}, the def2-SVP basis \cite{weigend2005balanced}, a C-PCM model of water \cite{truong1995new, barone1998quantum, cossi2003energies}, and D3 dispersion \cite{grimme2010consistent}. As described below, we trained on transition dipole moments, transition energies, and transition energy gradients. Since the gradient of the transition energy is the difference of two energy gradients, we computed both $S_0$ and $S_1$ energy gradients with CAM-B3LYP. 

Along with the PBE0 \cite{adamo1999toward} functional, CAM-B3LYP is among the most reliable functionals for TDDFT vertical excitation energies of organic molecules \cite{jacquemin2009extensive}. Previously we used the PBE0 functional, and found that it reproduced the absorption spectrum of azobenzene quite well \cite{axelrod2023mapping}. However, we used CAM-B3LYP here because, as a range-separated hybrid, it is presumably more accurate at describing excitations with charge-transfer character. It is therefore more likely to be reliable across the space of azobenzene derivatives. We have also found that it reproduces the absorption spectrum of azobenzene well (see Methods).

Training geometries were sampled randomly from metadynamics simulations of equilibrium \textit{cis} and \textit{trans} configurations. These were performed during NN conformer generation in active learning. To increase the number of red-shifted geometries in the training set, we first trained an initial model on these random geometries, and then used it to predict the absorption energies of geometries without TDDFT labels. We then selected the geometries with the highest predicted absorption wavelengths and performed TDDFT on them. We used 83,481 geometries in total.

\subsubsection{Training}
Parameters used for training are given in CSI Tables \ref{tab:training_details} and \ref{tab:loss_coefs}. We trained on both transition energies and their gradients to improve transition energy predictions. Previously we found that training on transition energies alone led to disappointing performance for red-shifted species \cite{axelrod2022thermal}. 

One training complication is that quantum chemical transition dipole vectors have arbitrary signs. This is because the wavefunction of each electronic state has an arbitrary sign. These cancel for properties of a given state, since matrix elements are taken with respect to the same state, but do not cancel for transition properties. Therefore, as we did previously for non-adiabatic couplings \cite{axelrod2022excited}, we multiplied the model output by $\pm 1$ for each geometry in each training batch. The sign was chosen separately for each geometry to minimize the prediction error for that configuration.

\subsubsection{Model performance}
CSI Table \ref{tab:excited_model_accuracy} shows the model accuracy for 50 species outside of the training set. 10 geometries were sampled from NN-driven metadynamics for each compound. The performance is excellent for both azobenzene and azonium derivatives. The error in the absorption wavelength is 3.5 nm for azobenzenes and 7.0 nm for azonium species. Both are significantly lower than typical errors in TDDFT itself \cite{greenman2022multi}. The $R^2$ value is above 0.97 for both the absorption wavelength and the transition dipole moment for all groups of compounds.

\subsubsection{Quantum chemical validation}
Hits were validated with TDDFT by computing the absorption wavelength of the lowest energy conformer. The difference between the TDDFT result and the model result was used to shift the NFF-computed spectrum.

\subsection{Quantum yield}

\label{sec:qy}
\begin{table*}[t!]
\centering
\begin{tabular}{c|c||c|c||c|c}
    \hline
    Base compound & Data points & $\lambda$ MAE ($\downarrow$) & $\lambda$ $R^2$ ($\uparrow$) & $\vec{\mu}$ MAE ($\downarrow$) & $\vec{\mu}$ $R^2$ ($\uparrow$) \\
    \hline
    Azobenzene & 170 & 3.51 & 0.98 & 0.05 & 0.98  \\
    Azonium & 329 & 7.03 & 0.97 & 0.16 & 0.98  \\
    All & 499 & 5.83 & 0.98 & 0.12 & 0.98  \\
    \hline
\end{tabular}
\caption{Performance of the excited state absorption spectrum model for species outside the training set. $\lambda$ is the absorption wavelength and $\vec{\mu}$ is the transition dipole vector. $\vec{\mu}$ was multiplied by $\pm 1$ for each geometry to minimize the error relative to DFT. Units are nm for $\lambda$ and Debye for $\vec{\mu}$. }
\label{tab:excited_model_accuracy}
\end{table*}

Certain substituents can inhibit photoisomerization, e.g. through hydrogen bond-induced locking \cite{bandara2010proof} or other mechanisms \cite{bandara2011short}. We therefore performed NN excited-state dynamics on our top candidates to predict whether they would isomerize under light. Previously we trained an excited-state NFF across the chemical space of azobenzene derivatives \cite{axelrod2022excited}. Combining it with surface-hopping MD \cite{tully1990molecular}, we showed that it can predict quantum yields that are in moderate agreement with experiment. We developed an architecture based on diabatic electronic states \cite{mead1982conditions}, and used it to improve the accuracy of NN fitting and dynamics for unseen species.

Here we applied a similar workflow to our top candidates. We started with our pre-trained model from Ref. \cite{axelrod2022excited}, which was trained on over 600,000 calculations from over 8,000 species. We then refined it with quantum chemistry data from the top molecules. We performed active learning similar to CSI Fig. \ref{fig:barriers_al}. We ran non-adiabatic excited state MD with the NFF, and then chose new geometries for quantum chemistry calculations either randomly or based on uncertainty, the $S_0$/$S_1$ energy gap, or excited state barriers. Further details on these sampling techniques can be found in Ref. \cite{axelrod2022excited}. We began with active learning for unsubstituted azobenzene while we waited for screening results. We then focused on our top 20 candidates, performing calculations on geometries from equilibrium and TS metadynamics for each of the hits. We then sampled 250 geometries per species in each round of active learning, and removed non-hits from the training set as FEP results came back. The final dataset had 6,985 geometries.

Analytic gradients are implemented for MRSF-TDDFT, but analytic non-adiabatic coupling vectors (NACVs) are not. We therefore could not use our diabatic architecture, since it requires NACVs for training. Instead we used a simple adiabatic model that predicts ground- and excited-state energies as separate outputs. We used the same dispersion-corrected PaiNN architecture as for thermal barriers, but with two energy outputs instead of one. We generated training data using MRSF-TDDFT with the 6-31G* basis \cite{hehre1986p}, the BHHLYP functional \cite{becke1993new}, and Grimme's D3-BJ dispersion correction \cite{grimme2010consistent, grimme2011effect}. We used a C-PCM model for water \cite{truong1995new, barone1998quantum, cossi2003energies}. Details of the training process are given in CSI Tables \ref{tab:training_details} and \ref{tab:loss_coefs}. Performance statistics are given in CSI Table \ref{sm_tab:namd_accuracy}. 

Surface hopping was performed with the Zhu-Nakamura method \cite{yu2014trajectory}, which requires only energies and gradients, not NACVs. Initial geometries were sampled from 15 ps of MD at 298.15 K in the NVT ensemble. 200 surface hopping trajectories were then performed for 4 ps each. Hops were restricted to geometries with $S_1$/$S_0$ gaps below 0.5 eV. The quantum yield was computed as the proportion of trajectories that isomerized, excluding those that stayed in the excited state. Further details of our implementation can be found in Ref. \cite{axelrod2022excited}, and the code is available at \url{https://github.com/learningmatter-mit/NeuralForceField}.
\begin{table*}[t]
\centering
    \begin{tabular}{c|c|ccccccccc}
        \hline
        Sampled by & Data points & Metric &  $E_0$ & $E_1$ & $\Delta E_{01}$ & $\vec{F}_0$ & $\vec{F}_1$  \\ 
        \hline
        ZN & 150 & MAE ($\downarrow$)  & 0.57 & 0.80 & 0.71 &  1.55 & 1.60   \\
        && $R^2$ ($\uparrow$) & 1.00  & 0.99 & 0.90 & 0.98 & 0.97   \\
        \hline
        Barrier & 149 & MAE ($\downarrow$) & 0.63 & 0.69 & 0.62 & 1.02 & 1.08  \\
        && $R^2$ ($\uparrow$) & 1.00 & 1.00 & 0.99 & 1.00 & 0.99  \\
        \hline
        Random & 214 & MAE ($\downarrow$) & 0.60 & 1.46 & 1.59 & 1.17 & 2.00 \\
        && $R^2$ ($\uparrow$) & 1.00 & 0.99 & 0.95 & 0.99 & 0.95  \\
        \hline
        Uncertainty & 208 & MAE ($\downarrow$) & 1.01 & 2.94 & 2.92 & 2.65 & 7.04   \\
        && $R^2$ ($\uparrow$) & 1.00 & 0.98 & 0.91 & 0.79 &  $-0.48$ \\
        \hline
    \end{tabular}
    \caption{Performance of the quantum yield model. Geometries were sampled from NAMD performed with the final trained model. Results are divided by the method used to select samples. ``ZN'' refers to Zhu-Nakamura sampling to select low-gap geometries, ``barrier'' refers to sampling of excited-state barriers, and ``uncertainty'' refers to sampling based on model uncertainty. Details of these sampling methods can be found in Ref. \cite{axelrod2022excited}. Energies are given in kcal/mol, and forces are given in kcal/mol/\AA. }
    \label{sm_tab:namd_accuracy}
\end{table*}
 
\begin{table*}[t!]
\centering
\begin{tabular}{c||c|c||c|c||c|c}
    \hline
    Iteration & $\Delta G^{\dagger}$ MAE ($\downarrow$) & $\Delta G^{\dagger}$ $R^2$ ($\uparrow$) & $\lambda$ MAE ($\downarrow$) & $\lambda$ $R^2$ ($\uparrow$) & p$K_\mathrm{a}$ MAE ($\downarrow$) & p$K_\mathrm{a}$ $R^2$ ($\uparrow$)  \\
    \hline
    1 & 2.4 $\pm$ 0.2 & 0.54 $\pm$ 0.05 & 21 $\pm$ 1 & 0.60 $\pm$ 0.04 & 1.38 $\pm$ 0.03 & 0.62 $\pm$ 0.03 \\
    2 & 2.3 $\pm$ 0.1 & 0.58 $\pm$ 0.04 & 16 $\pm$ 1 & 0.74 $\pm$ 0.03 & 1.49 $\pm$ 0.07 & 0.70 $\pm$ 0.03 \\
    \hline
\end{tabular}
\caption{Performance of the graph-to-property models after each round of exploration and exploitation. We report the mean and standard deviation of each metric from three models. Units are kcal/mol for $\Delta G^{\dagger}$ and nm for $\lambda$. }
\label{tab:graph_to_property}
\end{table*}
\begin{table*}[t!]
\centering
\begin{tabular}{c|c|c|c}
    \hline
    Iteration & $\Delta G^{\dagger}$ & $\lambda$ & p$K_\mathrm{a}$  \\
    \hline
    1 & 4,855 & 6,109 & 4,500  \\
    2 & 9,207 & 11,825 & 9,947 \\
    \hline
\end{tabular}
\caption{Combined size of the training, validation, and test set for graph-to-property models in each iteration. We used a train/validation/test split of 0.9/0.05/0.05. }
\label{tab:graph_to_property_dataset_size}
\end{table*}

\section{Graph-to-property models}
We trained graph-to-property models to predict isomerization barriers, absorption wavelengths, and p$K_\mathrm{a}$ values generated by our NFF. We used the Attentive FP architecture for each model \cite{xiong2019pushing}. This architecture combines an attention mechanism \cite{noam_lr} with neural network message-passing \cite{gilmer2017neural} to generate an embedding for each atom. These node embeddings are then converted to graph embeddings with attention-based pooling, before being mapped to the final property with a neural network. This model substantially improves on previous state-of-the-art results for nearly all properties in standard benchmarks \cite{xiong2019pushing}.

We trained an ensemble of three models for each property. Barrier models were trained on activation free energies of both azo and azonium compounds. Absorption models were trained on the maximum absorption wavelength at 5\% of the peak intensity. We converted these wavelengths to frequencies in eV before training, which we found to significantly improve model performance. The p$K_\mathrm{a}$ models were trained on per-azonium p$K_\mathrm{a}$ values. That is, they were trained to map the azonium SMILES to the p$K_\mathrm{a}$ associated with that azonium isomer. There were usually two azonium isomers per azo compound, for the two nitrogen protonation sites, unless they were equivalent by symmetry. 

Models were trained with an initial learning rate of $10^{-3}$ for a maximum of 1,000 epochs. The learning rate was dropped by a factor of two if the validation loss had not improved in 10 epochs. Training was stopped if the learning rate fell below $10^{-5}$. The best model was chosen as the one with the lowest validation loss. We used a batch size of 16 and train/validation/test splits of 0.9/0.05/0.05. Models were created and trained using the DeepChem library \cite{Ramsundar-et-al-2019}.

Model performance is shown in CSI Table \ref{tab:graph_to_property}. The models are reasonably accurate, with $R^2$ values between 0.54 and 0.74 depending on the property and generation. The $R^2$ value increased for each model going from iteration 1 to 2. MAEs were around 2 kcal/mol for $\Delta G^{\dagger}$, 15 to 20 nm for $\lambda$, and 1.5 p$K_\mathrm{a}$ units for p$K_\mathrm{a}$. Dataset sizes are given in CSI Table \ref{tab:graph_to_property_dataset_size}.

\bibliography{main}

\end{document}